\documentclass[useAMS,usenatbib,usegraphicx]{mn2e}
  
\usepackage{aas_macros}
\usepackage{threeparttable}
\usepackage{grffile}
\usepackage{wallpaper}
\usepackage{mathbbol}
\usepackage{rotating}

\usepackage{tocloft}
\newcommand{\vek}{\underline}
\newcommand{\fat}{\textbf}
\newcommand{\ita}{\textit}

\newcommand{\beq}{\begin{equation}}
\newcommand{\eeq}{\end{equation}}  
\newcommand{\RNum}[1]{\uppercase\expandafter{\romannumeral #1\relax}} 
         
\title[MCs and magnetic fields]{Impact Of Magnetic Fields On Molecular Cloud Formation \& Evolution}
  \author[K\"ortgen \& Banerjee]
  {Bastian~K\"ortgen\thanks{bkoertgen@hs.uni-hamburg.de} and Robi~Banerjee  \\
  Hamburger Sternwarte, Universit\"at Hamburg, Gojenbergsweg 112, 21029 Hamburg, Germany\\
  }

\date{Released 2015}

\pagerange{\pageref{firstpage}--\pageref{lastpage}} \pubyear{2015}

\begin{document}

\label{firstpage}

\maketitle

\begin{abstract}
We use magnetohydrodynamical simulations of converging flows to
investigate the process of molecular cloud formation and evolution out
of the magnetised ISM. Here, we study whether the observed {\em subcritical} HI clouds can become 
supercritical and hence allow the formation of stars within them. To do so, we vary the turbulent Mach number of the 
flows, as well as the initial magnetic field strength. We show that dense cores are able to build up under all conditions, but that star formation in these cores is either heavily delayed or 
completely suppressed if the initial field strength is $B>3\,\mu\mathrm{G}$. To probe the effect of magnetic diffusion, we introduce a tilting angle $\varphi$ between the flows and the uniform background magnetic 
field, which mimics non--ideal MHD effects. Even with highly diffusive flows, the formed cores are devoid of star formation, because no magnetically supercritical regions are build up.  
Hence we conclude, that the problem of how supercritical cloud cores
are generated still persists. 
\end{abstract}

\section{Introduction}
Stars and stellar systems form within the densest regions of molecular clouds, in gravitationally unstable cores which reside at the junctions of filaments \citep[e.g.][]{Andre13,Andre14}. Prior to 
gravitational collapse the build--up of 
filaments and the respective substructures is primarily controlled by
magnetic fields and supersonic turbulence
\citep[e.g.,][]{Shu87,MacLow04,Crutcher10}. But, the importance of
magnetic fields for star formation is still debated \citep[see
e.g.,][]{Li14, Padoan14}. 
On the one hand, the idea of supersonic turbulence controlling the star formation process assumes less important magnetic fields and thus primarily supercritical states. In such a scenario, the magnetic field lines are 
dragged along with the flow and density enhancements will collapse as soon as they become Jeans unstable. Furthermore, the turbulence is then not only supersonic but also \ita{superalfv\'{e}nic} \cite[e.g.,][]{Padoan99,Padoan99b}. This leads to highly 
twisted field lines and the resulting molecular clouds and clumps will not be coherent entities. The morphology instead 
will be influenced by the statistics/nature of the turbulence.\\
On the other hand, \citet[][]{Mestel56} first quantified 
the influence of magnetic fields on star formation by introducing the mass--to--magnetic flux ratio $\mu\equiv M/\Phi$ as a measure of the relative importance of gravitational and magnetic energies. Usually, this quantity 
is normalised to its critical value $\mu_c\simeq 0.13/\sqrt{G}$ (or $\mu_c\simeq 0.16/\sqrt{G}$ for more sheet--like clouds\citep[][]{Nakano78}). If the magnetic field is strong enough, accretion onto the cloud complex is mediated by the Lorentz force and mainly parallel to the field lines \citep[e.g.][]{Kudoh07,Inoue08,Kudoh10,Hennebelle13}.
In the cloud interior, strong fields stabilise the filaments and clumps against gravity. This also results in a reduced fragmentation efficiency.\\ 
Observationally, it has been shown in recent years that the magnetic field is indeed crucial for the star formation 
process  \citep[e.g.,][]{Beck01,Crutcher10,Li10,Crutcher12,Li14,Pillai14}. 
\citet[][]{Li10} used sub--mm polarisation measurements to retrieve the morphology of the magnetic field in molecular clouds and Galactic spiral arms. The authors have shown that the overall morphology of the field does not 
change significantly from the large scales down to the inner parts of molecular clouds. By using HI, OH, and CN Zeeman measurements \citet[][see also \cite{Crutcher12},fig. 7]{Crutcher10} have shown that nearby 
molecular clouds and cloud cores can be separated into two regimes according to their column density and magnetic field strength. At low column densities, the magnitude of the field almost does not change for roughly 
two orders of magnitude with a median value of $B_{LOS}\approx 5-6\,\mu\mathrm{G}$. This regime also coincides with magnetically subcritical HI clouds. As was pointed out by \citet[][and references therein]{Crutcher12}, these data are 
primarily diffuse HI clouds that are \ita{not self--gravitating}, but are rather in pressure equilibrium with their 
surroundings. Note, that the total magnetic field strength will even
be larger. E.g. recent studies give average values of $5-15\,\mu\mathrm{G}$ \citep[][]{Beck01,Crutcher10}.\\
At higher column densities the field strength increases close to linear. 
with increasing column density. At this stage, almost all 
measurements indicate (super-)criticality. Conversion of these data points to volume density shows that the scaling in the latter region is $B\propto n^{0.65}$, which perfectly fits to conditions of frozen--in magnetic field 
lines and isotropic collapse (which gives a relation $B\propto n^{0.66}$).\\ 
Numerically, the issue of magnetic fields and their relevance for molecular cloud formation has been investigated by many authors \citep[e.g.,][]{Inoue08,Price08,Price09,Kudoh10,Vazquez11a,Inoue12,Chen14}. Most of them 
concentrated on the initial stages of the formation process. Already at this early temporal stage, the magnetic field was shown to be crucial. \citet[][]{Price08} conducted smoothed particle hydrodynamics simulations of a 50 M$_\odot$ 
molecular cloud of radius $R=0.375$pc including magnetic fields of different strength (parameterised by critical mass--to--flux ratios of $\infty$, 20, 10, 5, and 3). They found that strong fields tend to suppress fragmentation on the one hand and the formation of stars on the other hand. However, as they point out, strong fields generate voids 
within the molecular cloud, which are magnetically supported with plasma--$\beta>1$. In addition, 
on very small scales, magnetic tension is able to prevent multiple fragments from merging, thus \ita{promoting}
 fragmentation. \\
More consistent with this study is the work by \citet[][]{Heitsch09a}. They have used MHD simulations of converging flows 
to analyse the impact of magnetic field strength and orientation on the formation of (molecular) clouds. Specifically, 
they looked at the extreme cases of the magnetic field being either aligned with or perpendicular to the flow direction. 
The flows were driven continuously due to the choice of inflow boundary conditions. Hence, the mass--to--flux ratio in 
their study would approach infinity in the limit of infinite timescales. Note that the authors have not included 
self--gravity in their simulations. Thus, every overdense substructure is pressure confined. However, they identify 
filaments and clumps that form due to turbulent compression, with clouds becoming more filamentary if magnetic fields 
are included. But, it is the alignment of the magnetic and (initial) velocity field that controls the formation of dense 
structures. As the authors point out, clouds are able to condense out of the WNM, if the fields are aligned. In case the 
magnetic field is perpendicular to the inflows, magnetic pressure suppresses the formation of dense structures, which 
could be termed molecular. However, there exist regions, which merge to form a filamentary network of \ita{diffuse} gas.\\
\citet[][]{Inoue09} studied the evolution of the shocked slab between two converging flows in the ISM by means of 
two--fluid MHD simulations in a $30\,\mathrm{pc}\times 10\,\mathrm{pc}$ box. The authors varied the angle 
between the mean magnetic field and and the flows. From 
analytical estimates they found a critical velocity, which depends on the magnetic field strength and the mentioned angle.
If the flow velocity is larger than the critical velocity only H\,I clouds are able to form because of dominating magnetic 
pressure. If it is less than the critical velocity, dense molecular clouds condense out of the WNM within the shocked slab. 
As the authors also point out, the dependence on the angle is crucial for the evolution of the gas within the slab, since 
the critical velocity goes to zero for angles approaching 90$^{\circ}$.\\
Most recently, \citet[][]{Chen14} studied the formation of prestellar cores due to the convergence of gas flows within 
molecular clouds. In detail, their simulation box was about 1 pc, representing a collapsing molecular clump. In order to 
analyse the core formation process, they used ideal MHD as well as non--ideal MHD via ambipolar diffusion (AD). 
In all of their models core formation was initiated by the collision of gas streams along the background magnetic field.
With AD only the later stages were seen to differ from the ideal MHD models, since the density regimes 
where AD is becoming efficient are build up via accumulation of gas by colliding flows. The mass--to--flux ratio of the 
cores formed in their simulations is in the range $\mu/\mu_c\sim0.5-7.5$ with a median value of $\mu/\mu_c\approx3$. 
Thus, most of the cores are supercritical\footnote{Note that the authors use inflow boundary conditions for the two 
converging flows. Hence, the mass--to--flux ratio of the simulation domain will grow with time and so it will for the cores as they accrete mass from an practically infinite mass reservoir.}.\\
However, it is important to conduct large scale simulations in order to take into account the whole evolutionary track of the 
gas from the diffuse ISM to the dense cores. This was achieved by \citet[][]{Vazquez11a}, who analysed molecular cloud formation subject to magnetic fields of different initial strength. It has been shown that 
stronger fields tend to delay the onset of star formation. 
The authors \citep[see also][]{Hartmann01,Vazquez06,Vazquez07} point out that the diffuse gas becomes molecular, self--gravitating \ita{and} magnetically supercritical 
\ita{at the same time}. This simple approach can explain the subcriticality of the diffuse HI clouds shown in 
\citet[][]{Crutcher10,Crutcher12}. \citet[][]{Heitsch14} mention that the supercritical state can be reached via gas accretion 
along the magnetic field lines. 
But as was stated by \citet[][]{Hartmann01}, the accumulation length to become 
magnetically supercritical is 
\beq
L_\mathrm{c}\approx 470\left(\frac{B_0}{5\,\mu\mathrm{G}}\right)\left(\frac{n}{1\,\mathrm{cm}^{-3}}\right)^{-1}\mathrm{pc}.
\eeq
\citet[][]{Vazquez11a} argue that, since the magnetic field lines in the Galactic plane describe closed circles, this length 
scale is easily overcome. This also indicates that the mass--to--flux ratios are lower limits and the data points shown in 
\citet[][]{Crutcher10} are only a temporal stage of subcriticality. But, as \citet[][]{CN14} point out, flow lengths of 
$L>500\,$pc are too large in order to sustain a large scale coherent flow. Bulk motions of this order of magnitude should rather fragment 
due to supersonic turbulence and thus diminish. Hence, the build--up of supercritical clouds would be delayed or 
even suppressed completely. The process, how molecular clouds achieve the transition from sub-- to supercritical states 
is thus still an open question.\\
In this study we tie in with the work of \citet[][]{Vazquez11a} by determining molecular cloud formation under different initial conditions. In section \ref{sec2} we therefore introduce our numerical model and the initial conditions. Section 
\ref{sec3} deals with the formation and evolution of clouds formed by head--on converging WNM streams under varying initial conditions. The following section \ref{sec4} then introduces the tilt of one flow with 
respect to the magnetic field and discusses in detail the evolution of the clouds and their subsequent star formation activity. This study is closed by a brief summary in section \ref{sec5}.

\section{Numerical setup and initial conditions}
\label{sec2}
\subsection{Details of the numerics}
For this study we use the finite volume AMR code FLASH \citep{FLASH00,Dubey08}. During each timestep a Riemann problem is solved at the cell interfaces, yielding the respective fluxes for the hyperbolic partial differential equations. The MHD fluxes are computed by a multiwave 
Riemann solver developed by \citet[][]{Bouchut07,Bouchut09} and implemented in FLASH by \citet[][]{Waagan11}, which preserves positive states for density and internal energy.\\
We apply periodic boundary conditions for the (magneto--)hydrodynamics and isolated ones for gravity.
 In addition to the basic (ideal) MHD equations, we include selfgravity as well as heating and cooling. 
The latter is treated as a source term in the energy equation and we follow the recipe by \citet[][with modifications by \cite{Vazquez07}]{Koyama00} for the radiative cooling of the gas. Since we are interested in the 
process of star formation, we also include sink particles to follow truely collapsing regions \citep{Federrath10} and the local Jeans length is resolved with at least ten grid cells to fulfill the Truelove criterion \citep{Truelove97}. In order to replace a certain gas volume by a sink particle, the gas has to pass several checks, which are described in great detail in \citet[][]{Federrath10}. We use a density 
threshold of $n\approx2\times10^6\mathrm{cm}^{-3}$. Once a sink particle has been created, it is only allowed to 
accrete gas. Feedback is \ita{not} included.\\
The numerical grid is refined when the local Jeans length is resolved with less than ten grid cells and derefined if it 
consists of more than 100 cells.
\subsubsection{Treatment of Ambipolar Diffusion}
We have conducted one simulation including the non--ideal MHD effect of ambipolar diffusion (AD). The AD module 
was implemented in FLASH and extensively tested by \citet[][]{Duffin08}. It uses the strong coupling approximation 
(like \citet[][see also \cite{Chen14}]{Chen12}), which was shown to be valid in the physical regime we are analysing \citep[see appendix in][]{Vazquez11a}. However, since 
we are using a slightly different density threshold than in \citet[][]{Vazquez11a} and different turbulent Mach numbers, the 
validity has to be proven again:\\
 Taking a typical length scale of $l=0.0625$ pc (which corresponds to the accretion/softening 
radius of the sink particles in our simulation with 11 levels of refinement) and a typical velocity at these scales 
of 0.5 km/s (taken from fig. \ref{fig2}) the ratio 
\beq
M_\mathrm{A}^2/R_\mathrm{AD}(l)\approx 1.5\times 10^{-8}.
\eeq
Here $M_\mathrm{A}$ is the Alfv\'{e}n Mach number and $R_\mathrm{AD}(l)$ is the AD Reynolds number at scale $l$.
Hence, according to \citet[][see also \cite{Vazquez11a}]{LiPS06} the strong coupling approxi\-mation is satisfied very well 
in our simulations.\\
The fluxes are computed using a central differencing scheme and the numerical timestep is primarily controlled by  AD. 
Note that the implementation by \citet[][]{Chen12} uses super--timestepping to speed up the simulation \cite[see also][]{Choi09}.
\subsection{Initial conditions}
\begin{figure}
 \includegraphics[width=0.5\textwidth]{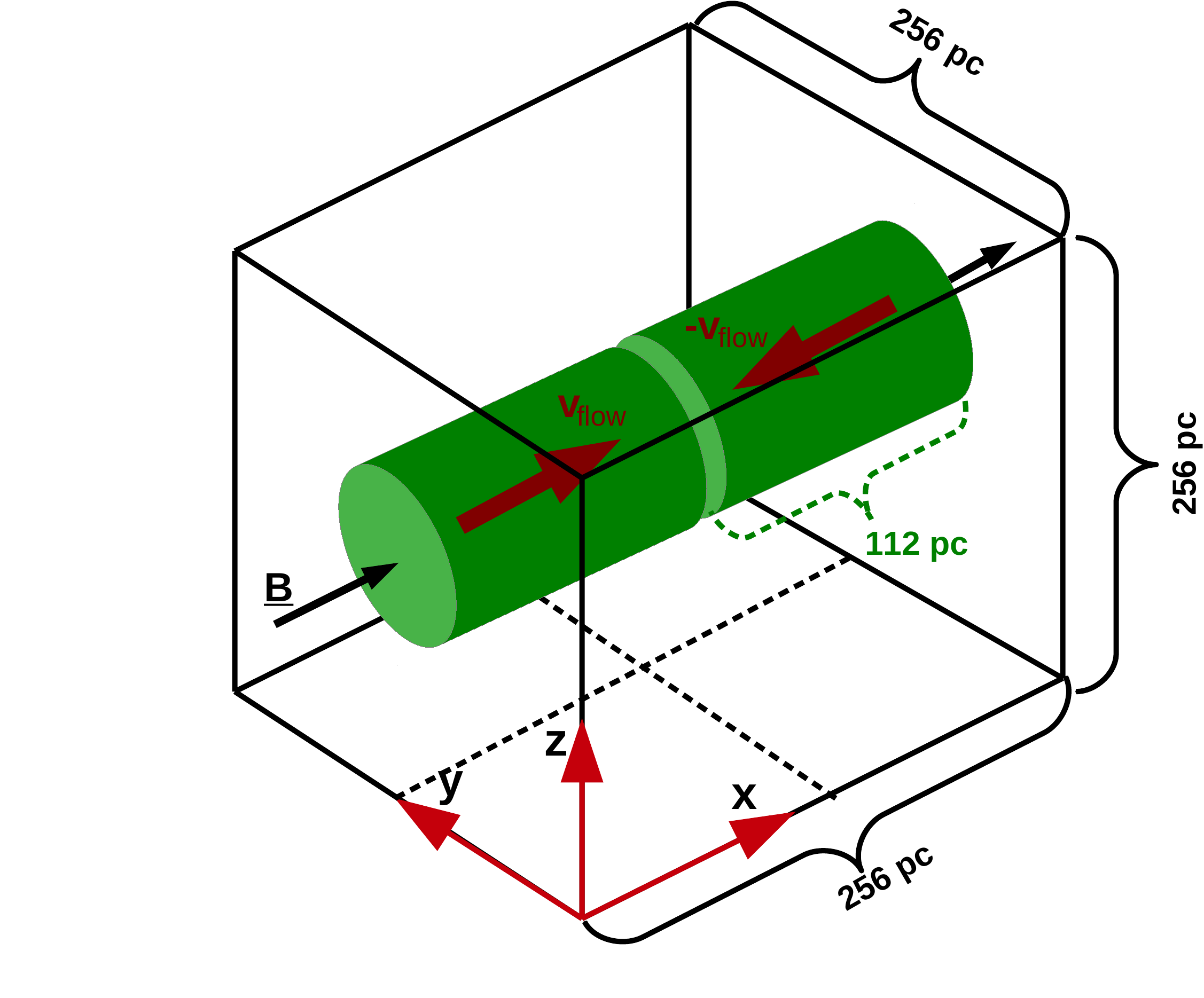}
 \caption{Setup of the initial conditions. The black dashed lines denote the flow axes and the position of the collision layer, respectively. Adapted from \citet[][]{Vazquez07}.}
 \label{fig1}
\end{figure}
Our numerical setup is very similar to this of \citet[][see also
\citet{Vazquez11a} and \citet{Banerjee09a}]{Vazquez07}. The physical size of the numerical box is $256\,\mathrm{pc}$ and the two cylindrical 
flows collide at the centre of the domain, that is, at $\mathrm{L}_\mathrm{x}=128\,\mathrm{pc}$. Each flow has a linear dimension of $l=112\,\mathrm{pc}$ and a radius of $r=64\,\mathrm{pc}$, thus twice as large as 
in the aforementioned studies (see table \ref{tab1} for the runtime details). The heating and cooling prescription by \citet{Koyama00} gives a thermally unstable regime in the density range $1\,\mathrm{cm}^{-3}\leq n\leq10\,\mathrm{cm}^{-3}$, corresponding 
to a temperature interval of $500\,\mathrm{K}\leq T \leq5000\,\mathrm{K}$ if thermal equilibrium conditions are applied. According to these restrictions we have chosen the initial density to be 
$n=1\,\mathrm{cm}^{-3}$ and the temperature as $T=5000\,\mathrm{K}$. With this temperature, we can define a sound speed of the warm neutral medium and we choose the velocity of every single flow in such 
a way that the resulting \ita{isothermal} sonic Mach number is $\mathcal{M}_\mathrm{f}=2$. In addition we add a turbulent velocity field to the flows to mimic the general turbulent behavior of the ISM \citep{MacLow04}. 
The turbulent fluctuations are calculated in Fourier space with a Burger's type spectrum, i.e. $E(k)\propto k^{-2}$ for $k\geq k_\mathrm{inj}$, where $k_\mathrm{inj}$ is the energy injection scale.
Furthermore, these turbulent fluctuations trigger the onset of dynamical instabilities such as the non-linear thin-shell instability (NTSI, \cite[][]{Vishniac94}) or the Kelvin--Helmholtz instability 
\citep[e.g.][]{Heitsch05,Heitsch08b}. The initially uniform magnetic field has a strength of $\left|\vek{B}\right|=\left\{3,4,5\right\}\,\mu\mathrm{G}$ and is aligned with the flows, that is, 
$\vek{B}\propto\hat{\vek{x}}$ where 
$\hat{\vek{x}}$ is the unit vector in the x direction. Using the description for the critical mass--to--flux ratio by \citet{Nakano78}, i.e. $\mu_{\mathrm{crit}}=0.16/\sqrt{G}$, the two streams in total are initially 
subcritical, but can become supercritical very fast due to accretion of mass along the field lines (see table \ref{tab1}). The numerical resolution is adjusted to give a maximum refinement level of 
$\mathcal{L}_{\mathrm{max}}=13$, which corresponds to a maximum physical resolution of $\Delta x_{\mathrm{max}}\approx0.007\,\mathrm{pc}\approx 1600\,\mathrm{AU}$.
\begin{table*}
 \caption{Overview of the conducted simulations with varying flow and magnetic field parameters.  
 $\phi$ is the inclination of one of the flows and $\mathcal{M}_\mathrm{f}$ is the isothermal Mach number of the converging WNM streams. 
 $\mathcal{M}_{\mathrm{RMS}}$ denotes the Mach number of the turbulent fluctuations, $\mathcal{M}_\mathrm{A}$ is the turbulent Alfv\'{e}n Mach number, $\beta$ is the ratio of thermal to magnetic pressure and $\kappa^{*}$ indicates the ratio of turbulent to numerical diffusion with 
 the turbulent diffusion coefficient calculated according to \citet[][]{Lazarian12} (their equations (3) and (4)). Numerical diffusion is evaluated as $\kappa_\mathrm{num}=\Delta x\times v_{\Delta x}$, where $v_{\Delta x}$ 
 is the velocity at grid scale $\Delta x$. $\chi$ denotes the effective ratio of magnetic to numerical diffusion (see appendix \ref{appa}). $\mu/\mu_\mathrm{crit}$ is the normalised mass--to--magnetic flux ratio.}
 \begin{center}
  \begin{tabular}{p{2cm} p{0.7cm} p{0.5cm} p{0.4cm} p{0.8cm} p{0.8cm} p{0.7cm} p{0.7cm} p{0.7cm} p{1cm} p{1.5cm} }
  \hline
  \hline
  Run Name	&$\phi$ &$\left|\vek{\mathrm{B}}\right|$		&$\mathcal{M}_\mathrm{f}$	&$\mathcal{M}_\mathrm{RMS}$	&$\mathcal{M}_\mathrm{A}$	&$\kappa^{*}$ &$\chi$ &$\beta$	&$\mu/\mu_{\mathrm{crit}}^{a)}$ &Min. $\Delta\mathrm{x}^{b)}$\\
			&$\left(^{\circ}\right)$ &$\left(\mu \mathrm{G}\right)$&				&				& &	&		&  &$\left(\mathrm{pc}\right)$\\
  \hline
  B3M0.4I0	&0		 &3	&2	&0.4	&0.39 &2.53 &0.00&1.93	&0.79	&0.03	\\
  B3M0.8I0	&0		 &3	&2	&0.8	&0.79 &21.04 &0.00&1.93	&0.79	&0.03	\\
  B3M1.2I0	&0		 &3	&2	&1.2	&1.18 &42.67 &0.00&1.93	&0.79	&0.03	\\
  \hline
  B4M0.4I0	&0		 &4	&2	&0.4	&0.29 &1.04 &0.00&1.08	&0.59	&0.03	\\
  B4M1.5I0	&0		 &4	&2	&1.5	&1.10 &42.67 &0.00&1.08	&0.59	&0.03	\\
  \hline
  B5M0.5I0	&0		 &5	&2	&0.5	&0.29 &1.04&0.00&0.69	&0.47	&0.0075	\\
 \hline
  B3M0.5I30	&30		&3	&2	&0.5 &0.49 &5.02 &6.91&1.93	&0.79	&0.0075\\ 
  B3M0.5I50	&50		&3	&2	&0.5 &0.49 &5.02 &16.22&1.93	&0.79	&0.0075\\
  B3M0.5I50a	&50$^{c)}$	&3	&2	&0.5 &0.49 &5.02 &16.22&1.93	&0.79	&0.0075\\
  B3M0.5I60	&60		&3	&2	&0.5 &0.49 &5.02 &20.73&1.93	&0.79	&0.0075\\
  B3M0.8I60	&60		&3	&2	&0.8 &0.49 &5.02 &20.73&1.93	&0.79	&0.0075\\
  \hline
  B4M0.5I30	&30		&4	&2	&0.5 &0.36 &1.99 &9.22&1.08	&0.59	&0.0075\\
  B4M0.5I60	&60		&4	&2	&0.5 &0.36 &1.99 &27.63&1.08	&0.59	&0.0075\\
 \hline 
  B5M0.5I30	&30		&5	&2	&0.5 &0.29 &1.04 &11.52&0.69	&0.47	&0.0075\\
  B5M0.8I30	&30		&5	&2	&0.8 &0.47 &4.43 &11.52&0.69	&0.47	&0.0075\\
  B5M0.5I40	&40		&5	&2	&0.5 &0.29 &1.04 &19.03&0.69	&0.47	&0.0075\\
  B5M0.5I50	&50		&5	&2	&0.5 &0.29 &1.04 &27.03&0.69	&0.47	&0.0075\\
  B5M0.5I60	&60		&5	&2	&0.5 &0.29 &1.04 &34.55&0.69	&0.47	&0.0075\\
  B5M0.8I60	&60		&5	&2	&0.8 &0.47 &4.43 &34.55&0.69	&0.47	&0.0075\\
  B5M0.5I60Mf4	&60		&5	&4	&0.5 &0.29 &1.04 &34.55&0.69	&0.47	&0.0075\\
  \hline
  B5M0.5I60AD$^{d)}$	&60		&5	&2	&0.5 &0.29 &1.04 &34.55&0.69 &0.47	&0.0075\\
   \hline
  \hline
  \end{tabular}
  \newline
  \begin{flushleft}
  \tiny{\fat{Remarks:}}\\
  \tiny{\ita{a}) According to the prescription by \citet[][]{Nakano78} (i.e. $\mu_\mathrm{crit}\simeq0.16/\sqrt{G}$).}\\
  \tiny{\ita{b}) Maximum allowed resolution in the simulations.}\\
  \tiny{\ita{c}) Simulation with a different initial turbulent seed field.}\\
  \tiny{\ita{d}) Run with ambipolar diffusion. Simulation was stopped at $t\approx 12\,$Myr.}

\end{flushleft}
 \end{center}
 \label{tab1}
\end{table*} 

\section{MC Formation by head--on colliding flows}
\label{sec3}
Colliding streams of gas are ubiquitous in the ISM \citep[e.g. due colliding supernovae shells,][]{Inoue08} as well as in the interior of molecular clouds 
\citep[e.g. in filaments or the junctions of filaments,][]{Hennebelle08b,Chen14}. Therefore the dynamics and the structure can vary 
significantly, depending on the galactic or local environment or the respective driving mechanism \citep[e.g.][]{Inoue08,Inoue12}. In this section we summarise the evolution of molecular clouds, which are being 
formed by head--on colliding flows. For more thorough analyses, we refer the reader to the studies of e.g. \citet{Banerjee09a,Vazquez06,Vazquez11a,Hennebelle99,Hennebelle08b,Heitsch08b}. An overview of the main initial 
physical parameters of the respective simulations is given in table \ref{tab1}.

\subsection{Varying the turbulent velocity}
\begin{figure*}
 \centering
 \includegraphics[height=6.4cm,angle=-90]{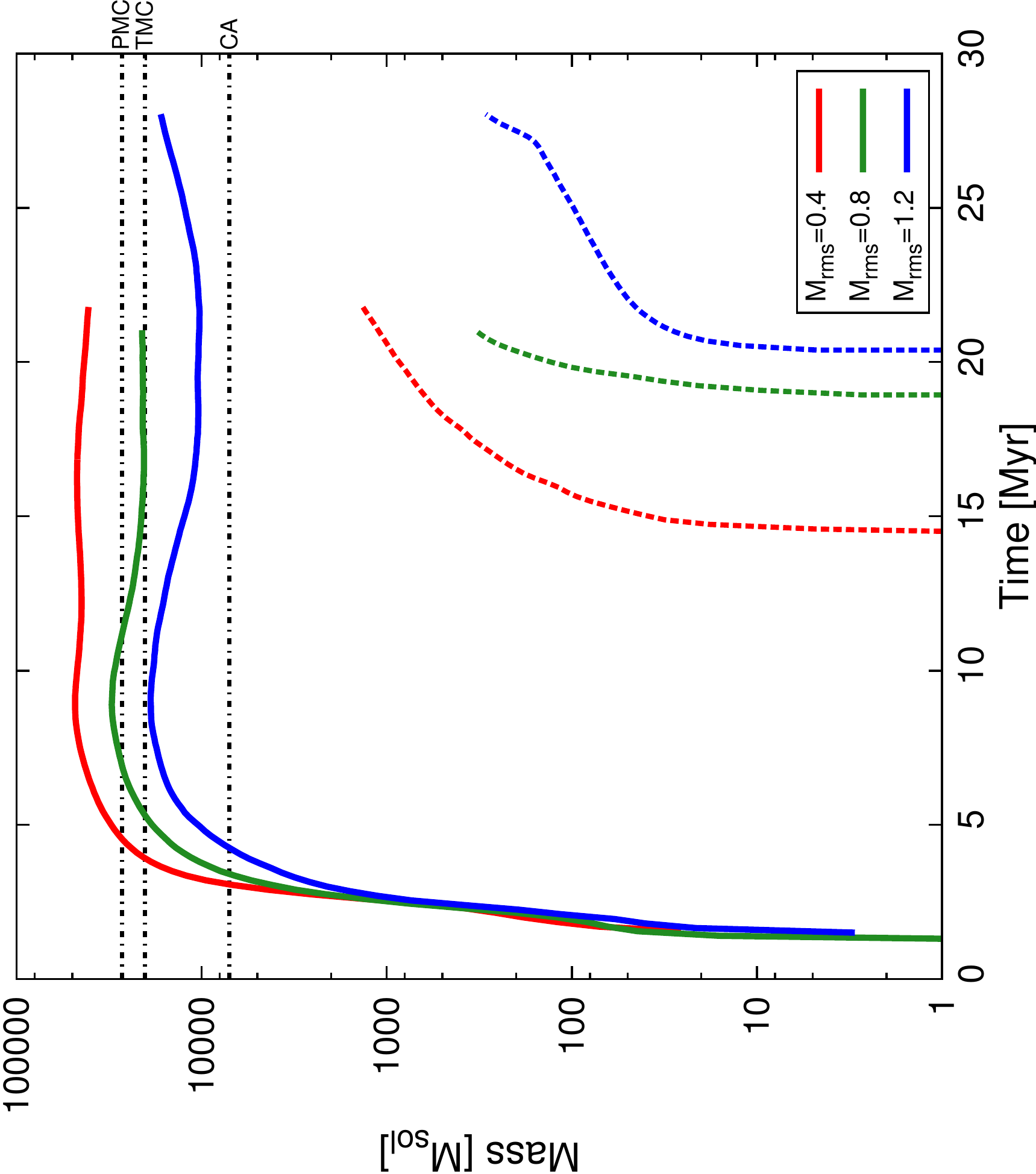}\includegraphics[height=6.0cm,angle=-90]{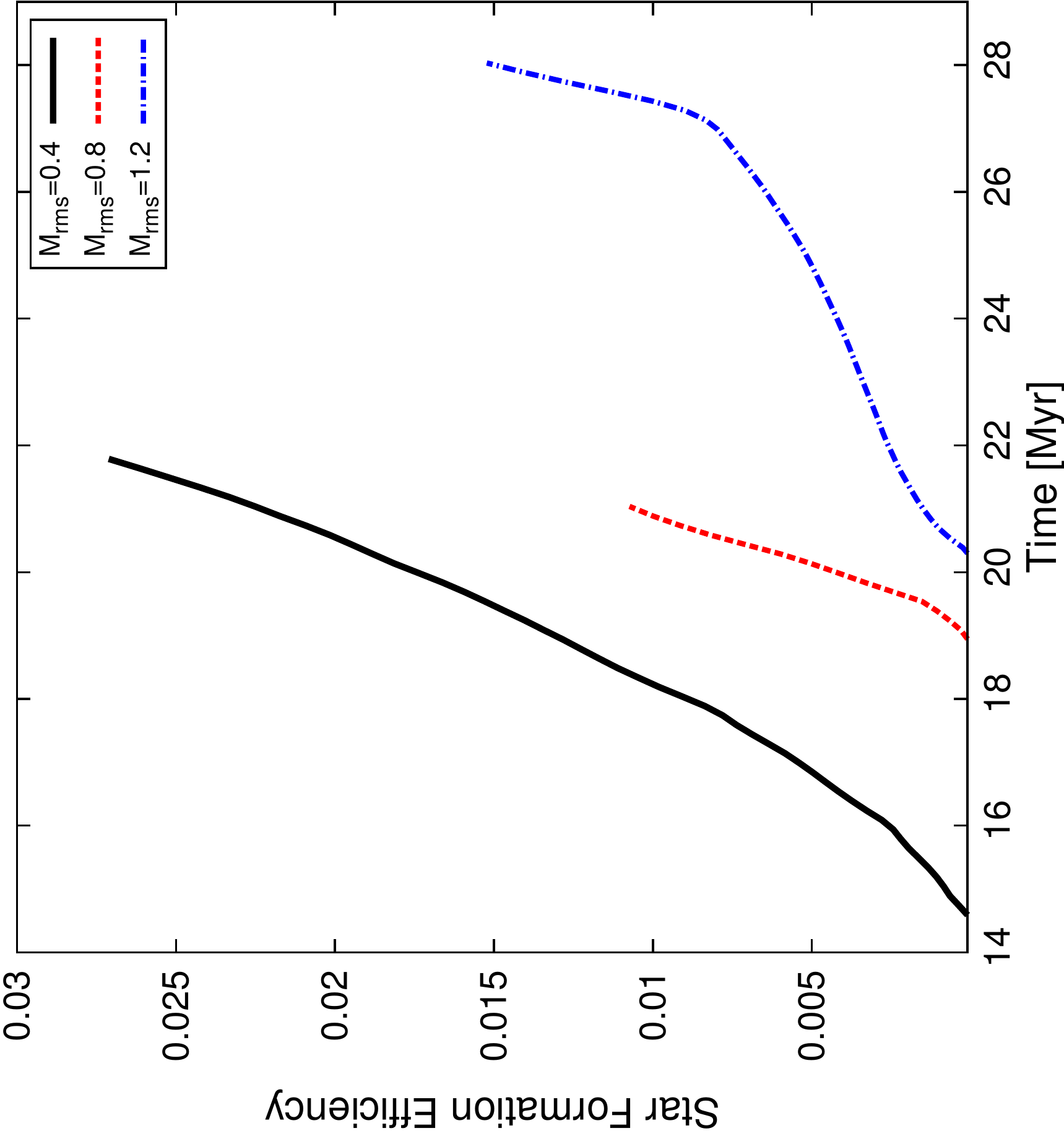}\,\,\includegraphics[height=5.8cm,angle=-90]{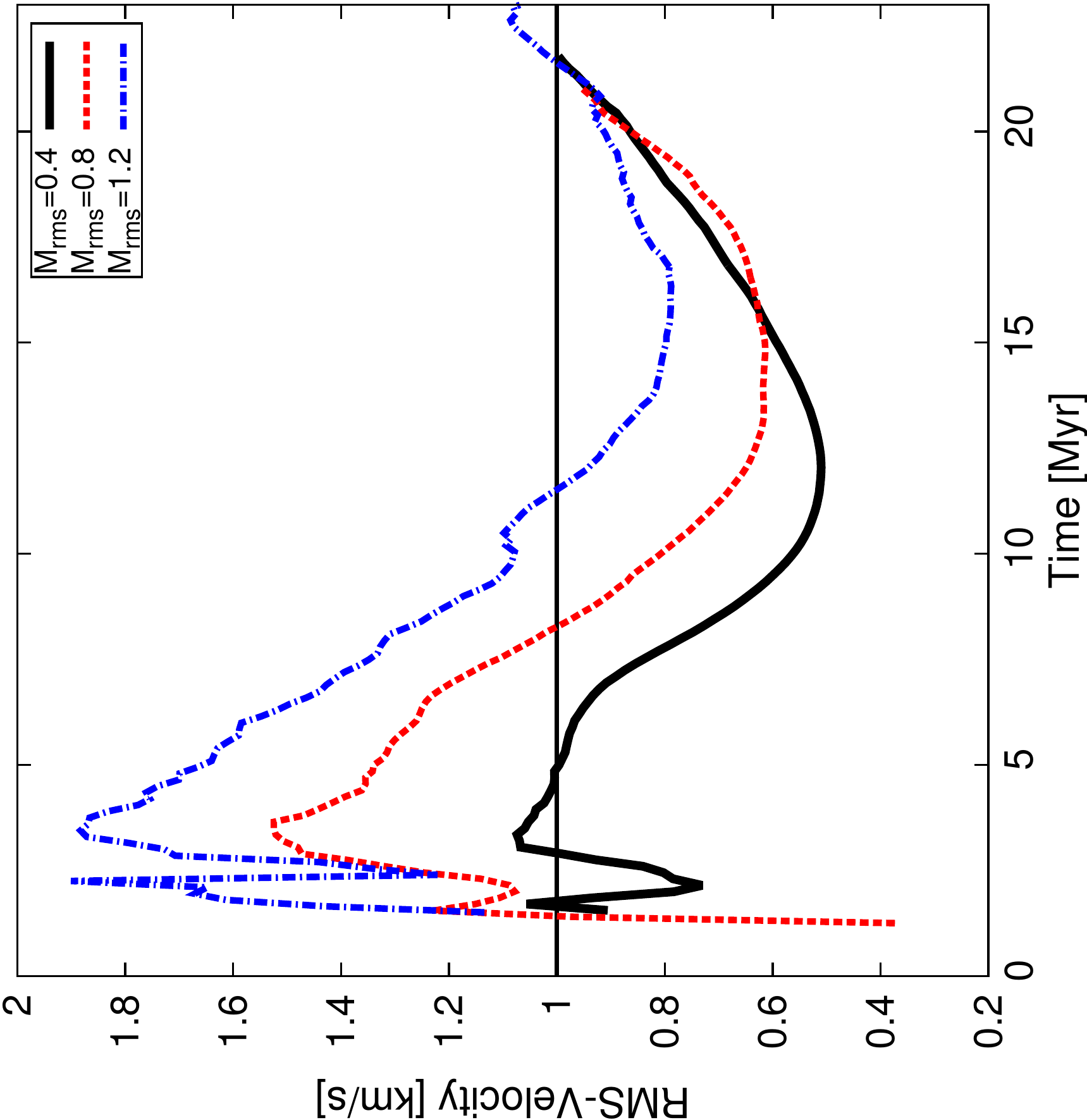}
 \caption{\ita{Left:} Evolution of cloud (solid) and sink particle (dashed) masses for three turbulent velocity fields (for a colour version see online manuscript). The onset of star formation 
 is clearly seen to be delayed due to the action of turbulent motions that keep dissolving dense structures. The horizontal dash-dotted, black lines denote 
 the masses of observed molecular clouds, like the Perseus MC \citep[PMC,][]{Lombardi2010}, the Taurus MC \citep[TMC,][]{Lombardi2010}, and the Corona Australis complex \citep[CA,][]{Alves14}.\ita{Middle:} Star formation efficiency for $\mathrm{B3M}\dots\mathrm{I}\dots$ runs. \ita{Right:} Corresponding root mean square velocity of the dense gas. The RMS--velocities converge after turbulence has decayed and global collapse of the cloud has begun. Before this point, the amplitude of the resulting turbulent velocities is determined by the initial conditions.}
 \label{fig2}
\end{figure*}

\begin{figure*}
\begin{tabular}{llll}
 \fat{B3M0.4I0}	&\fat{B3M0.8I0}	&\fat{B3M1.2I0}	&\fat{B5M0.5I0}\\
 \includegraphics[height=4cm]{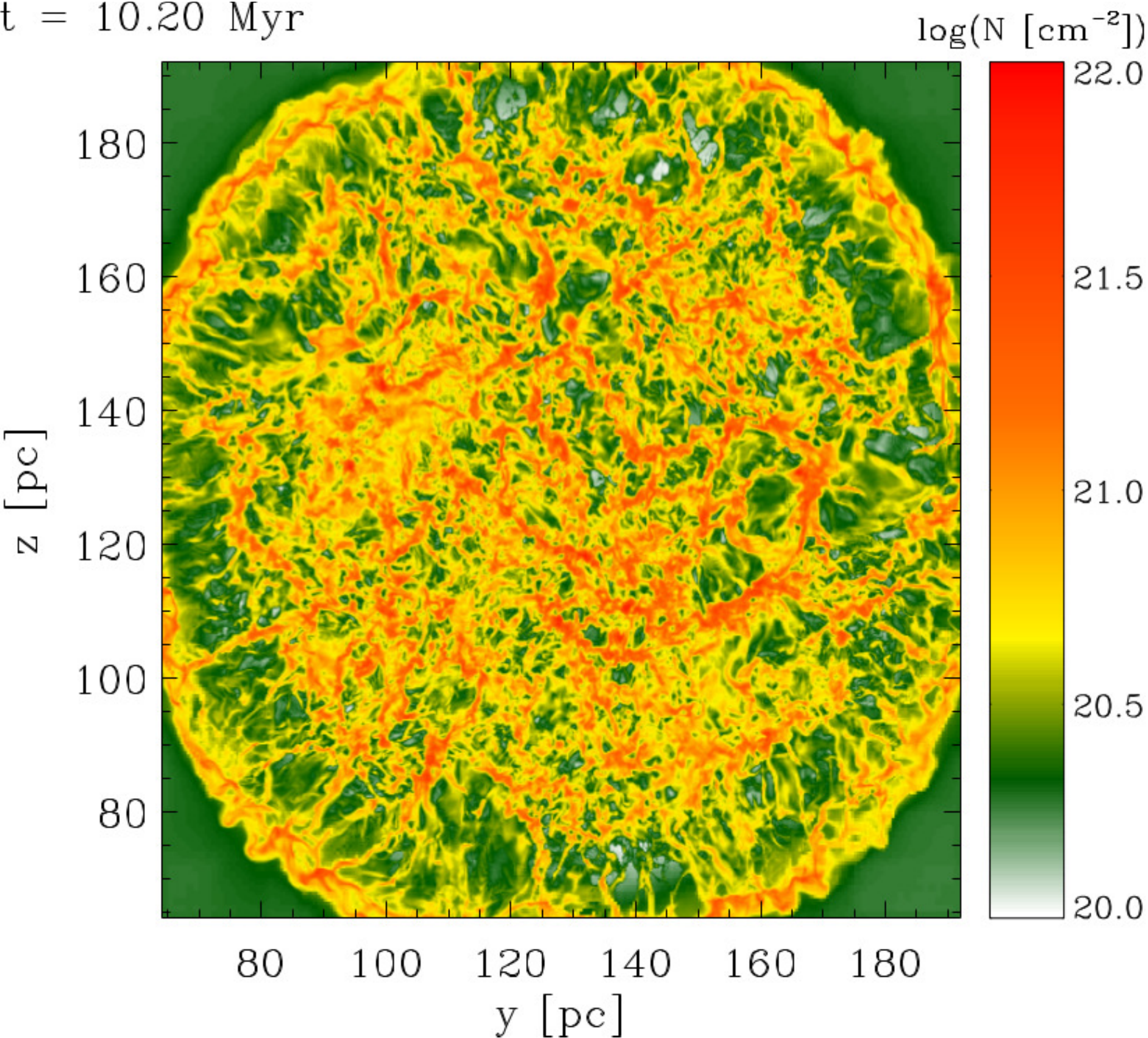}&\includegraphics[height=4cm]{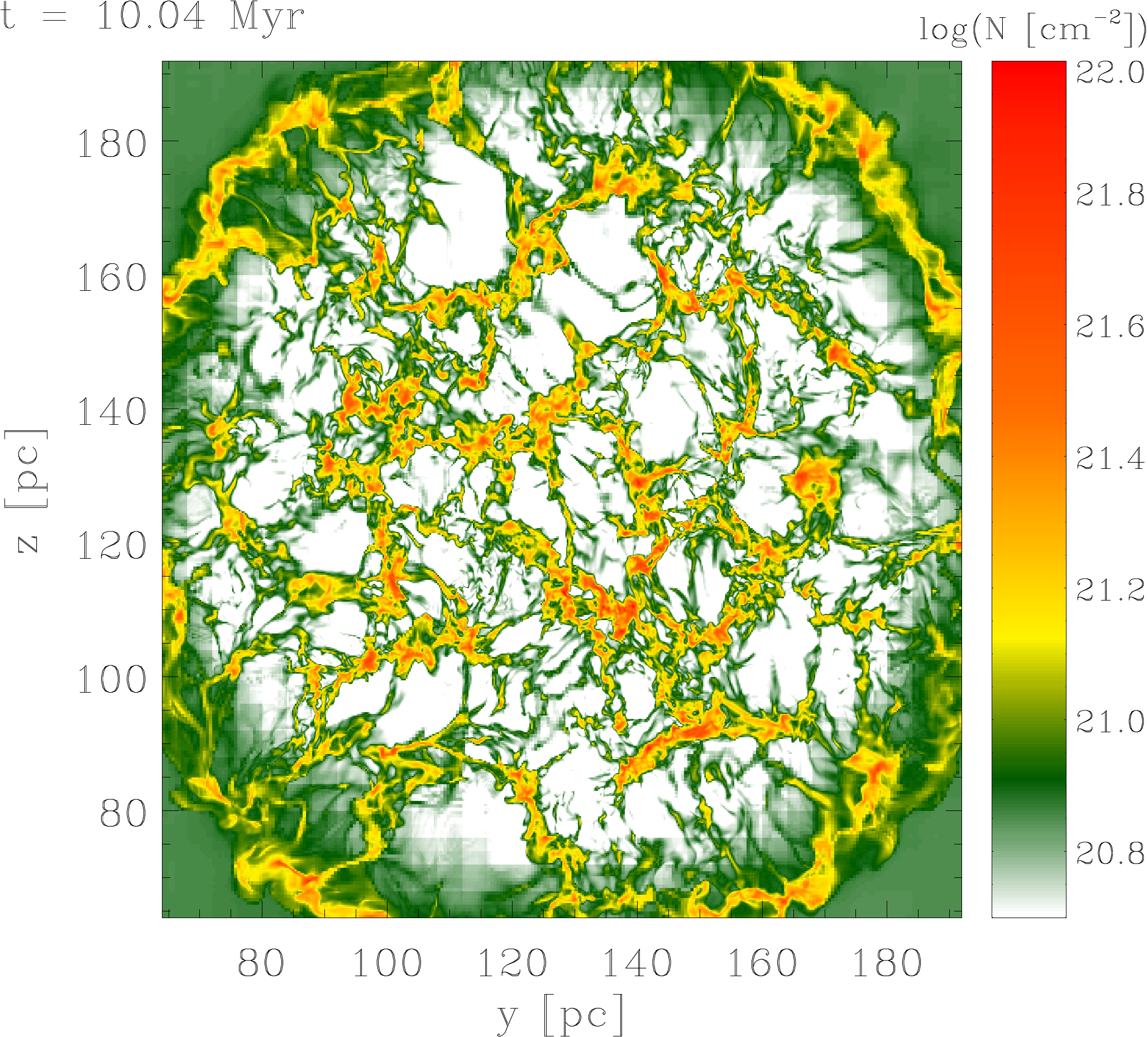}&\includegraphics[height=4cm]{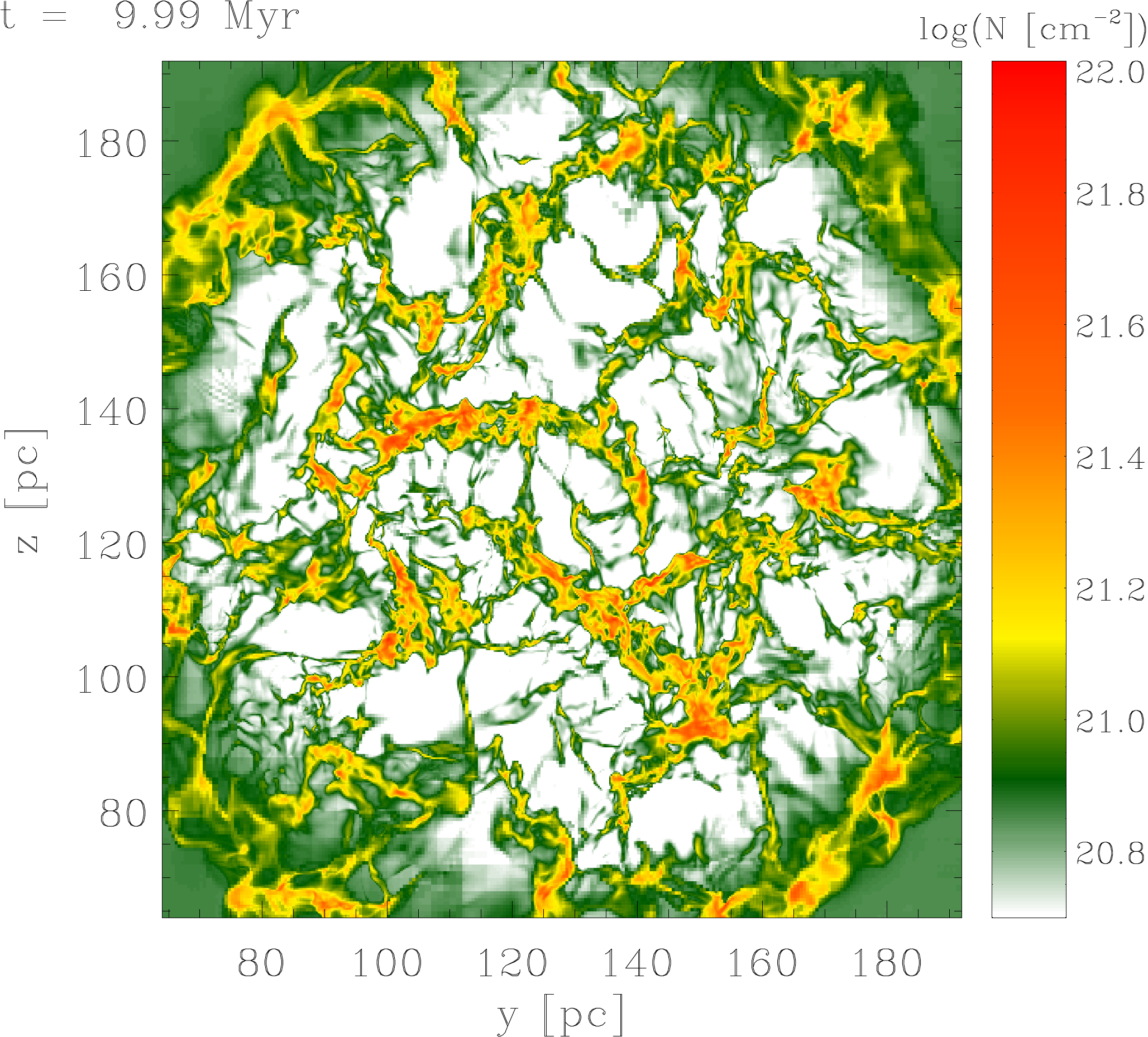}&\includegraphics[height=4cm]{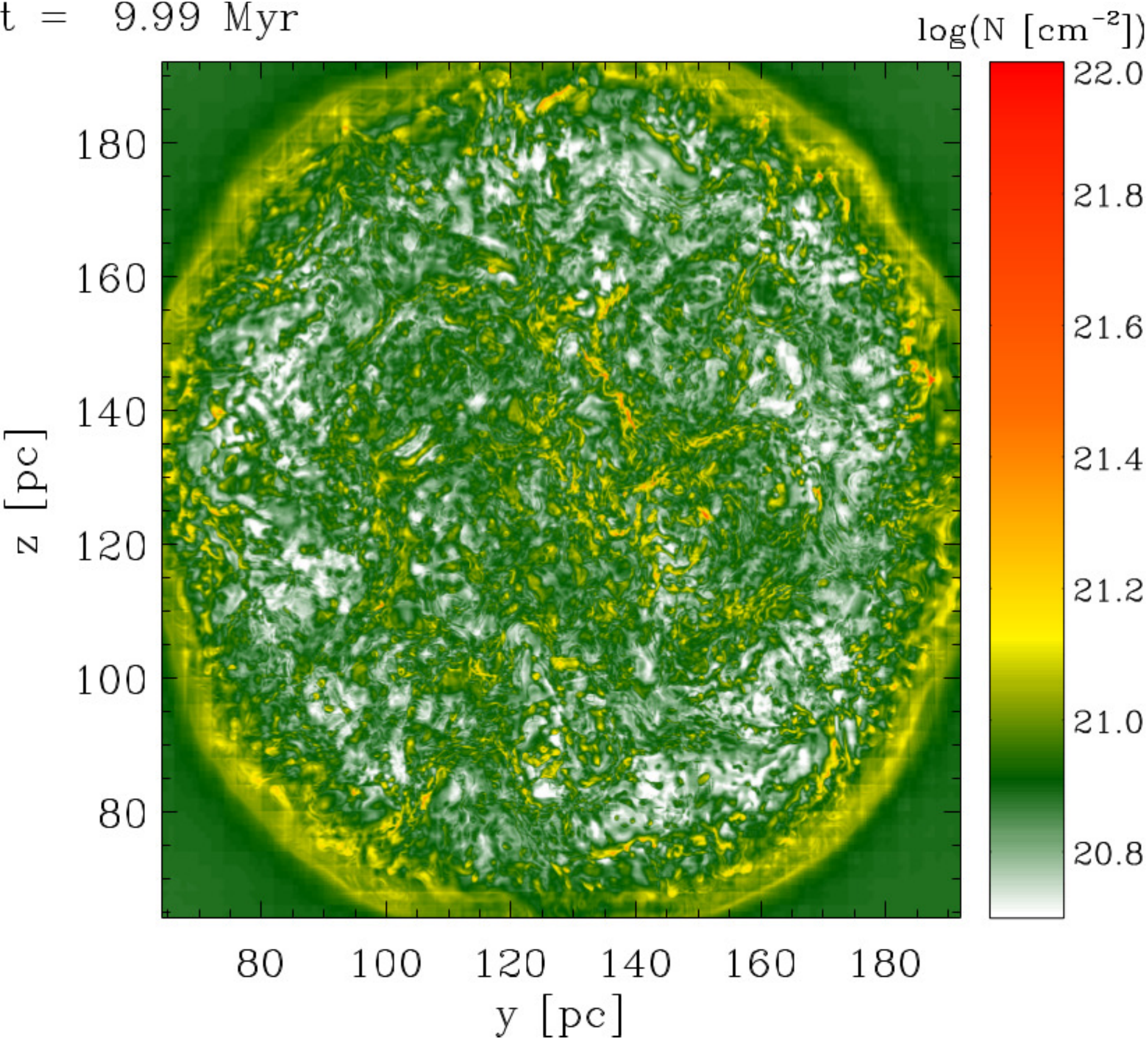}\\
 \includegraphics[height=4cm]{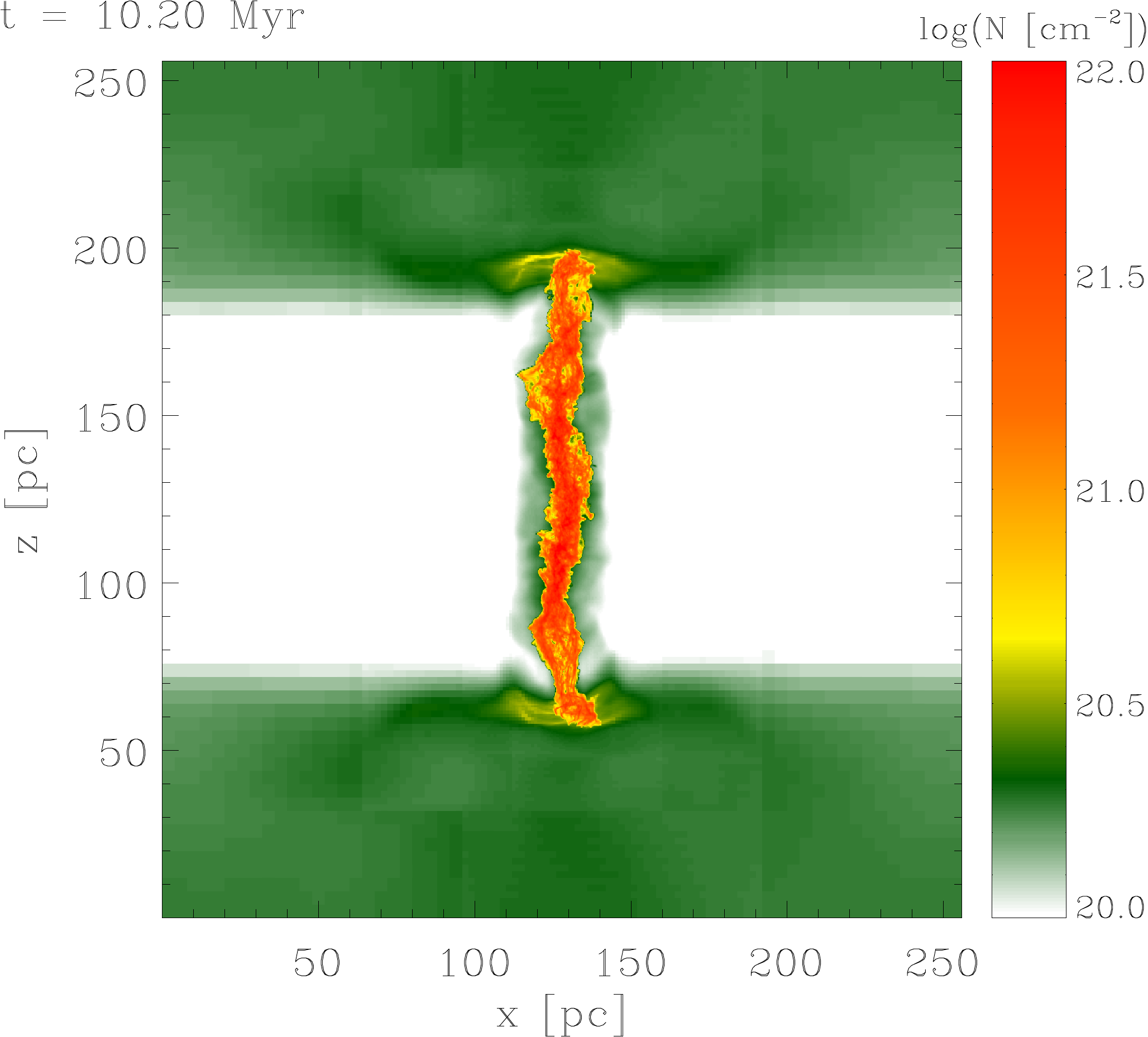}&\includegraphics[height=4cm]{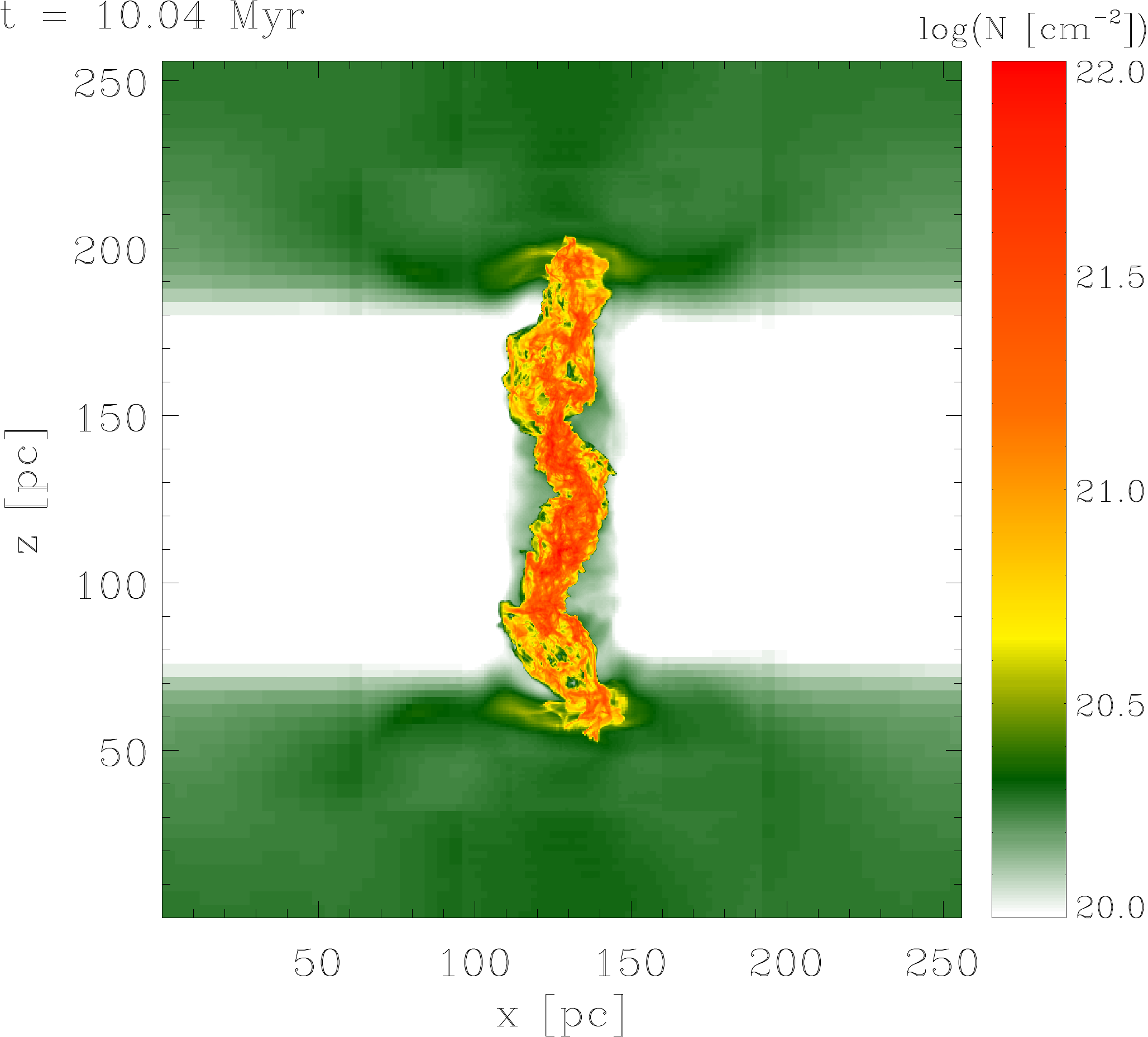}&\includegraphics[height=4cm]{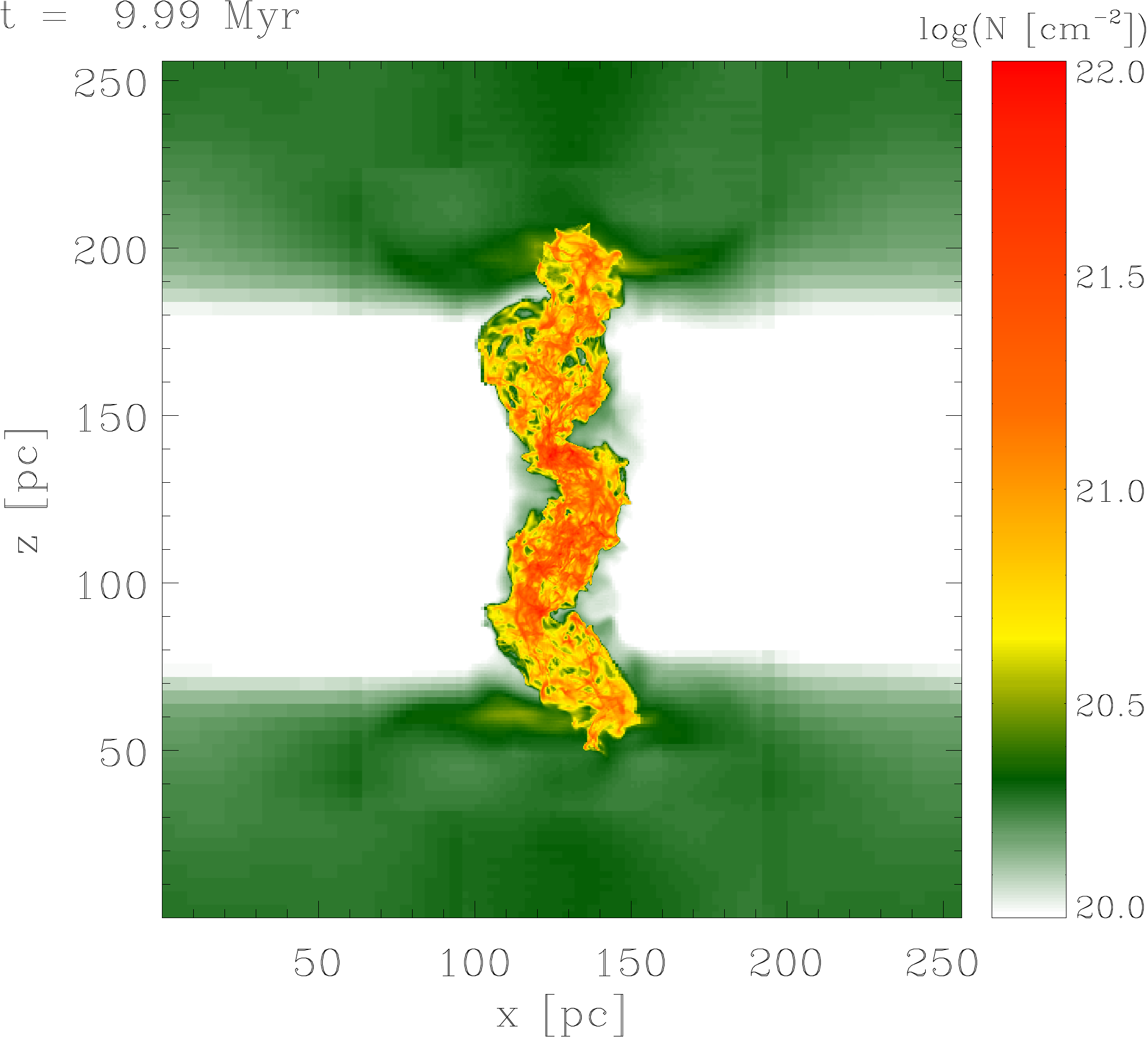}&\includegraphics[height=4cm]{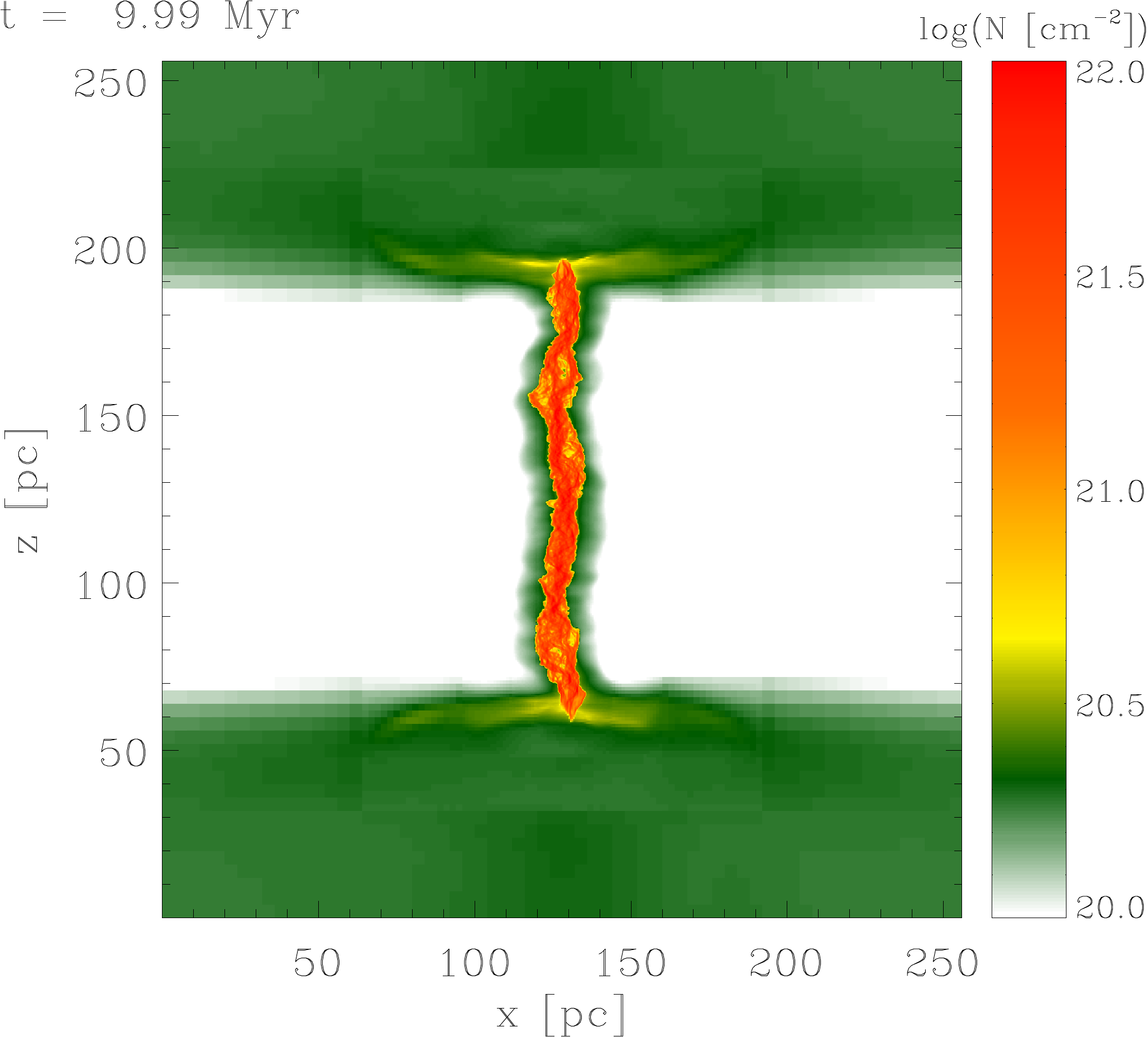}
\end{tabular}
\caption{Slices of column density along two different axes for different simulations at t$\approx$10\,Myr. Top: Along x--axis (i.e. face--on view). Bottom: Along y--axis (i.e. edge--on view).}
\label{fig3}
\end{figure*}

The dynamics of molecular clouds which formed in the compression zone of two colliding streams strongly depend on the initial kinematics of the individual flows. On the one hand, the flows are supersonic with respect 
to the WNM and thus generate strong shocks and compressions \citep[][]{Vazquez07,Banerjee09a}. On the other hand, large scale instabilities as well as stellar feedback inject energy into the ambient ISM. This energy, if 
not already in the form of kinetic energy, can be converted to kinetic energy and thus a turbulent regime is produced, where the turbulence cascades down until it is dissipated on atomic/molecular scales. This turbulence is 
primarily supersonic \citep[e.g.,][]{MacLow04}. These random motions generate a certain level of anisotropy within the bulk flows and the respective contribution to 
the process of molecular cloud formation is two--folded. Firstly, turbulence contributes an effective ram pressure, which can help to stronger compress fluid elements. Secondly, the inhomogeneous velocity field distorts the 
overall bulk flow and reduces the mass flux, which then directly translates to the build up of less massive clouds (see 
fig. \ref{fig2}).\\
If the collision of the WNM streams is along the magnetic field lines, the early stages ($t \leq 3\,\mathrm{Myr}$) of cloud formation can be understood as being nearly independent of the magnetic field. The first phases during the 
collision are thus controlled by the bulk and turbulent velocity (see tab. \ref{tab1}).\\ 
Fig. \ref{fig2} shows the evolution of the dense gas ($n\geq100\,\mathrm{cm}^{-3}$) for different initial turbulent Mach 
numbers. The compression by the flows induces the formation of a molecular cloud by the combined action of dynamical and thermal instability 
\citep[e.g.,][]{Vazquez07,Heitsch08b,Banerjee09a}. Due to the onset of runaway cooling of thermally unstable gas, the cloud becomes more massive with time. At the same time it assembles mass by accretion of gas along the 
field lines. Independent of the degree of turbulence, the onset of dense gas formation starts at the same time indicating the dominance of the ram pressure by the 
bulk flows. Only at slightly later times around $t\approx2-3\,\mathrm{Myr}$ the effects of different turbulent Mach numbers are seen. Flows of higher turbulent Mach numbers reduce the mass flux and hence reduce the final mass of the cloud. This is seen in fig. 
\ref{fig3}. The stronger the turbulence, the less compact is the resulting cloud. The mass concentrates in pressure confined filaments, which are further immersed in a diffuse, warm medium, with a steep density and 
temperature gradient between the WNM and CNM that can be interpreted as a phase--transition front rather than a contact discontinuity due to the ambient mass flux across the transition layer \citep[][]{Banerjee09a}.\\
At later stages, the initial turbulence has decayed and the presence of turbulent motions is due to self--gravity \citep[see e.g.][]{Ballesteros07}. Once self--gravity dominates, certain regions then 
proceed to collapse to form a star. The onset of star formation is clearly delayed by the presence of stronger initial turbulence (see fig. \ref{fig2}). 

\subsection{Dependence on the magnetic field strength}
In the previous section we have neglected the possible influence of the magnetic field. However, the ISM and molecular clouds are highly magnetised \citep[in the range $5-15\mu$G,][]{Beck01,Crutcher10,Crutcher12,Li10,Li14} and thus the magnetic 
field affects the overall evolution of molecular clouds in the ISM as well as their preceding condensation out of the latter \citep[e.g.,][]{Hennebelle99,Hennebelle13}. As was shown by \citet[][]{Crutcher10} using 
Zeeman measurements, the \ita{line--of--sight} component of the interstellar magnetic field can be approximated by an interval of nearly constant magnitude followed by a regime that consists of a linear increase of the 
field strength as function of (column--)density. Since Zeeman splitting provides information of one component only, the total magnetic field strength will be larger. Recent studies provide average values of the 
magnetic field of approximately $5-15\,\mu\mathrm{G}$ \citep[][]{Beck01,Crutcher10}. It is thus reasonable to investigate the influence of varying magnetic field strength on the molecular cloud formation process. This has 
recently been done by \citet[][]{Vazquez11a} for initial magnetic field strengths of $\left|\vek{B}\right|=\left\{2,3,4\right\}\mu\mathrm{G}$ (corresponding to $\mu=\left\{1.18,0.79,0.59\right\}$\footnote{Note, the values for 
the mass--to--flux ratio in \citet[][]{Vazquez11a} refer to the box length of 256\,pc, instead of the flow length of 112\,pc, which we here take care of.}) and the action of ambipolar diffusion. Our study covers the upper range of 
their values, namely the range of $\left|\vek{B}\right|=\left\{3,4,5\right\}\mu\mathrm{G}$. The choice of these field strengths gives thermally dominated ($\left|\vek{B}\right|=3\,\mu\mathrm{G}$) environments, regimes with an 
equipartition of thermal and magnetic energies ($\left|\vek{B}\right|=4\,\mu\mathrm{G}$), and completely magnetically dominated regions ($\left|\vek{B}\right|=5\,\mu\mathrm{G}$). 
\begin{figure}
 \centering
\includegraphics[height=8cm,angle=-90]{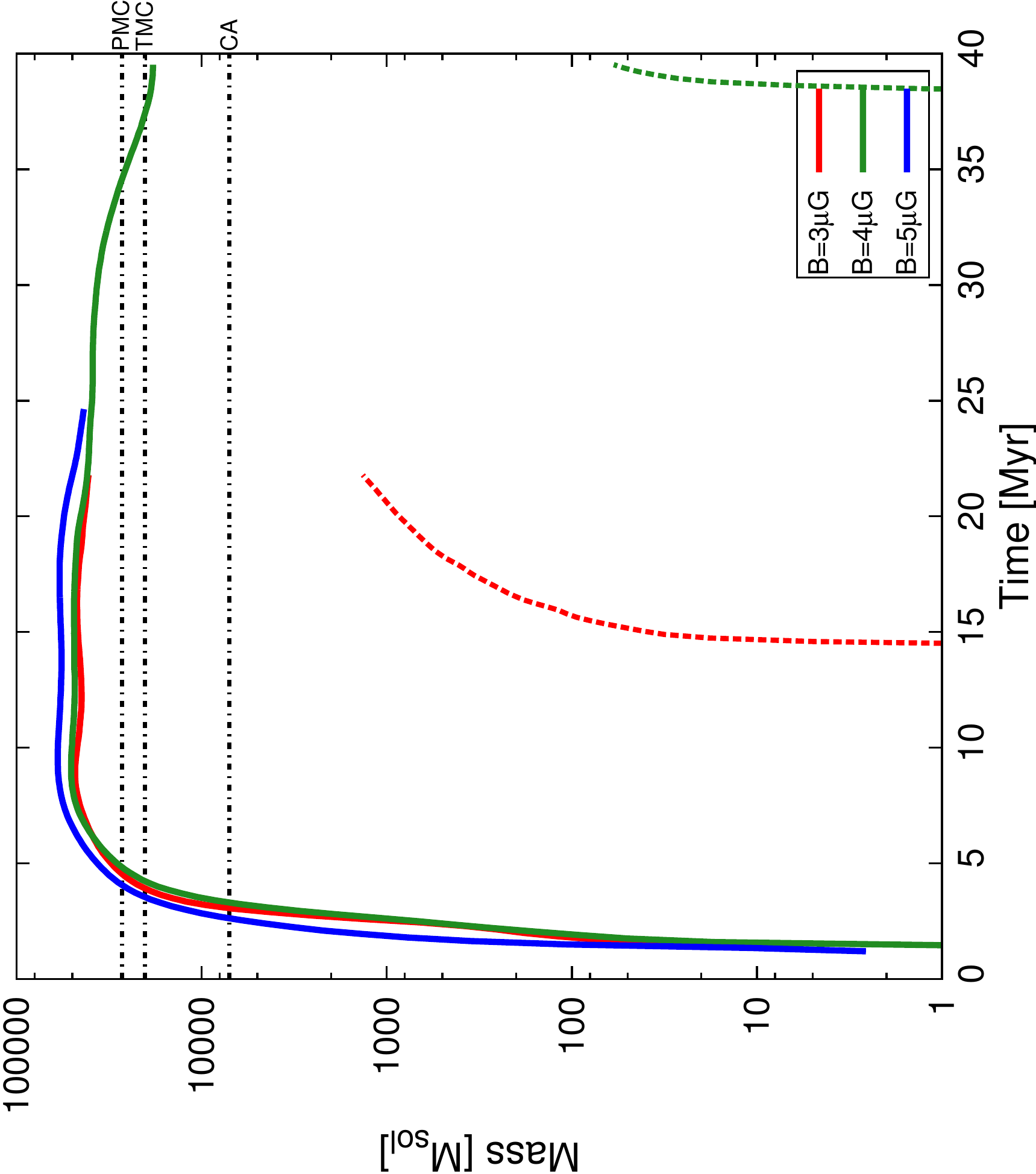}
 \caption{Evolution of cloud (solid) and sink particle (dashed) masses for three magnetic field strengths (for a colour version see online manuscript). 
 The clouds assemble similar final masses, but some slight differences in the accretion phase are seen. More prominent is the lack/delay of star formation for simulations with $\left|\vek{B}\right|>3\,\mu\mathrm{G}$. The horizontal dash-dotted, black lines denote the masses of observed molecular clouds, like the Perseus MC \citep[PMC,][]{Lombardi2010}, 
 the Taurus MC \citep[TMC,][]{Lombardi2010}, and the Corona Australis complex \citep[CA,][]{Alves14}.}
 \label{fig4}
\end{figure}
The evolution of the cloud and sink particle mass for different initial field strengths is shown in fig. \ref{fig4}. Here, the final cloud masses do not differ too much from each other, showing that the initial turbulent motions are more efficient in controlling the early phases of gas accumulation. But differences are seen in the early mass accretion. The 
cloud, which is embedded in a strong magnetic field is seen to be build up at a slightly earlier time after the start of the simulation. Furthermore the accretion of matter from the diffuse halo surrounding the cloud at 
early times differs for the strongest initial magnetic field. This fact can be explained by momentum and energy conservation. Since the flows collide 
head--on, the gas is compressed in the collision layer. The external ram pressure by the flows forces the gas to move perpendicular to the magnetic field lines in order to ensure conservation of linear momentum. At the 
same time gas compression enhances the magnetic field strength. Magnetic tension then acts as a restoring force and since the plasma-$\beta$ is less than unity, the 
dominant magnetic field is too stiff to be bend efficiently. This results in a less efficient gas motion perpendicular to the original bulk flow motion and an earlier compression of the gas. Thus, efficient accretion happens only along the field lines. The rightmost plot in fig. \ref{fig3} shows the column density after $t \approx 10\,$Myr for a strong 
magnetic field. The density gradients are smoother in comparison to the weaker field and the cloud is more compact, i.e. no clearly defined filaments condense out. At the same time the column density (in the face--on view) 
does not reach sufficiently large values. 
The critical column density to become magnetically supercritical can be written as \citep[see also][]{Vazquez11a} 
\beq
N_\mathrm{crit}\approx 2.92\times10^{20}\left(\frac{B}{1\mu\mathrm{G}}\right)\mathrm{cm^{-2}}.
\label{eq2}
\eeq
Although all different clouds assemble mass by accretion from the surrounding diffuse gas, the most striking difference is the complete lack of star formation for the higher magnetised clouds. 
\begin{figure*}
 \includegraphics[height=6cm,angle=-90]{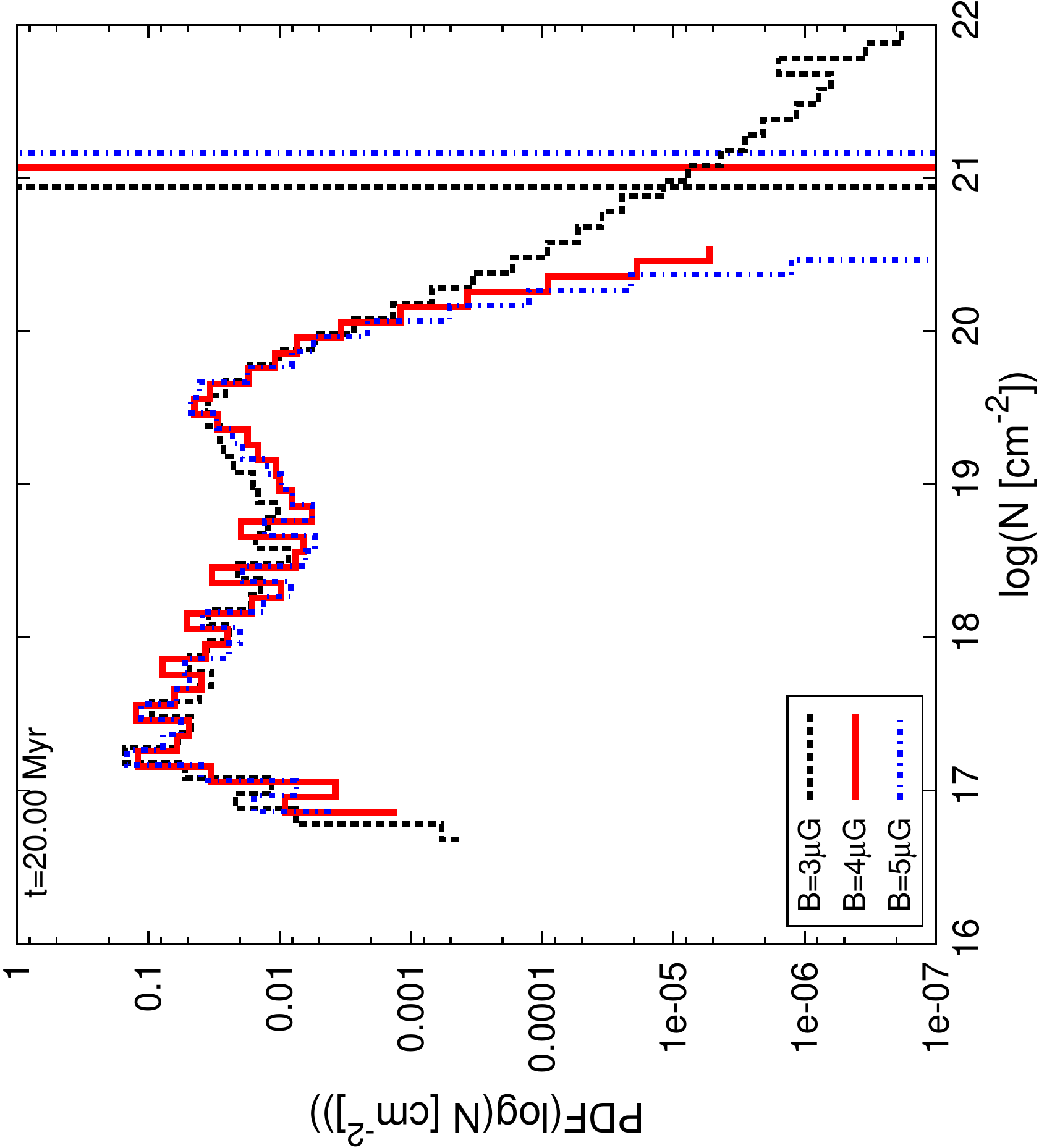}\includegraphics[height=6cm,angle=-90]{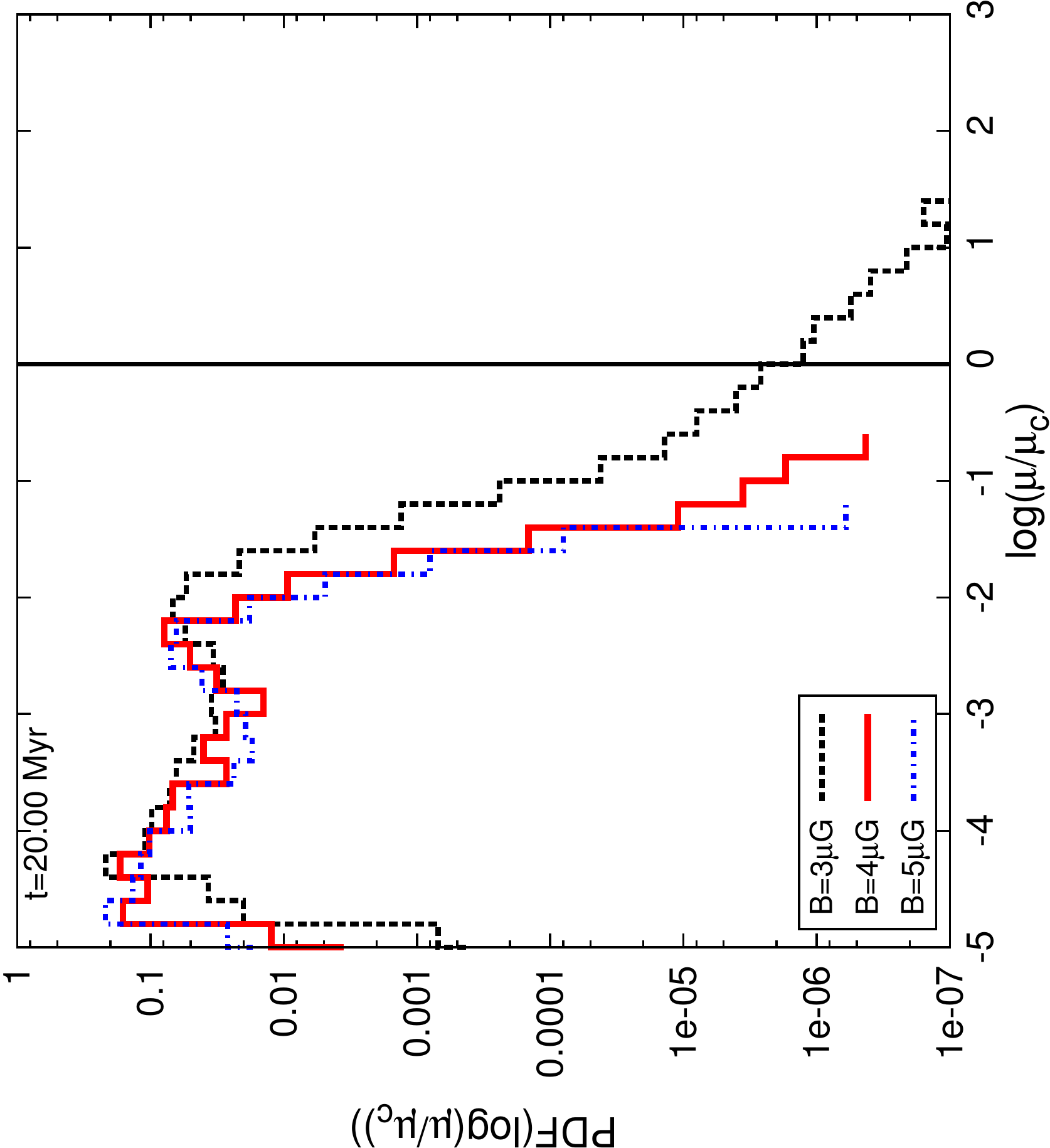}
 \caption{\ita{Left:} Column density PDF for different initial magnetisations at 20\,Myr. The weaker magnetic field allows for the development of a power--law tail, which indicates gravitational collapse. The vertical 
 lines denote the respective threshold column densities for magnetic criticality after eq. \ref{eq2}. \ita{Right:} Corresponding $\mu$--PDF. Here, the vertical line indicates the critical ratio.}
 \label{fig5}
\end{figure*}
In order to quantify the gas dynamics at a specific evolutionary stage, fig. \ref{fig5} shows the probability distribution function of the column density (left, hereafter N-PDF) and the mass--to--flux ratio (right, from now 
on $\mu$-PDF). As shown by 
\citet[][]{Vazquez00b}, the statistics of a gas can be analysed by using a density PDF. For isothermal turbulence, this PDF develops a lognormal distribution with its variance depending on the Mach number of the gas. 
More recently it has been demonstrated that the width of the distribution also depends on the plasma--$\beta$ as well as on the turbulent forcing parameter, i.e. if the driving of turbulence is purely solenoidal or 
compressive \citep[see e.g.,][]{Federrath12,Federrath13}. The same is also valid for the N-PDF, which is used in observational studies since the volume density is not accessible
\citep[e.g.,][]{Kainulainen11,Schneider13,Schneider14}. The shape of the N-PDFs in fig. \ref{fig5} is not lognormal. The reason is the multi--phase nature of the ISM. 
However, in the weaker magnetised case, a power-law tail at high 
column densities evolves, which is always seen in self--gravitating systems \citep[][]{Federrath13,Schneider13}, indicating the presence of gravitationally unstable regions. The vertical lines denote the threshold column 
densities according to eq. \ref{eq2}. The thermally dominated case shows a transition to supercritical states, whereas the maximum column density in the magnetically dominated gas is approximately a factor of five lower. 
At this time the WNM flows vanished. Increases in column density are only due to mass accretion from the environment.
Since the column density and the mass--to--flux ratio are coupled via
\beq
\mu\approx\frac{\Sigma}{\left|B_\mathrm{LOS}\right|}, 
\label{eq3}
\eeq
where $\Sigma$ is the column density in g/cm$^{2}$ \citep[][]{Nakano78}, the overall shape of the $\mu$-PDF should be very similar to the one of the N-PDF. This is indeed the case, as can be seen from fig. \ref{fig5}, right. The modifications are due to the additional dependence on the magnetic field. Here, again, the mass--to--flux ratio shows a similar distribution, but for the weaker field it is shifted towards higher values. Furthermore, the 
transition from subcritical to trans--critical regions is smoother, because of the lack of stiffness of the magnetic field. The 
dependence on column density then also implies the outcome of a power--law tail in the distribution, which continues up to values of $\log(\mu/\mu_c)\approx1$, showing the presence of highly unstable, dynamically dominated 
regions. The mass--to--flux ratio for runs B4M0.4I0 and B5M0.5I0 is similar distributed. This indicates that initial dynamical processes should be more energetic than observed in the simulations, since dynamic compressions 
always result in increasing magnetic energy, which at some stage starts to dominate over thermal and gravitational energy. This yields a re-expansion of compressed regions and a simultaneous stabilisation of these.\\
The results from simulations with higher magnetisation now raise the question, how stars can form in such highly magnetised media. 

\section{Inclined WNM flows}
\label{sec4}
 Here we probe the influence of inclined colliding flows. Inclined collisions are easily justified by assuming the emergence of a 
supernova shock wave and its propagation through a Galactic spiral arm or by non--uniform large scale gravitational forces. The motion of the flow at an inclination with respect to the magnetic field results in an enhanced diffusivity of the latter and this process thus can 
be thought of as a non--ideal MHD process \citep[e.g.][]{Heitsch05,Inoue08,Heitsch09a}. 


\subsection{The setup}
\begin{figure}
 \includegraphics[width=0.5\textwidth]{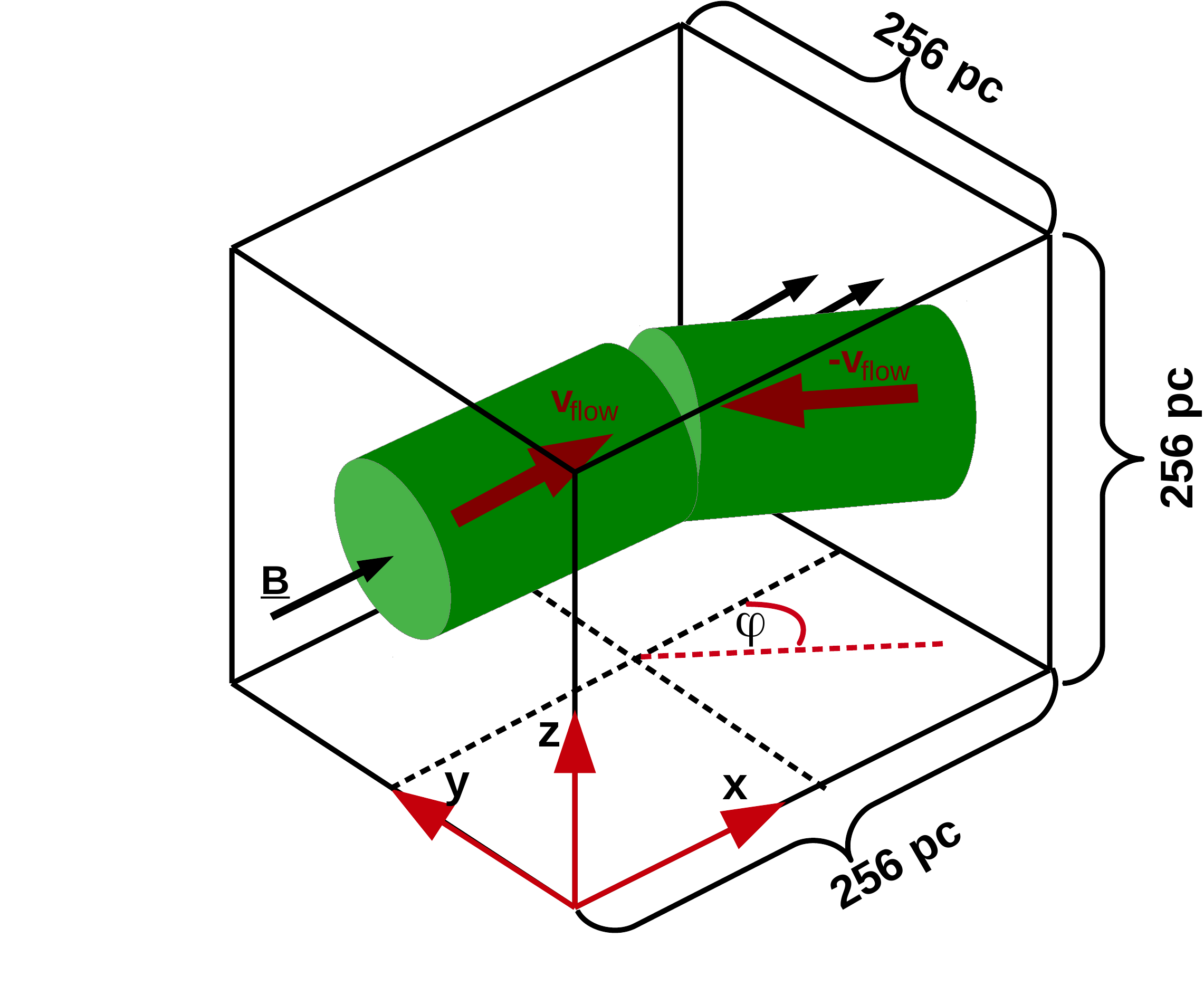}
 \caption{Setup of the initial conditions with inclined flow. The box and physical parameters are the same as in fig. \ref{fig1}, but now one flow is inclined at an angle $\varphi$ with respect to the background 
 magnetic field. The inclined axis is shown as the red dashed line. Note, the collision still occurs at the centre of the simulation domain, in contrast to what is schematically shown here.}
 \label{fig6}
\end{figure}
The initial geometry can be seen in fig. \ref{fig6}. The basics are the same as for the head--on case, but now one flow is 
inclined at an angle $\varphi$ with respect to the x--axis. The initial background magnetic field is kept constant and 
aligned with the x--axis. The figure may imply that there might be a region where quiescent gas resides, but this is not the 
fact. The two flows still collide in the centre of the simulation box and since the magnetic field is still uniform, the 
collision will induce a normal shock. Table \ref{tab1} lists the initial parameters for this study. 

\subsection{Magnetic Flux Reduction and Star Formation}
\begin{figure}
\includegraphics[height=8cm,angle=-90]{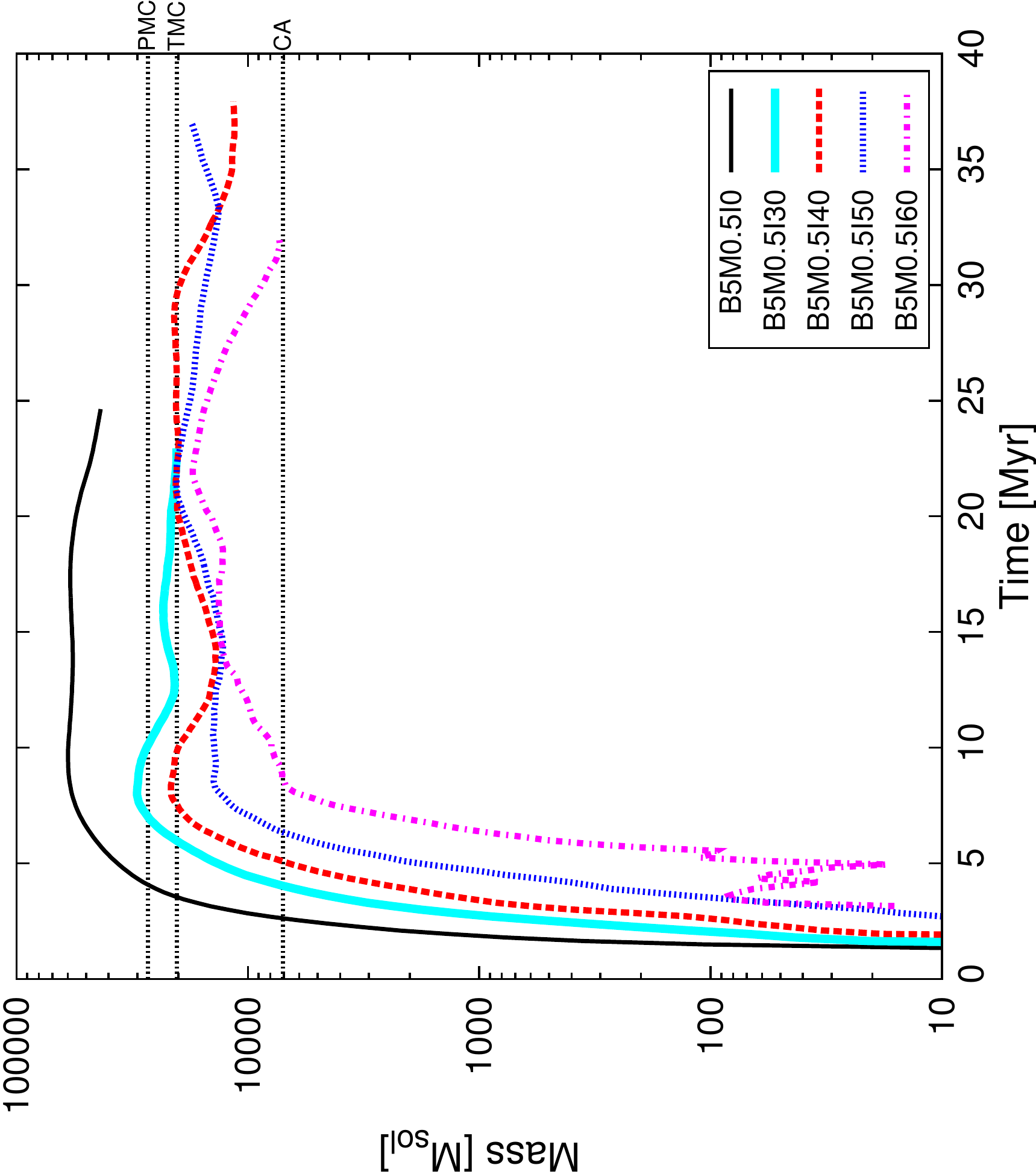}
\caption{Temporal evolution of mass for runs B5M0.5I$\dots$. The horizontal black,dash-dotted lines have the usual meaning.}
\label{fig7}
\end{figure}

\begin{figure}
\includegraphics[height=6cm]{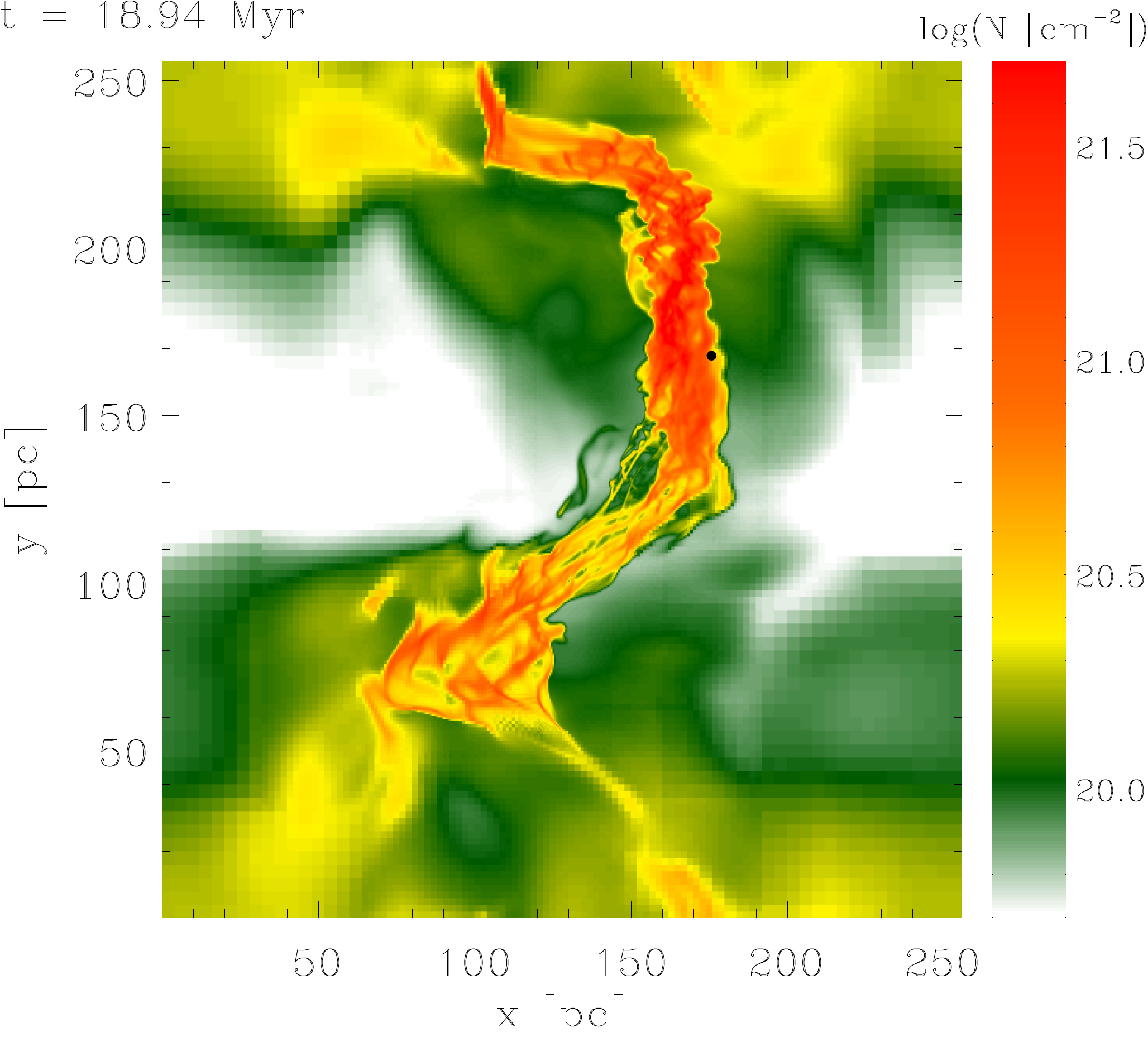}
\caption{Column density map for run B3M0.5I50. The total integration length is 60\,pc. The resulting global cloud structure 
is due to the initial compression by the flows.}
\label{fig8}
\end{figure}

\begin{figure}
\includegraphics[height=6cm,angle=-90]{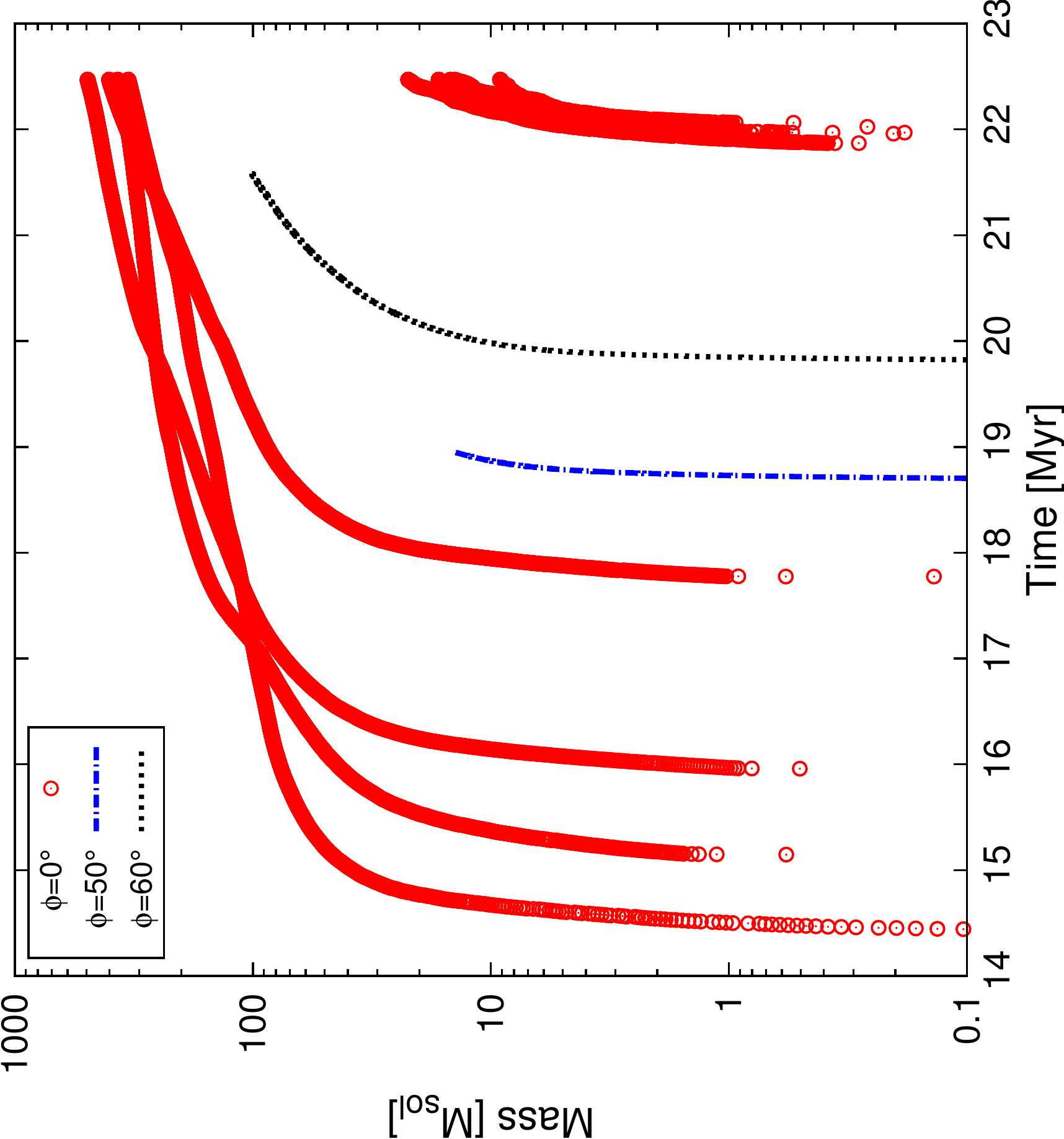}
\caption{Evolution of individual sink particle masses for three inclinations of runs B3M$\dots$I$\dots$. The higher the inclination, the later star formation begins. In addition, the star formation rate and efficiency 
are strongly influenced (see also fig. \ref{fig2}).}
\label{fig9}
\end{figure}

We commence with the thermally dominated case, i.e. $\left|\vek{B}_0\right|=3\,\mu\mathrm{G}$. The inclination of one flow increases the diffusivity (see tab. \ref{tab1} and appendix \ref{appa}) where larger angles lead to larger 
magnetic diffusivity. Fig. \ref{fig2} has already shown the evolution of the cloud mass as function of time. The mass crucially depends on the strength of the initial turbulent 
velocity fluctuations. In addition, if one applies an inclined flow, shearing motions and magnetic effects have to be taken into account. The magnetic field is able to slow down 
the inclined flow so that the collision will end soon and no gas is driven into the thermally unstable regime. But the final masses of the formed clouds are very similar, 
only varying by a factor of a few (see fig.\ref{fig7}). This indicates that at later times, the information of the initial conditions is completely lost. \\
More interesting is the \ita{way how the cloud evolves}. For small inclinations 
$\left(\varphi\leq30^{\circ}\right)$, 
 no significant distortions occur and the mass accumulation and the final mass are comparable to cloud masses formed by purely head--on collisions (within factors of 2-3). The inclined flow is aligned with the 
magnetic field very fast. For highly inclined streams the 
condensation from the WNM to the CNM sets in later due to the above mentioned processes. At the same time, mass growth is stopped and a short phase of \ita{mass loss} is evident as a direct consequence of strong shearing 
motions (see fig. \ref{fig7}).
But as soon as the strongest unstable fronts have 
vanished, the cloud turns back to a stabilised state with continuous accretion of matter. Figure \ref{fig8} shows the resulting molecular cloud structure for run B3M0.5I50. Shown is the column 
density along the z--axis, that is, perpendicular to the background magnetic field. The black dot resembles a sink particle. The global morphology of the cloud is mainly 
influenced by the geometry of the colliding WNM streams with additional impact by the misalignment of the flow. It resembles a sheet--like shape with trailing arms with the one at the near side of the tilted flow 
being more elongated. This elongation is due 
to the later collision of the flows when the bulk of the mass has already been compressed. The resultant motion of the cloud yields that the still streaming gas interacts with the outer edges of the compressed gas by 
'pushing' it away from the actual molecular cloud complex, thereby forming this observed elongated structure. At this time the flow is already too slow to significantly compress the gas, implying that the gas in the 
trailing arm is not able to sufficiently cool down by thermal instability. It is therefore not able to become gravitationally unstable. These shear flows and the resulting occurence of trailing arms are possibly seen in 
observations of e.g. the Taurus molecular cloud, i.e. the non--star forming low column density arm \citep[][]{Alves14}.\\
In contrast to the evolution of the total mass of the molecular cloud, the evolution of the stellar (sink particle) mass is greatly influenced by inclining one flow. The most obvious indication is the delay of star 
formation with increasing misalignment (see fig. \ref{fig9}).
Due to the misalignment, magnetic pressure and magnetic tension act as opposing agent against gravity. In addition, the shear flows disrupt density enhancements and thus the transition from the WNM to dense, cold 
structures is hampered. Once, the turbulence has fully vanished, the cloud is still subject to its fast bulk motion. Clumps within the complex can only grow by accretion of matter from the immediate environment, because of 
the lack of turbulent compression, which could provide the seeds for gravitational unstable cores. The shear due to the misalignment also yields a less compact cloud. The material is not fully compressed by the two flows. 
Instead, a great amount is at first compressed and enters a phase of oscillating motions and dispersion due to shearing 
motions. The accumulation of enough Jeans masses to render the gas gravitationally unstable is delayed and 
also very inefficient, since the denser regions are greatly scattered and do not possess enough mass.\\
As can be seen from fig. \ref{fig8}, there is at least one sink particle, indicating ongoing star formation. We stopped the 
simulation here, because star formation proceeds from there on (as can be seen e.g. from run B3M0.4I0).

\subsection{Comparing cloud dynamics in magnetically differing environments}

\begin{figure*}
 \centering
 \begin{tabular}{ccc}
 \fat{$\left|\vek{B}\right|=3\,\mu\mathrm{G}$}	&\fat{$\left|\vek{B}\right|=4\,\mu\mathrm{G}$}	&\fat{$\left|\vek{B}\right|=5\,\mu\mathrm{G}$}\\
\includegraphics[height=5cm]{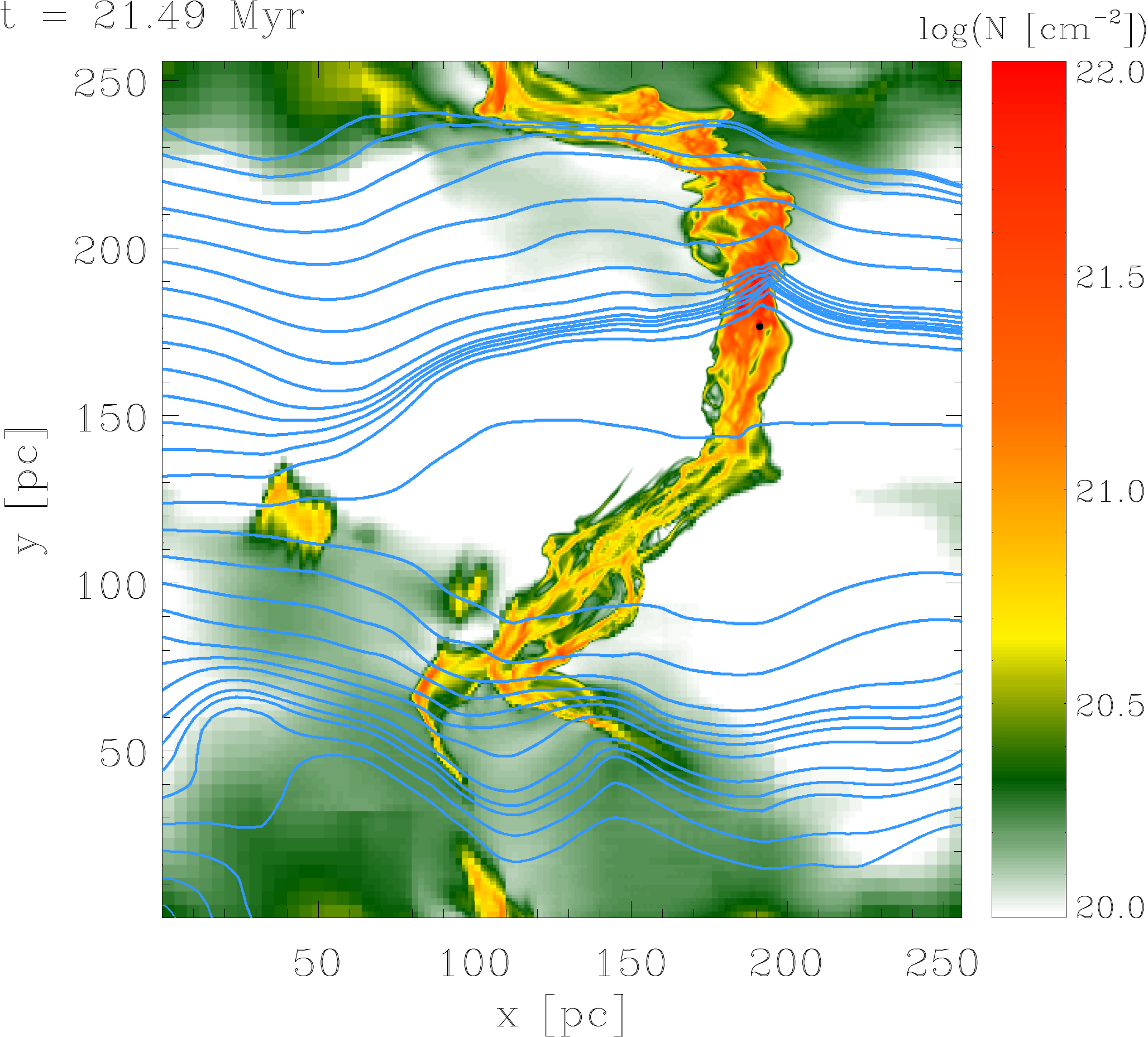} &\includegraphics[height=5cm]{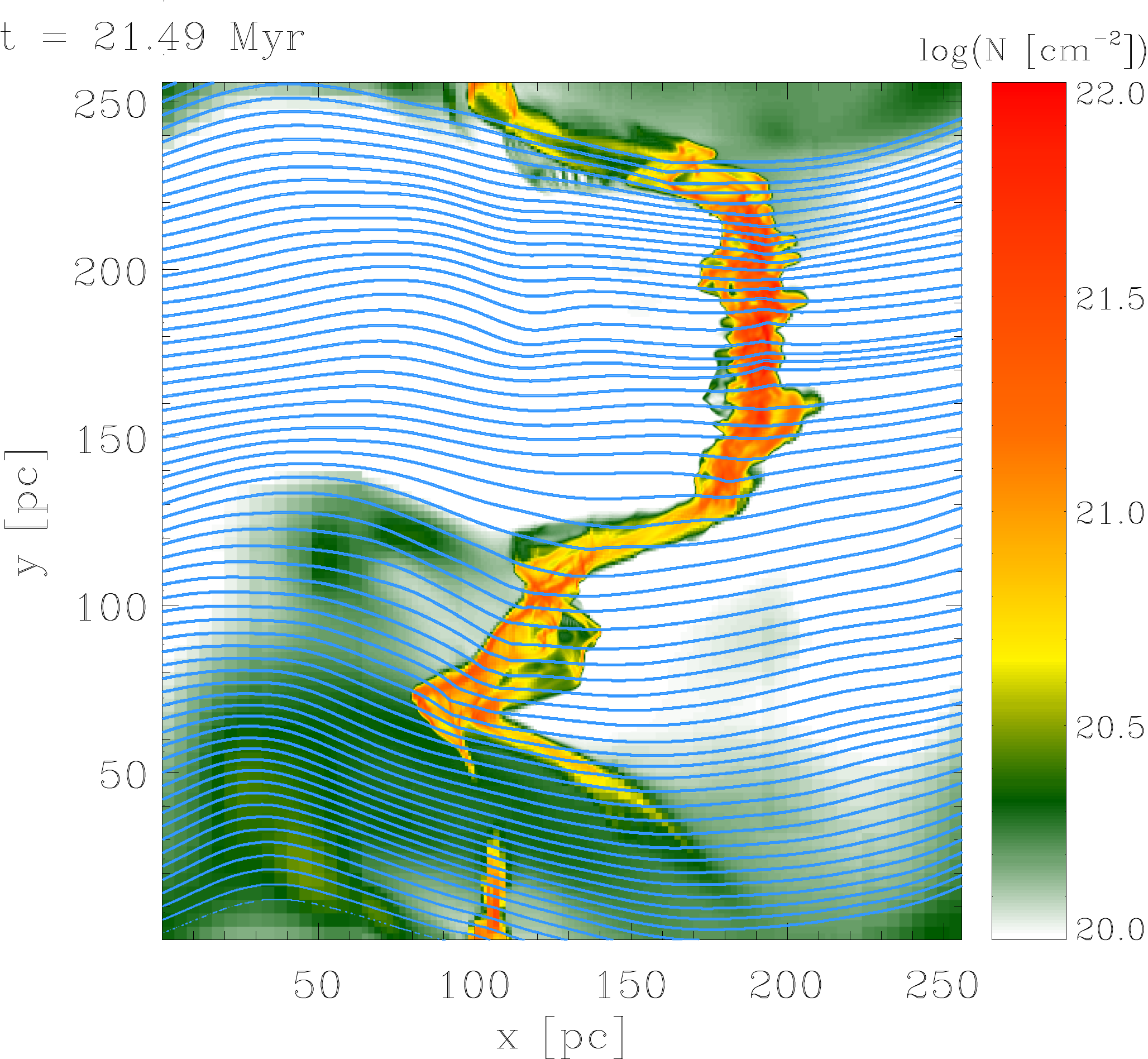} &\includegraphics[height=5cm]{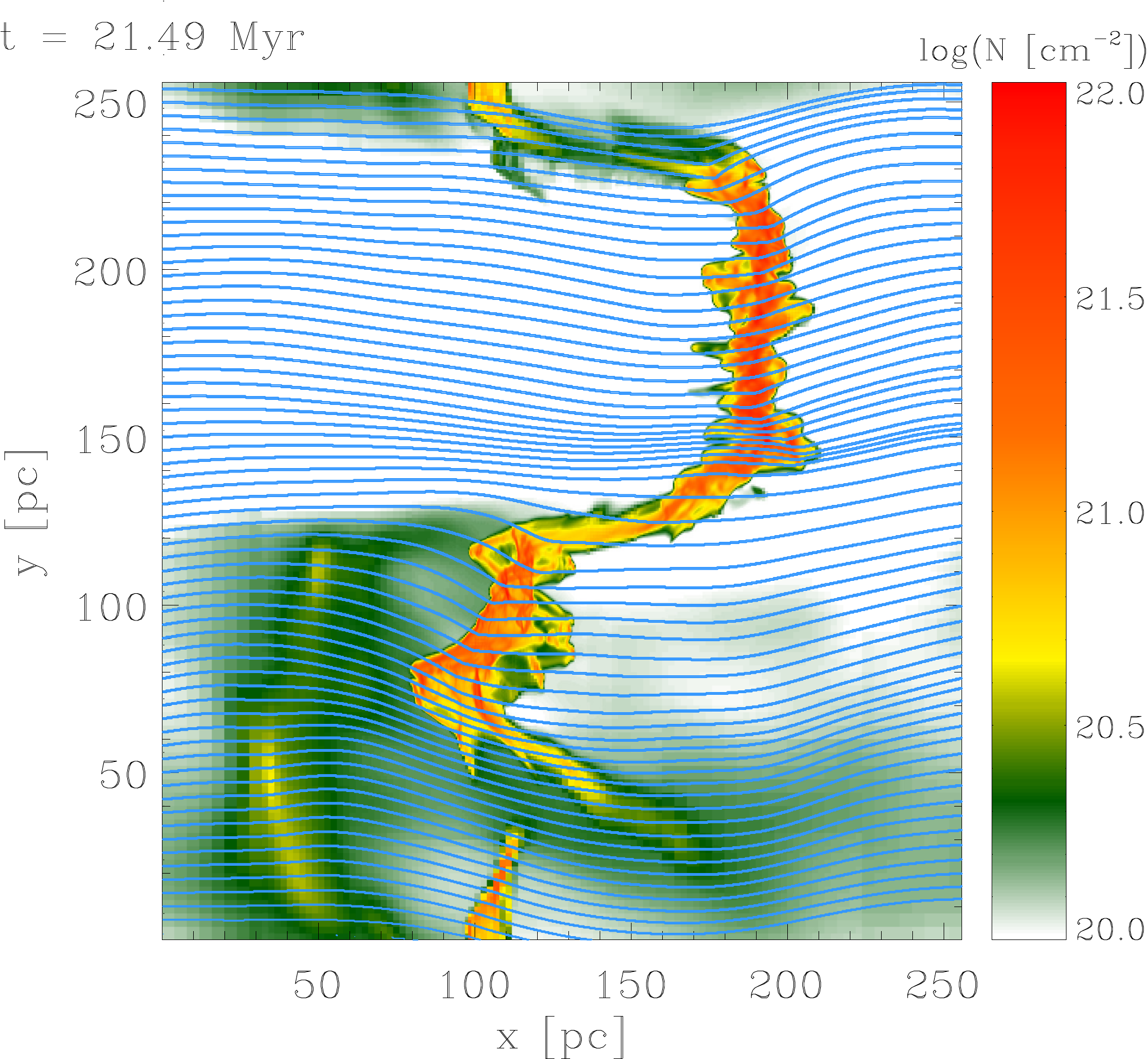}
 \end{tabular}
 \caption{Column density map with overlayed magnetic field lines for $\mathrm{B}=3\,\mu$G,$\mathrm{B}=4\,\mu$G, and $\mathrm{B}=5\,\mu$G from left to right. Initial inclination is $\Phi=60^{\circ}$.}
 \label{fig10}
\end{figure*}

Figure \ref{fig10} shows the column density in the direction perpendicular to the background magnetic field for three different initial magnetic field strengths (from left to right: $\left|\vek{B}\right|=3\,\mu\mathrm{G},
\left|\vek{B}\right|=4\,\mu\mathrm{G, and}\left|\vek{B}\right|=5\,\mu\mathrm{G}$) with an initial tilt of $\varphi=60^{\circ}$. The weakest field case shows a strong distortion of magnetic field lines as well as the onset of star 
formation. In comparison, the stronger fields show a more ordered magnetic field, which shows no clear deviation from its initial uniform alignment. For $\left|\vek{B}\right|=4\,\mu\mathrm{G}$ one can infer some large scale 
modulation of the field due to global dynamics as a resulting imprint of the large inclination. The morphology of all three molecular clouds is very similar, although some local differences occur. The main cloud (having a
sheet--like shape) is more compact for weaker fields, whereas the difference between the two strong magnetisations is negligible. This attribute results from the thermally dominated gas. The magnetic field 
does not control the gas dynamics and thus is forced to follow the motion of the fluid. Once, local density enhancements condense out, the magnetic field is dragged inwards together with accreting material. In the cases of 
more realistic fields, it is the magnetic field that dominates the fluid motion and that keeps the cloud coherent \citep[see e.g.,][]{Hennebelle13}. 
At the same time the trailing arms now occur to be slightly denser. These arms are magnetically supported and thus more stable against shear flows and mixing by large scale fluid instabilities.\\
In contrast to a 3\,$\mu$G--field, there is no star formation for the cases of 4\,$\mu$G and 5\,$\mu$G, yet, although fig. \ref{fig4} indicates that these highly magnetised clouds are also more massive. The greater total masses and 
the lack of star formation combine to a picture of a fragmented cloud (see fig. \ref{fig14}). Any intrinsically driven turbulence is 
subalfv\'{e}nic and the magnetic field thus stays coherent. Such a field configuration has also been observed via polarised emission from CO \citep[e.g.][]{Li10,Li14}. So, what is the basic impact of the magnetic field 
on the star formation process? As long as accretion happens along the magnetic field, the influence of the latter can be safely ignored. Once, the gas begins to fragment, subcritical regions are produced, as long as the 
parental fragment was only slightly supercritical. Thus, accretion along field lines has to continue in order to generate 
supercritical fragments. Otherwise, magnetic 
pressure will drive the gas out of the potential well and the fragments stay subcritical and star formation stops.\\
The temporal evolution of 
\begin{figure*}
\includegraphics[height=6cm,angle=-90]{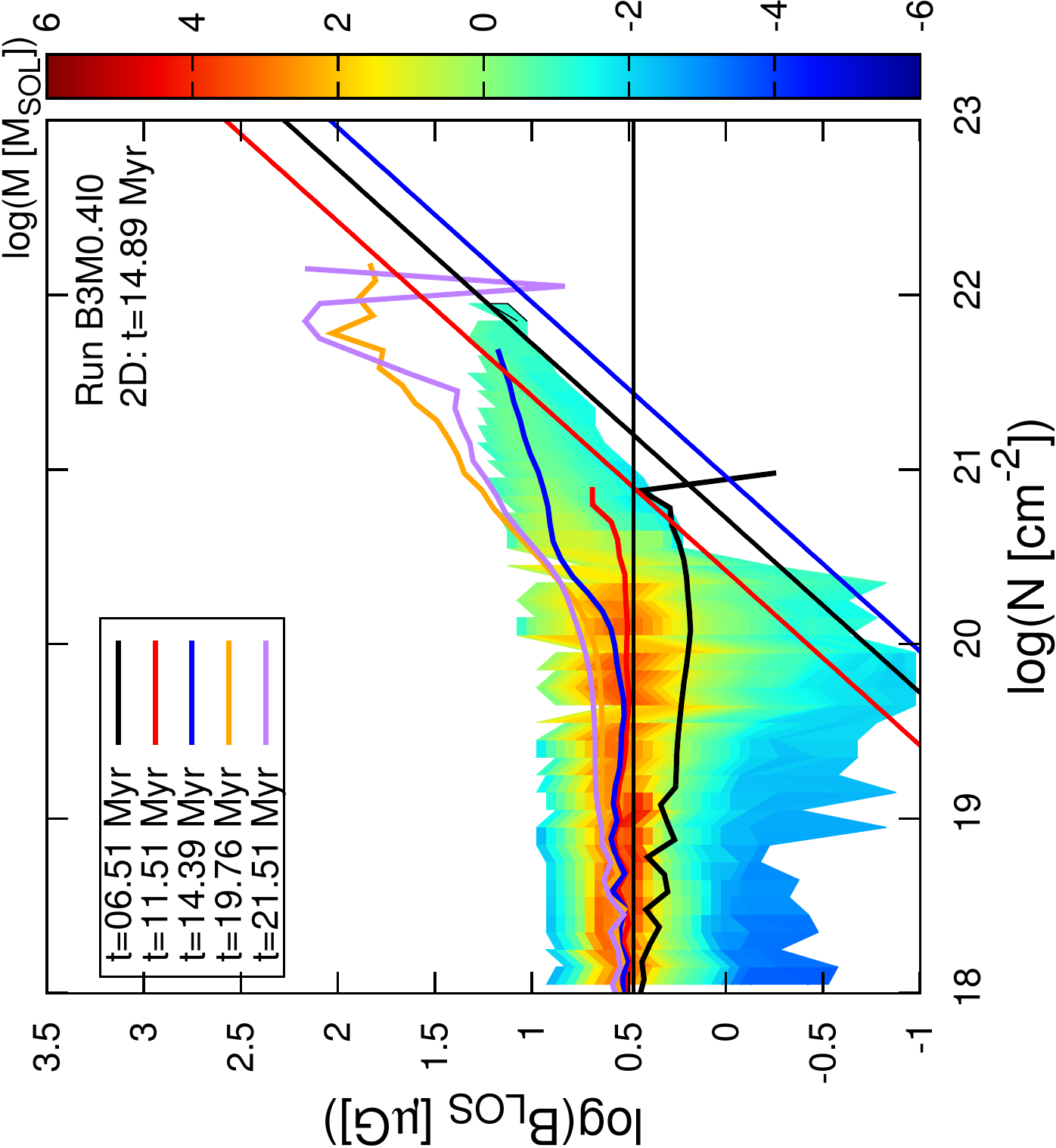}\,\includegraphics[height=6cm,angle=-90]{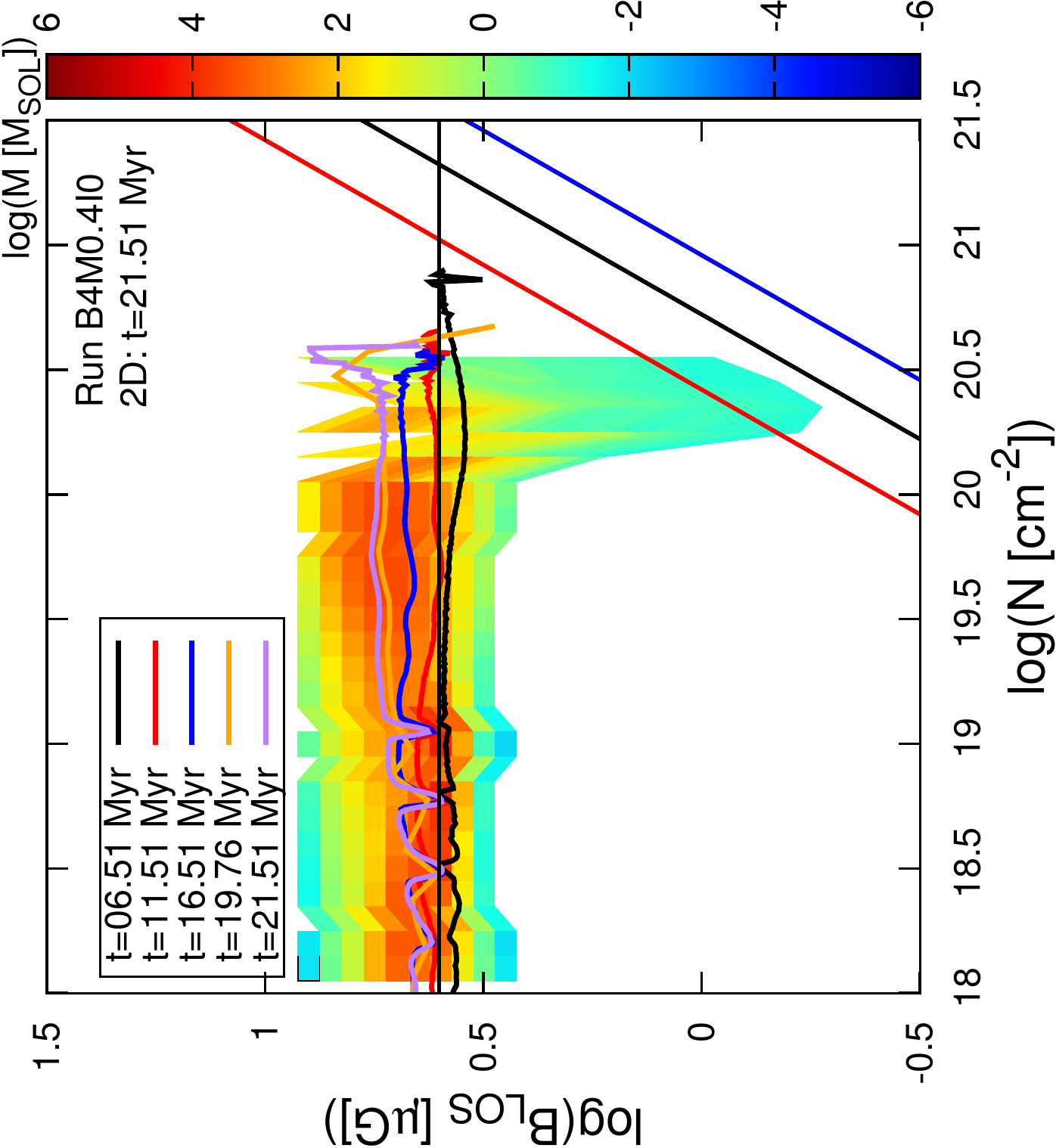}\,\includegraphics[height=6cm,angle=-90]{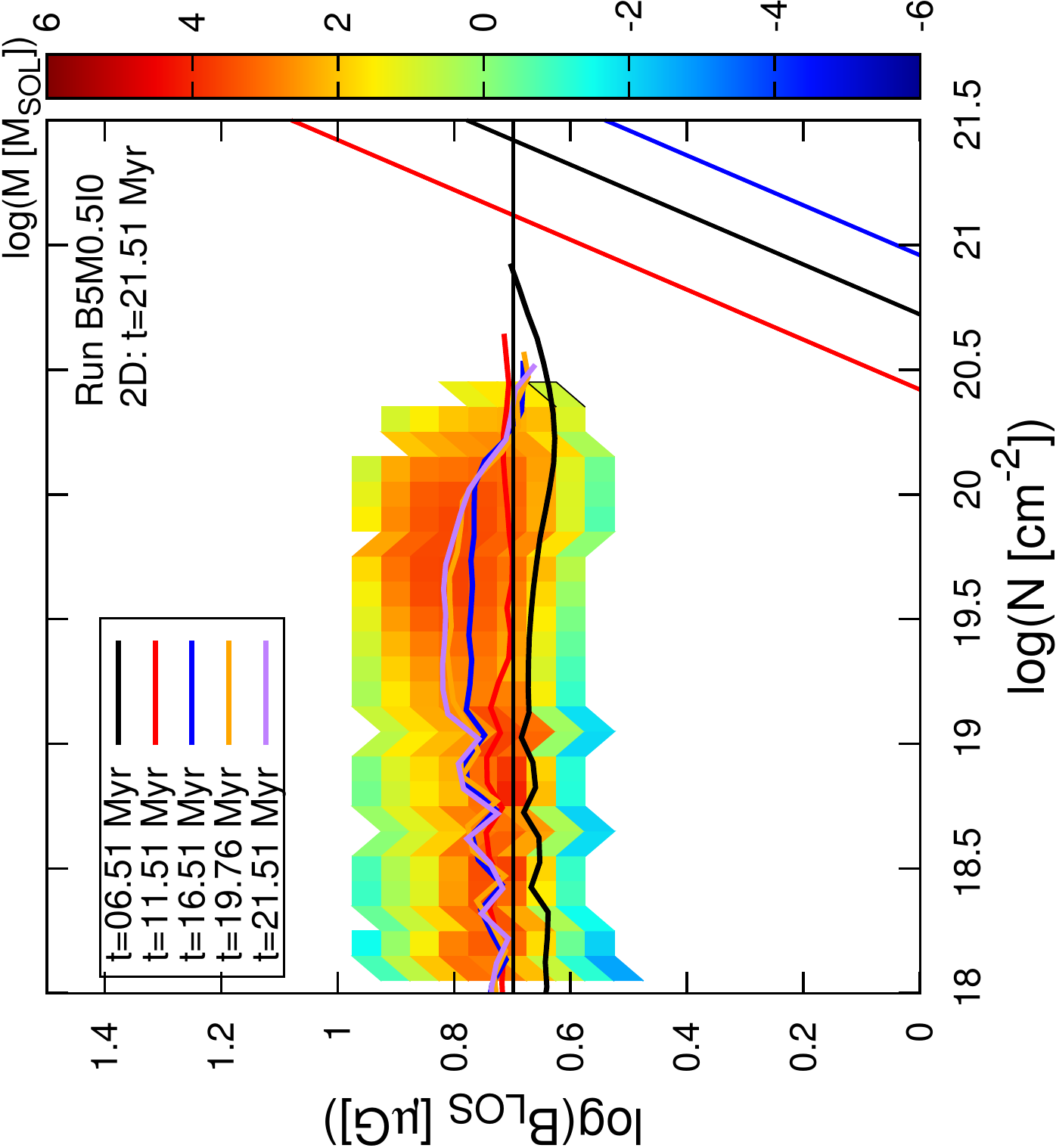}\\
 \includegraphics[height=6cm,angle=-90]{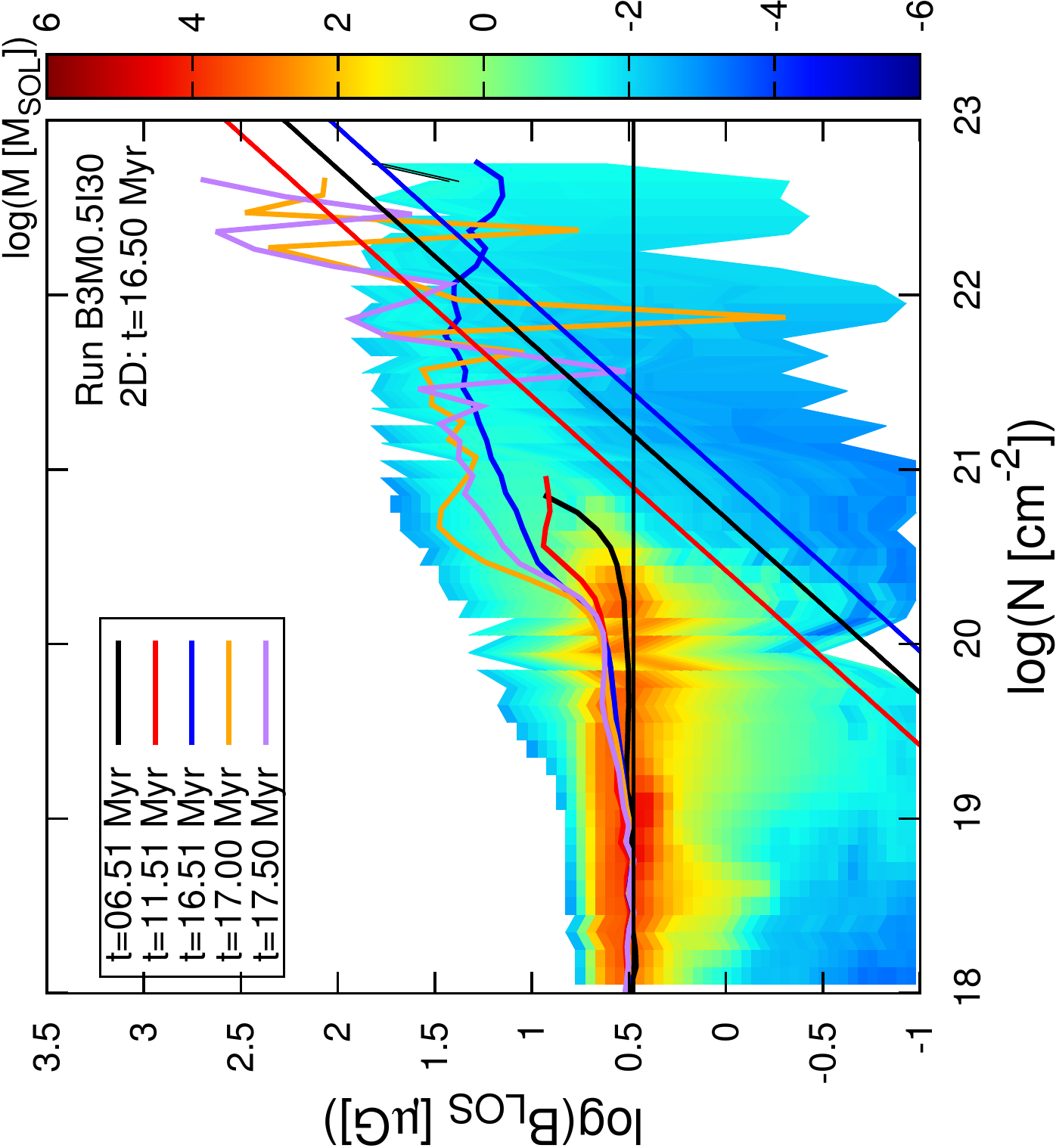}\,\includegraphics[height=6cm,angle=-90]{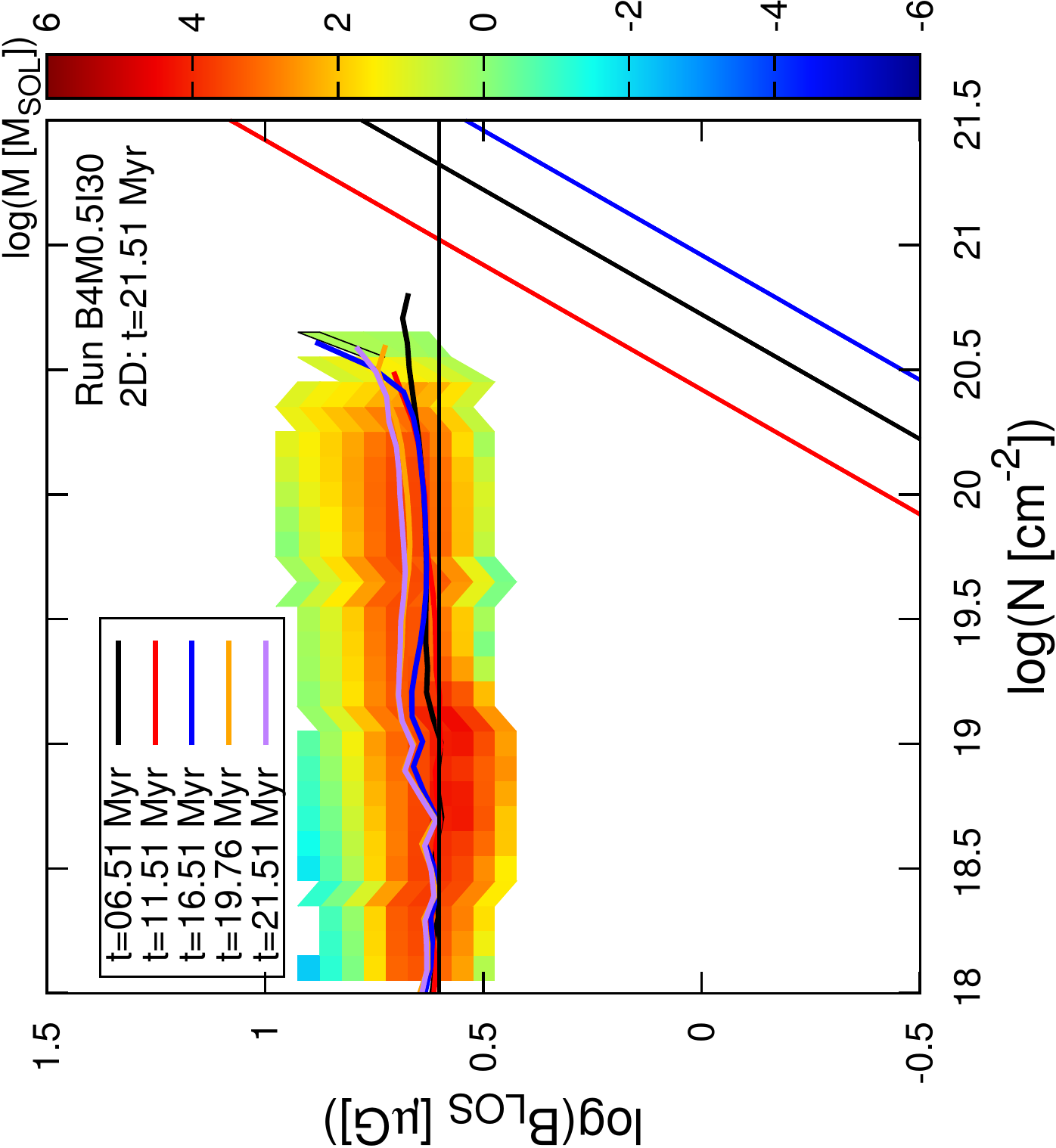}\,\includegraphics[height=6cm,angle=-90]{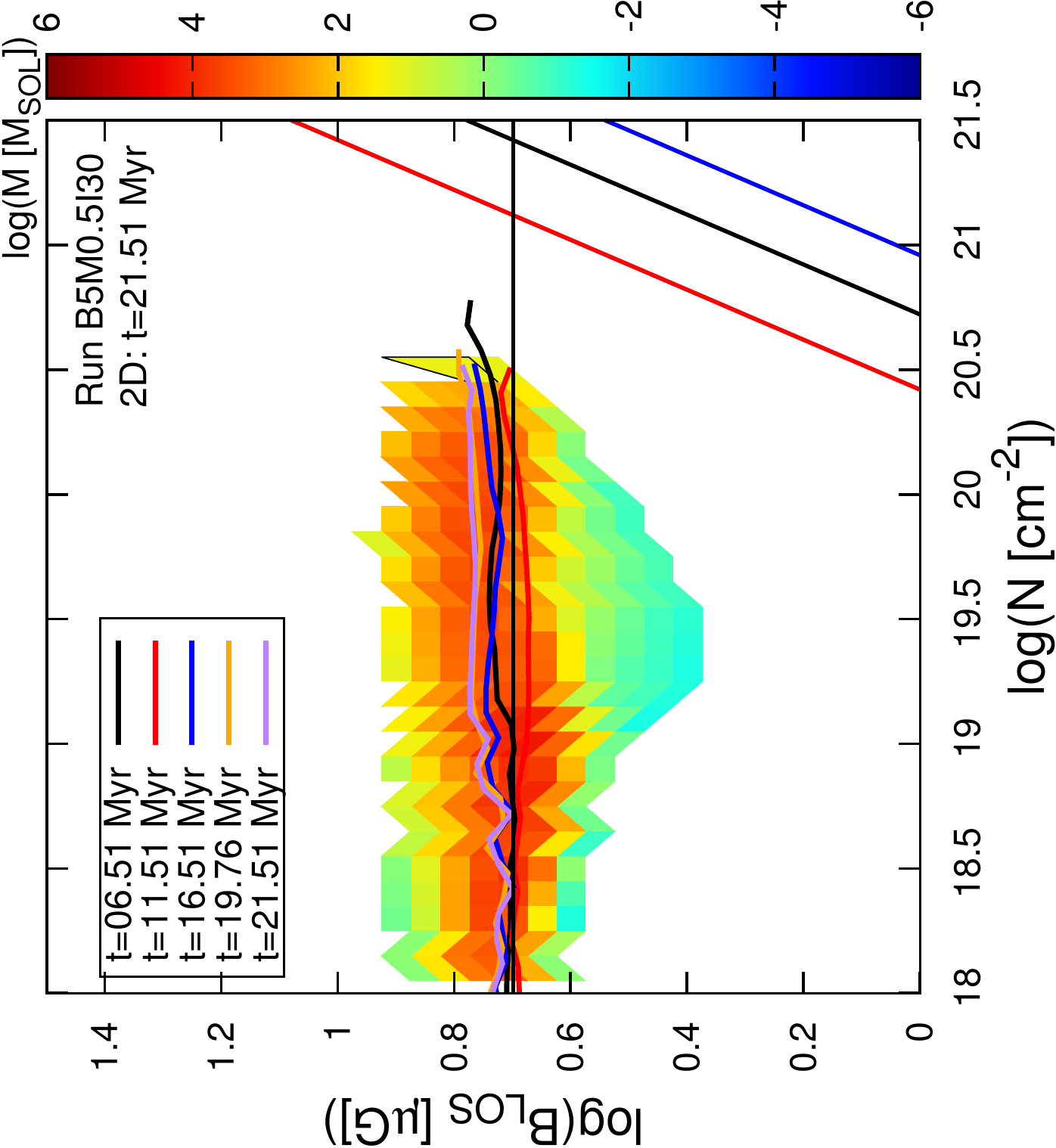}\\
 \includegraphics[height=6cm,angle=-90]{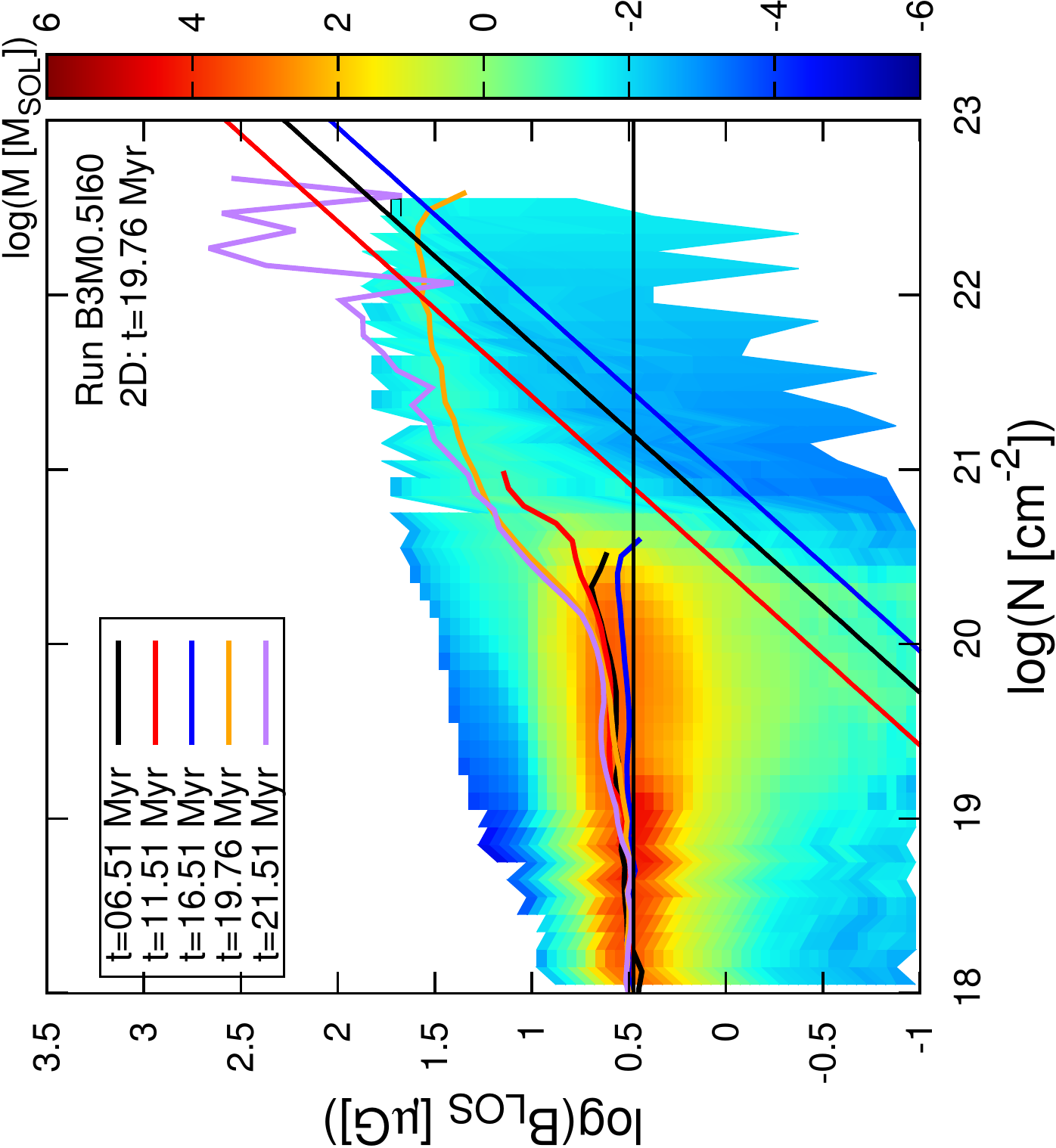}\,\includegraphics[height=6cm,angle=-90]{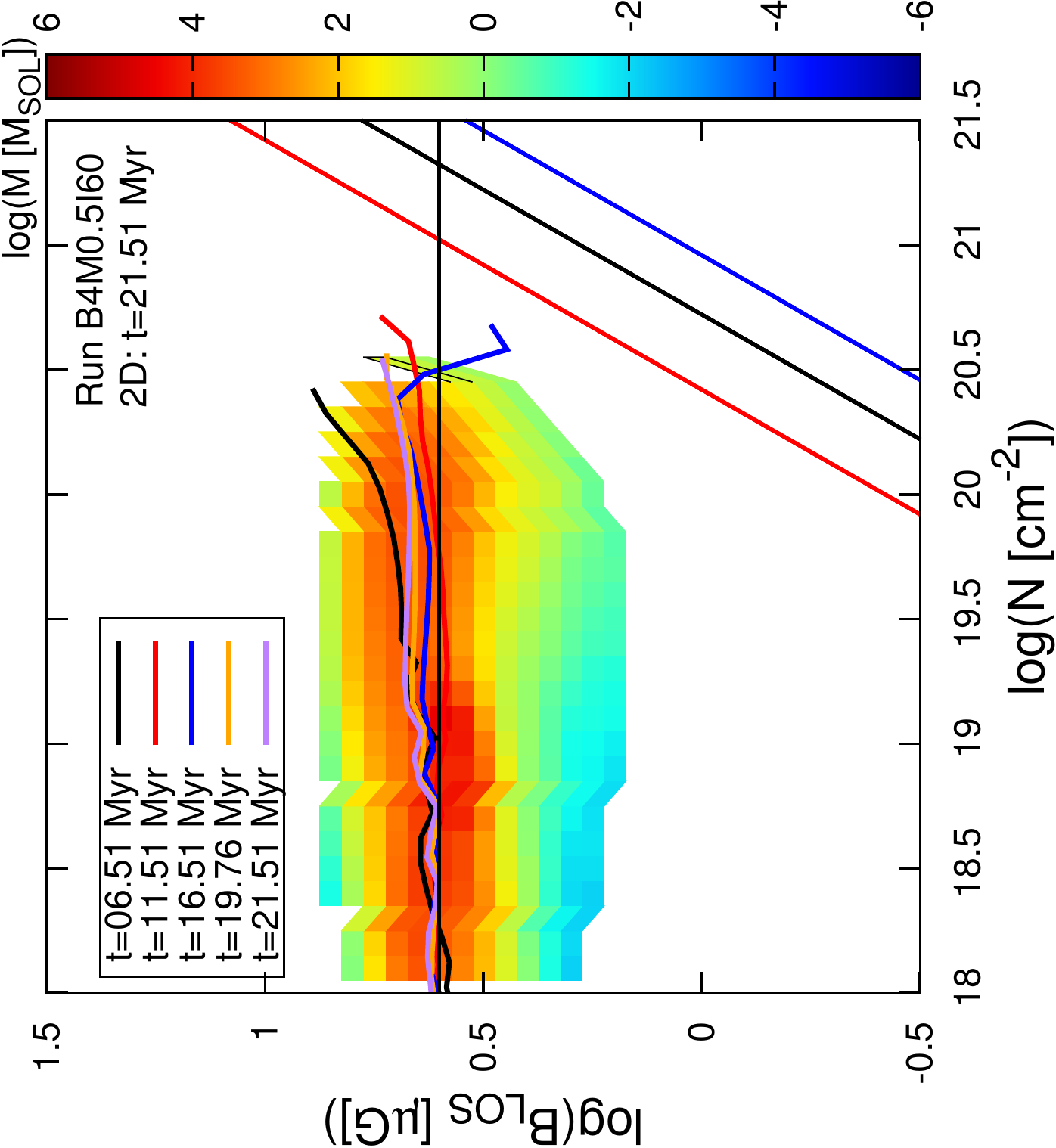}\,\includegraphics[height=6cm,angle=-90]{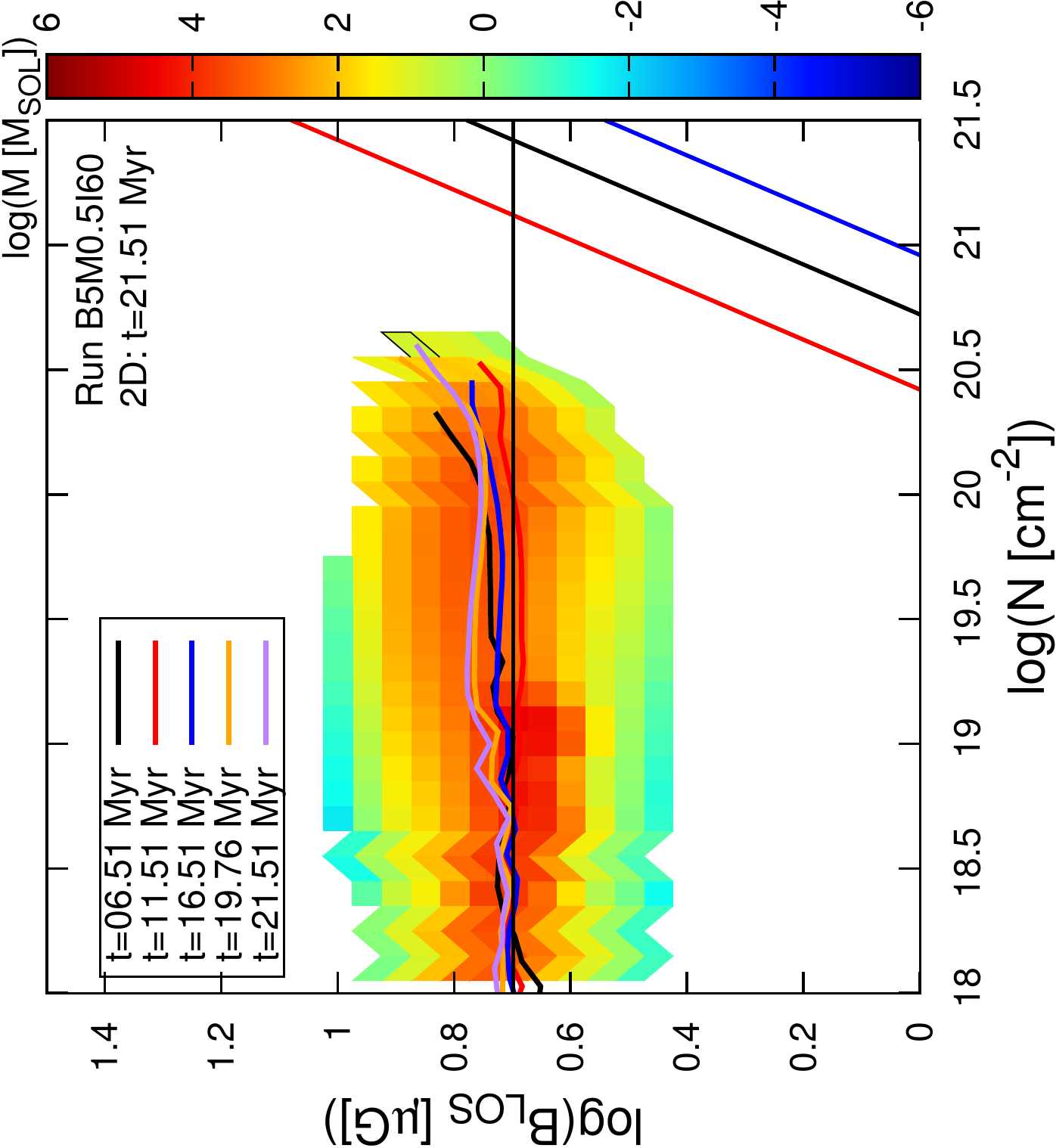}
\caption{Evolutionary path of the cloud in $\mathrm{N-B}_{\mathrm{LOS}}$ space. In this case, $B_\mathrm{LOS}=B_\mathrm{x}$. From left to right: $\mathrm{B}=3\mu$G,$\mathrm{B}=4\mu$G, and $\mathrm{B}=5\mu$G, respectively. From top to bottom: $\Phi=0^{\circ}$,$\Phi=30^{\circ}$,$\Phi=60^{\circ}$. Different colours denote different times. Note the different data range 
 for the stronger fields. Also shown are the criticality condition \citep[dot-dashed black line]{Crutcher10,Crutcher12}, corrected for projection effects \citep[solid straight black line]{Shu99}, and assuming equipartition 
 of turbulent and magnetic fields \citep[dashed black line]{McKee93}. Colour coded is the mass as function of column density, $N$, and line--of--sight magnetic field $B_\mathrm{LOS}$.}
\label{fig11}
\end{figure*}
the line--of--sight component of the magnetic field as function of column density is shown in fig. \ref{fig11}. Different colours or linestyles denote different evolutionary stages. The straight lines indicate different 
criticality conditions according to various studies and using different approaches for deriving this condition \citep[][]{McKee93,Shu99,Crutcher10,Li14}. From left to right the initial magnetic field becomes stronger and from 
top to bottom the inclination increases in steps of $30^{\circ}$, starting at $\varphi=0^{\circ}$. All cases have in common that the column density gradually increases at \ita{constant} magnetic field magnitude, 
indicating gas accumulation along the field lines \citep[see also][]{Crutcher10}. \\
The thermally dominated case shows signs of early fragmentation and field compression, giving rise to an increase of the magnitude at relatively low column densities. 
These effects render the whole cloud magnetically subcritical. The clouds then undergo different dynamical phases with varying contribution of the magnetic field, i.e. times of pure accretion along the field lines, twisting of the field 
by collapse and compression, and finally amplification of the field by large scale collapse and increasing column density. In between there exist stages, where the cloud shows signs of supercriticality. We here point out 
that the ordinate only shows the \ita{average} line--of--sight component, i.e. there exist indeed supercritical regions that are not significant in terms of mass or volume fraction (see 2D histogram). Comparison with the weakest field shows 
that star 
formation is immediately initiated, when the gas becomes supercritical. The collapse proceeds and more material is dragged into the potential well. The resulting magnetic field amplification is still too low and finally it 
diffuses out of the central region. After the sink particle has formed, some of the field lines relax, thereby decreasing the density in 
some regions. As time proceeds the cloud becomes more compressed due to its global gravitational collapse.\\
For better visualisation, the columns for $\mathrm{B}=5\,\mu$G and $\mathrm{B}=4\,\mu$G are shown in the column density range $19.5\leq\log{N}\leq21$ and 
$1\,\mu\mathrm{G}\leq\left|\vek{B}\right|_\mathrm{LOS}\leq25\,\mu\mathrm{G}$, since there occurs no significant amplification of the magnetic field during the evolution of the molecular cloud. This is indeed very intriguing, because observed 
magnetic fields are far larger in magnitude. We only see motion along the field lines, as has already been mentioned before, but we do also see \ita{no sign of gravitational contraction}. Only some small modulations are seen, especially in the case of the cloud formed by head--on collision and $\mathrm{B}=4\,\mu$G, but this amplification is less than a factor of two and thus not 
significant. At low column densities instead one can infer a slight 'global' amplification of a view percent. This can be accounted for accretion of mass from the diffuse halo surrounding the dense cloud. These data 
already indicate that the uniform component of the magnetic field is the leading component and no clear tangling of the field is observed. Furthermore, every process of gas accumulation 
perpendicular to the field lines is instantaneously balanced by magnetic forces (see fig. \ref{fig12}). The flow cannot become dynamically important in order to bend the field lines and to render the magnetic field supercritical. \\
Even in the case 
\begin{figure}
 \includegraphics[width=7cm,angle=-90]{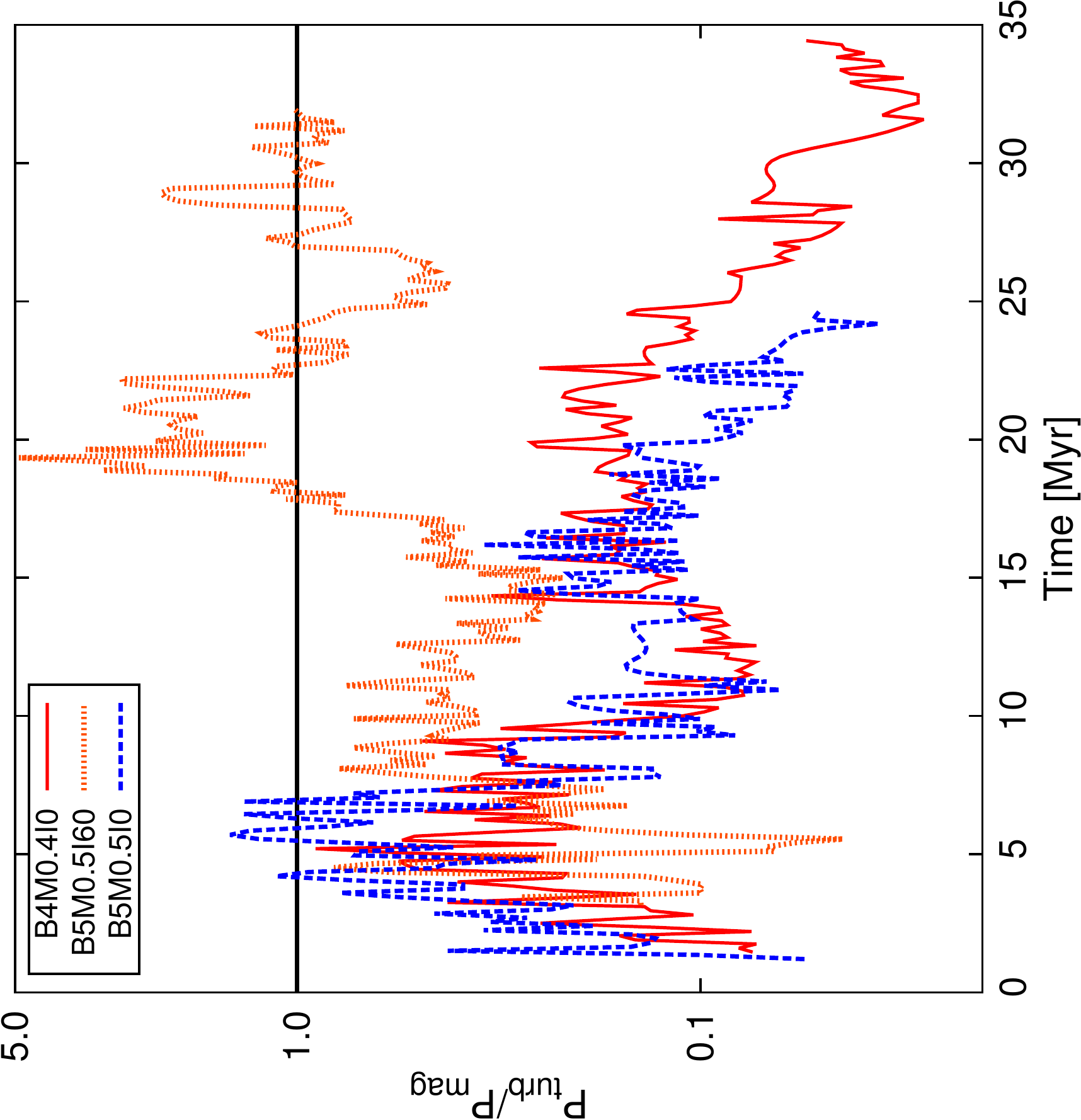}
 \caption{Temporal evolution of the ratio of turbulent ram pressure to magnetic pressure for three different runs. \ita{On average} the magnetic field dominates and the ram pressure is not sufficient to induce motions 
 perpendicular to the field lines.}
  \label{fig12}
\end{figure}
of high diffusivity or large inclination, there is no amplification and/or tangling seen, indicating that diffusion processes might play only a minor role in rendering the field supercritical. If one takes a look at 
fig. \ref{fig13}, it is obvious that the gas is highly subcritical. Shown are mass histograms as function of the normalised mass--to--flux ratio for the three magnetic fields at three late evolutionary stages. 
It is only for the weakest magnetic field that the gas 
\begin{figure*}
  \centering
  \begin{tabular}{ccc}
 \fat{$\left|\vek{B}\right|=3\,\mu\mathrm{G}$}	&\fat{$\left|\vek{B}\right|=4\,\mu\mathrm{G}$}	&\fat{$\left|\vek{B}\right|=5\,\mu\mathrm{G}$}\\
  \includegraphics[height=5cm,angle=-90]{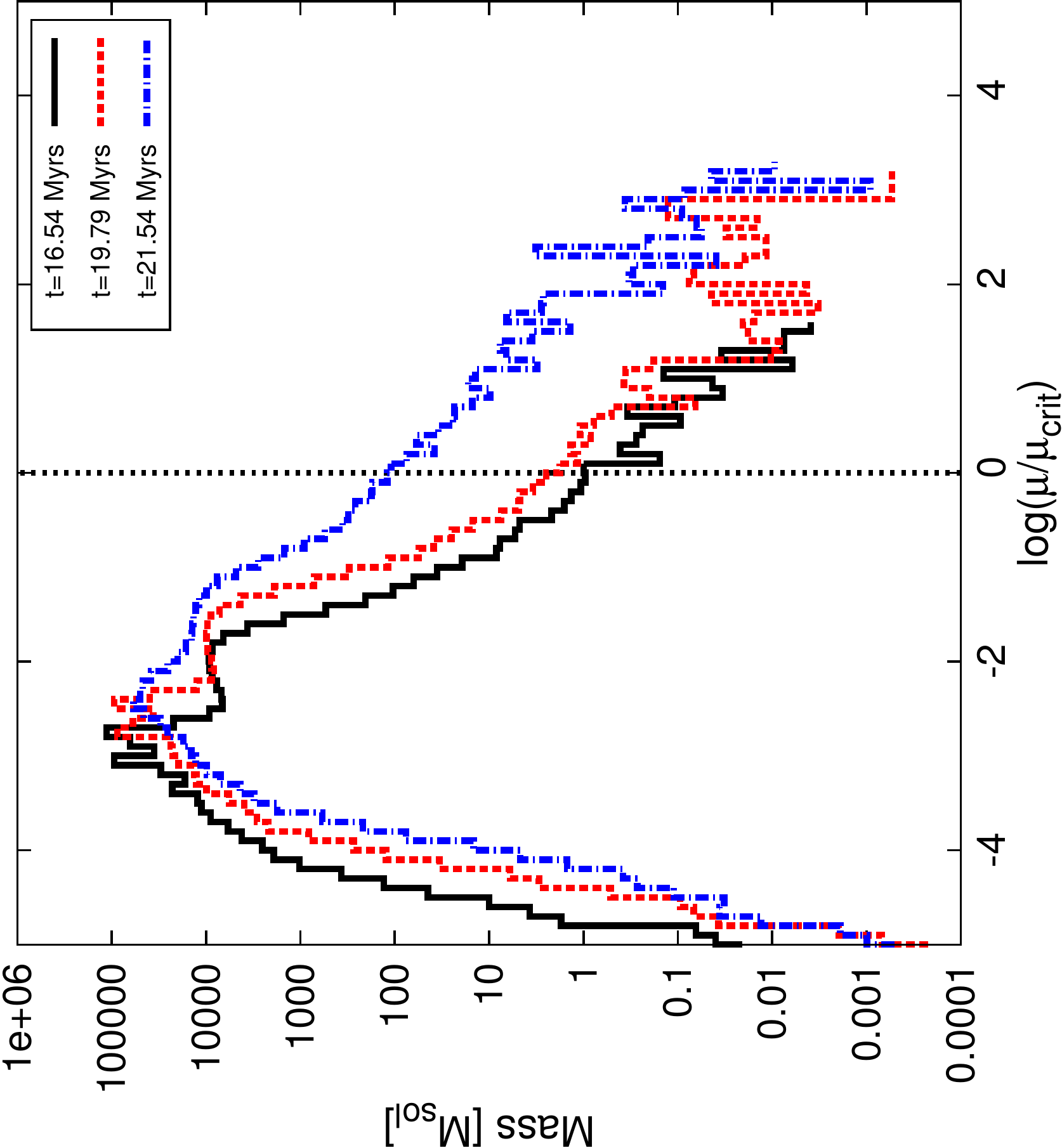} &\includegraphics[height=5cm,angle=-90]{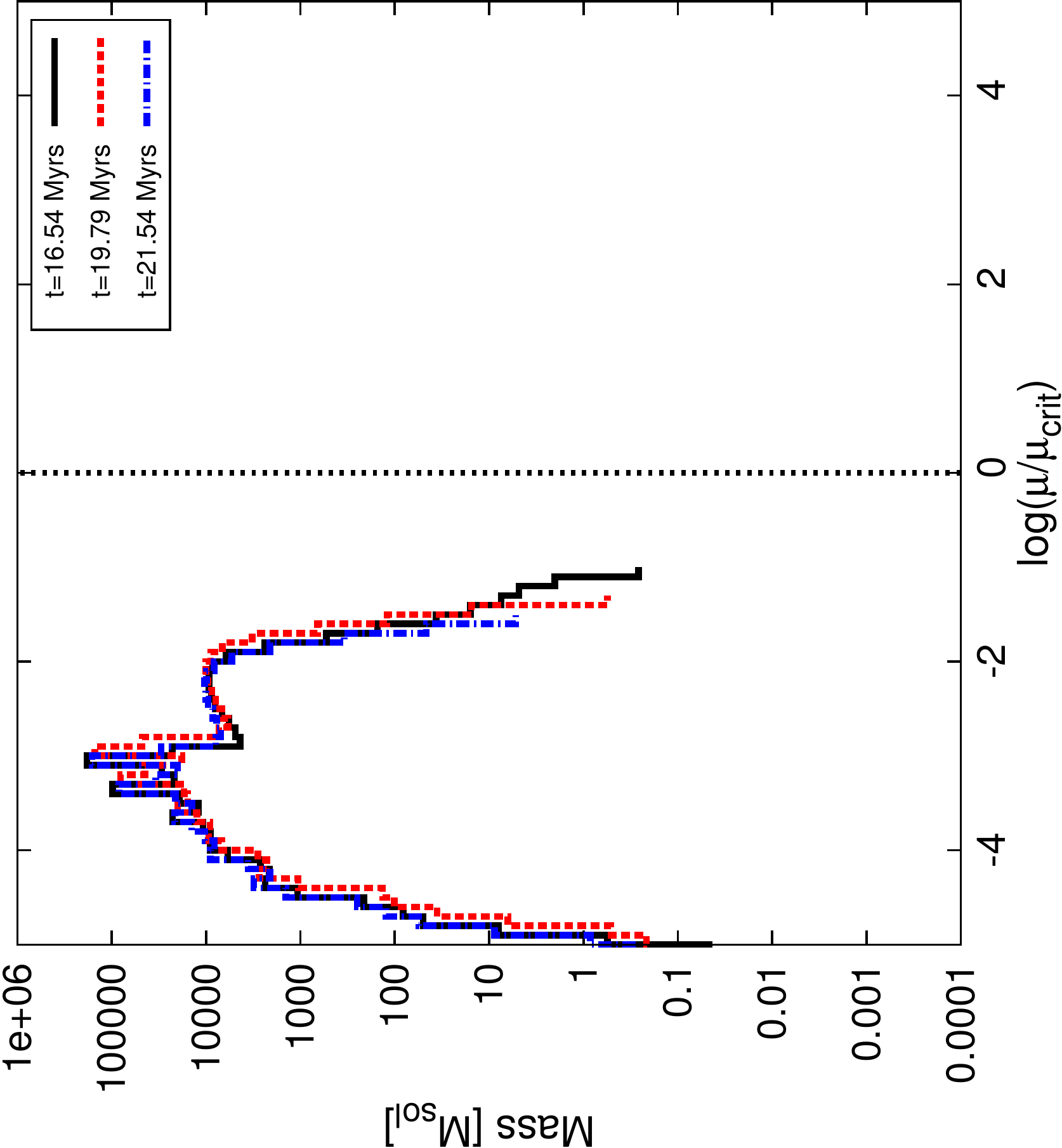} &\includegraphics[height=5cm,angle=-90]{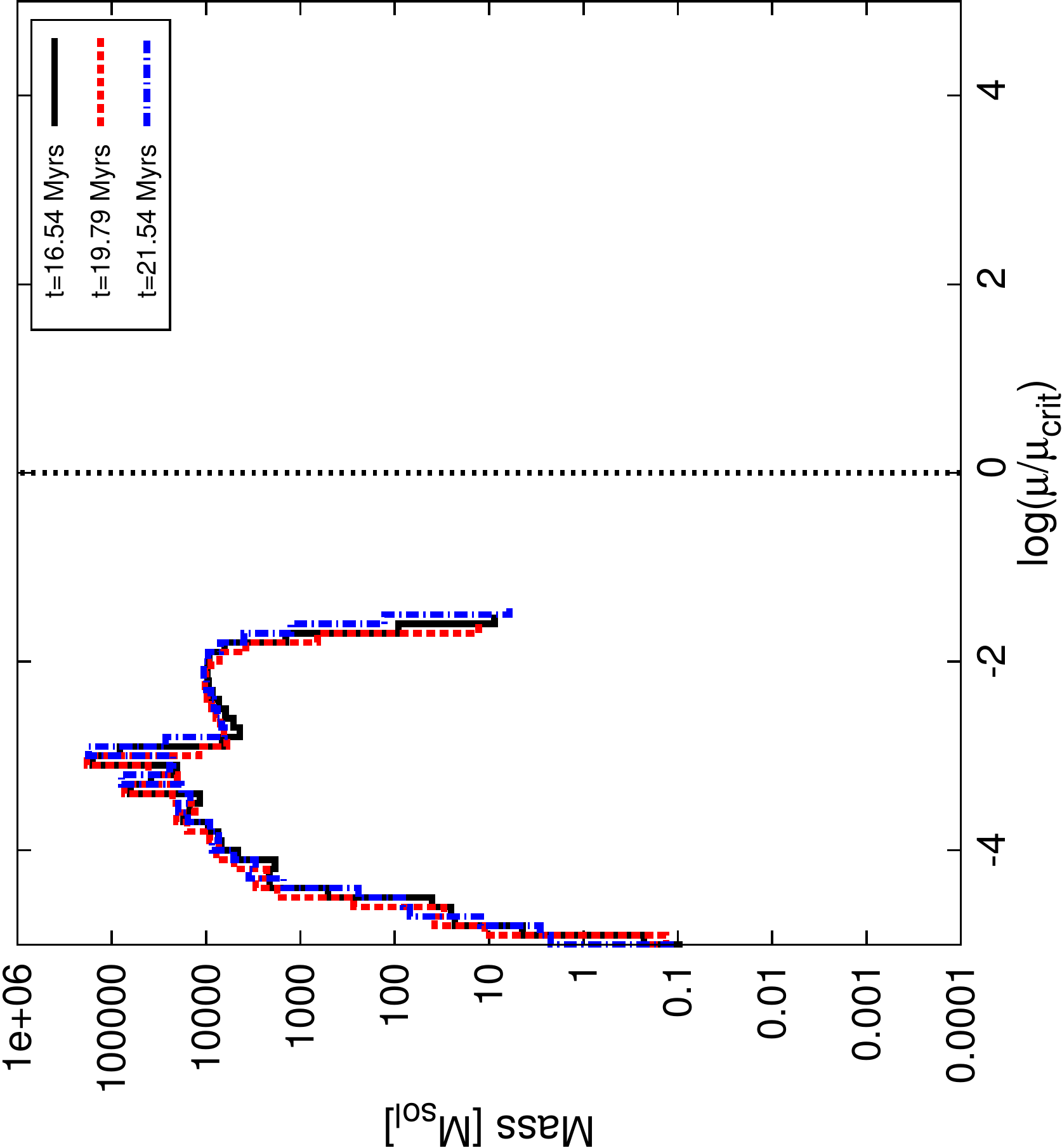}
  \end{tabular}
  \caption{Mass as function of the mass-to-flux ratio for B3M0.5I60, B4M0.5I60, and B5M0.5I60 (from left to right). Different colours denote different times. The vertical line again indicates the critical value.}
  \label{fig13}
\end{figure*}
shows some sign of evolution. One can clearly identify the power--law tail (which can be accounted for the $N$--PDF) and its growth as more mass enters the supercritical regime. In contrast, the higher magnetisation cases 
show roughly no evolution. Once a given distribution of the gas has developed it is seen to be globally stationary. The difference between the 4$\,\mu$G and 5$\,\mu$G cases are small, i.e. the 4$\,\mu$G case develops some 
larger mass--to--flux ratios. However, both regimes are far from being even critical.\\
Run B5M0.5I60AD includes the process of ambipolar diffusion in addition to an initial tilt. As was already mentioned in 
the remarks of table \ref{tab1} the simulation was stopped at $t\approx12\,$Myr. The subsequent evolution of the cloud 
showed no significant difference to the runs without ambipolar diffusion. Also in this case, the shear flows tend to suppress the formation of dense cores, where ambipolar diffusion would be most efficient.\\

\subsection{Dynamics of dense cores}
\begin{figure*}
 \centering
 \includegraphics[width=5.5cm,angle=-90]{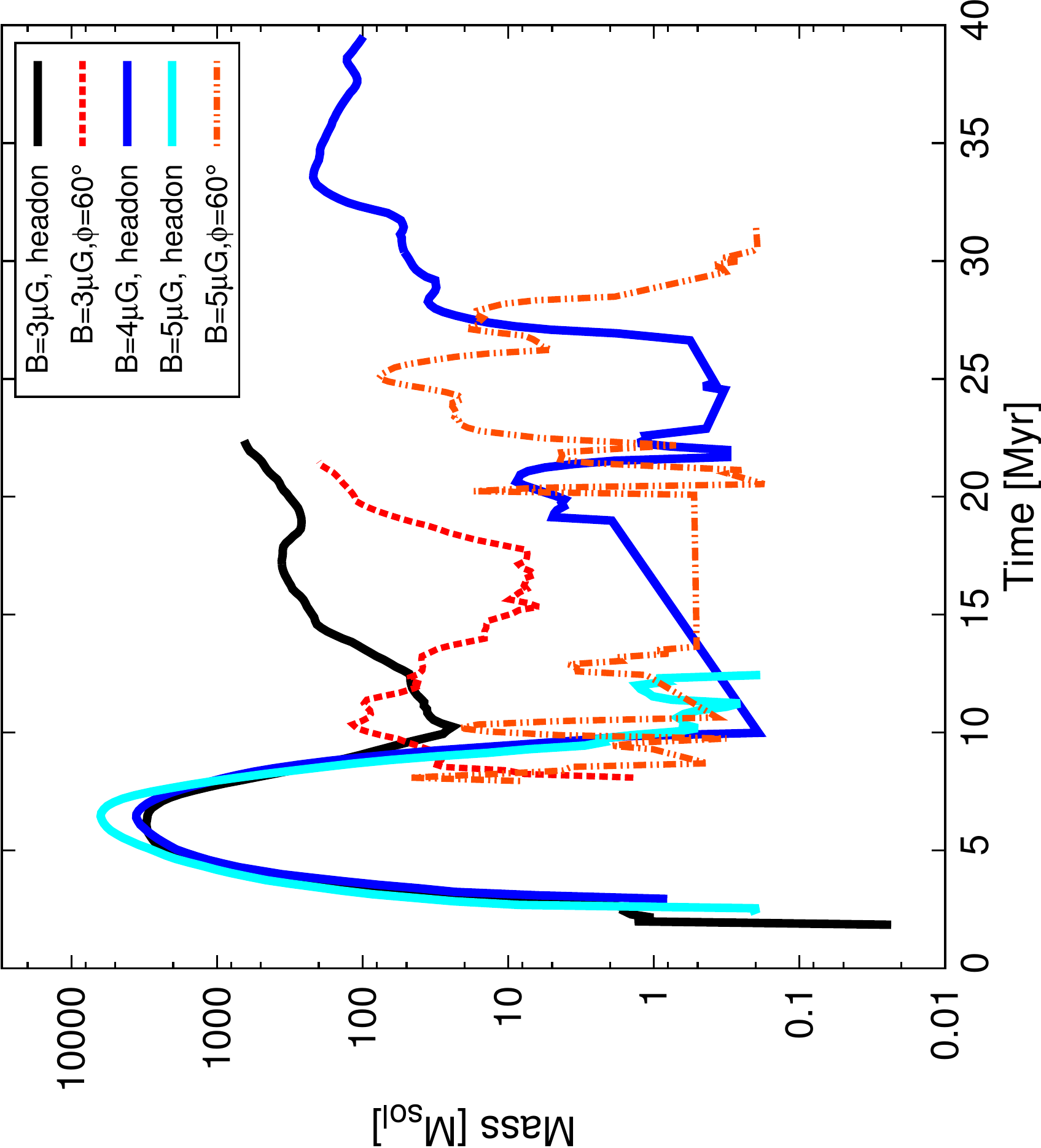}\,\includegraphics[width=5.5cm,angle=-90]{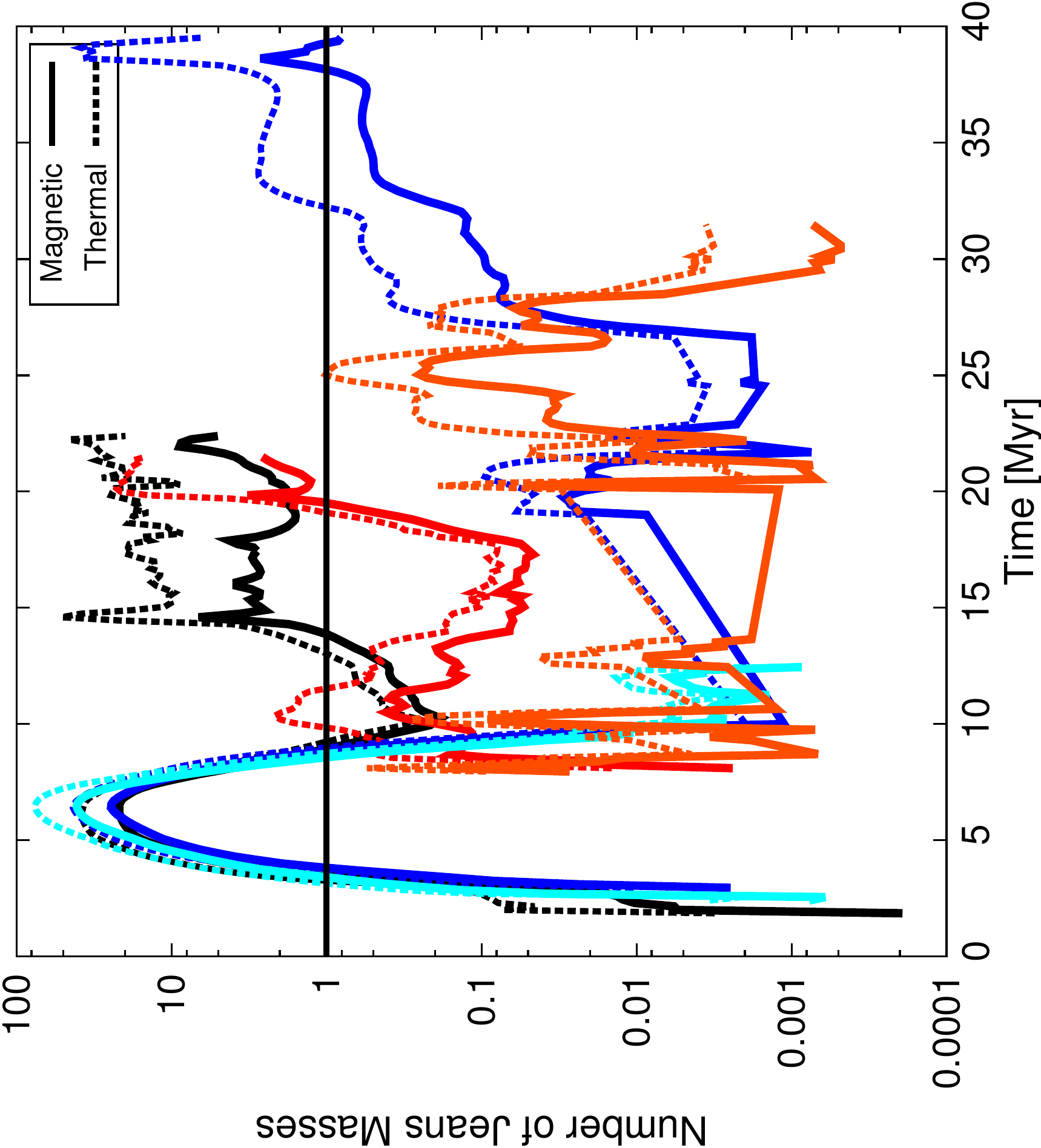}
  \caption{Temporal evolution of the gas with minimum density of $n=1000\,\mathrm{cm}^{-3}$, resembling the densest parts deeply embedded in the cloud.\ita{Left:} Temporal evolution of mass. Note that for runs B4M0.4I0 and 
  B5M0.5I60 there are stages where \ita{no dense material exists} [indicated by the linear increasing (B4M0.4I0) and the non--varying stages (B5M0.5I60). The former is just the line connecting two data points!] between 10 and 20 Myr.
  \ita{Right:} Evolution of the number of thermal and magnetic jeans masses. Colours correspond to the left figure. }
  \label{fig14}
\end{figure*}

\begin{figure}
 \includegraphics[width=7cm,angle=-90]{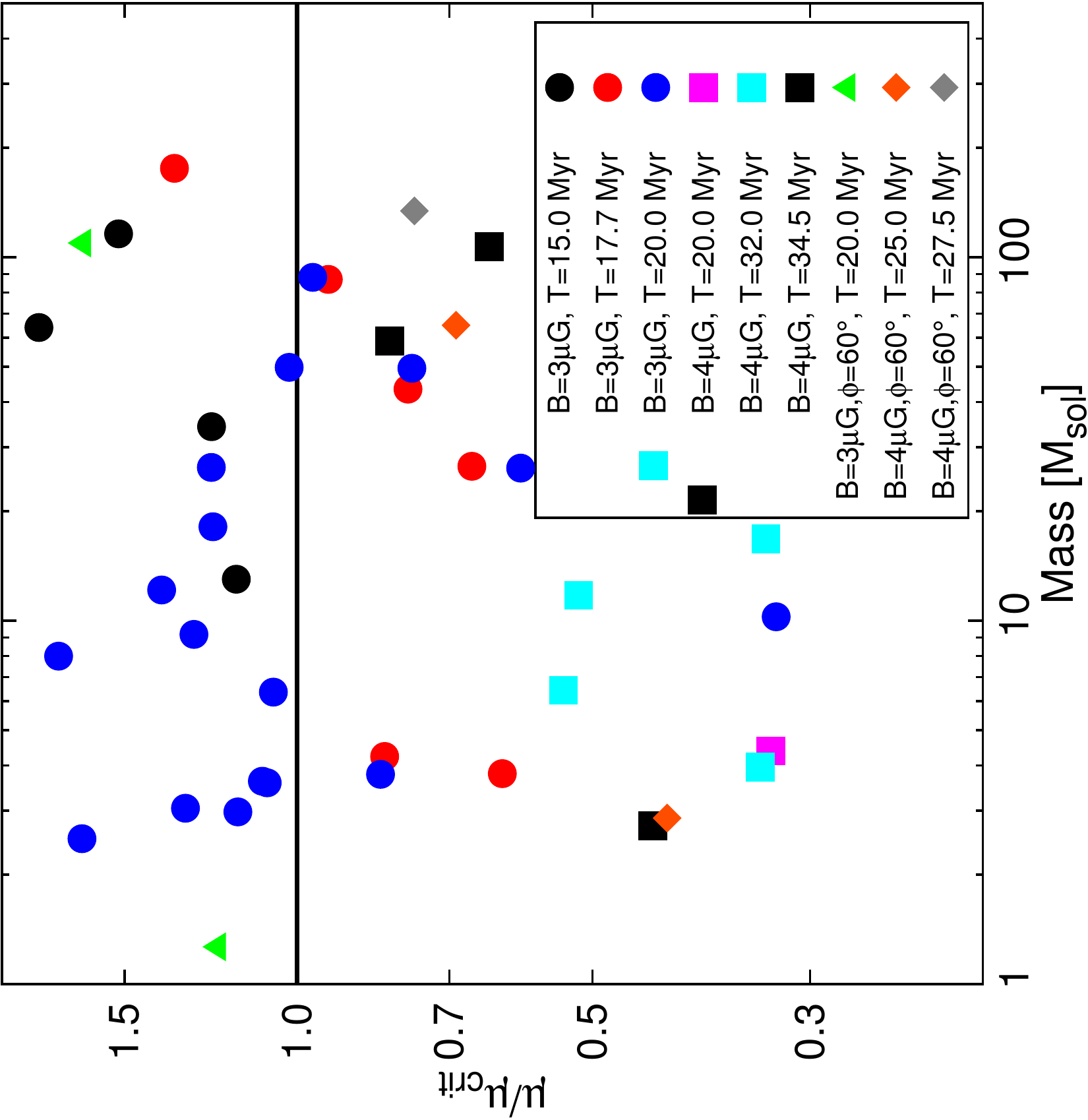}
 \caption{Normalised mass--to--flux ratio as function of core mass for 
  runs B3M0.4I0 (circles), B3M0.5I60 (triangles), B4M0.4I0 (squares), and B4M0.4I60 (diamonds) at different evolutionary stages. Note that a comparison of equal temporal stages is not possible due to the lack of cores 
  with densities of $n\geq1000\,\mathrm{cm}^{-3}$. For runs B5M$\dots$I$\dots$ no such cores were found! The 
mass--to--flux ratio is normalised to $\mu_c=0.16/\sqrt{G}$ \citep[][]{Nakano78}.}
  \label{fig15}
\end{figure}

Due to turbulence, overdensities occur which become gravitationally bound.
Figure \ref{fig14} shows the evolution of the densest regions within the formed molecular clouds, i.e. of these with minimum density of $n=1000\,\mathrm{cm}^{-3}$. The left image shows the evolution of mass, the right panel the evolution of thermal and magnetic Jeans numbers.
The temporal evolution is shown for the whole simulation, thus earlier ending graphs indicate the complete lack of gas with the respective minimum density from this time on.\\
For all runs with zero inclination, the compression by the two converging streams induces a transition to dense material with a few thousand solar masses. But as soon as the ram pressure of the confining flows becomes 
weaker these dense regions re--expand, showing that the regions were only pressure confined entities. For run B3M0.4I0 a phase of increasing mass follows, which is mainly due to 
accretion of matter along the magnetic field lines. In the end the densest regions of the molecular cloud reach a total mass of a few hundred solar masses. Comparison with run B3M0.5I60 shows a difference of only a factor 
of a few in the end of the simulation.\\
 The most striking difference is the first evolutionary phase, where the cores of run B3M0.5I60 undergo strong variations, because of the additional shearing motions, after they have 
firstly formed at far later times.\\ 
Run B4M0.5I0 already shows the influence of the stronger magnetic field. The decrease in external ram pressure by the converging flows also induces a re--expansion of the dense material within the cloud complex. But now 
the magnetic field is already strong enough to ensure a less efficient mass accretion. At around $t\approx 10\,$Myr \ita{no dense material exists}. This stage lasts until $t\approx 20\,$Myr, where the global collapse of the 
molecular cloud yielded strong enough compression to form dense material again\footnote{Note the linear increasing interval is simply the connecting line of two data points at 10 and 20 Myr.}. Strong internal variations of 
the cloud then lead to a highly varying mass evolution. In the end,
masses similar to run B3M0.4I0 are reached, and stars start to form. 
Interestingly it takes roughly 20 Myr for stars to form after the 
reoccurence of dense cores. This already indicates that for shearing 
flows the onset of star formation with magnetic and thermal energies in equipartition is further delayed to far later times. But during such a long evolution, the clouds would then be subject to large scale Galactic 
processes and our setup would not be appropriate.\\
Further increase of the initial magnetic field strength shows even more dramatic changes in the overall evolution of the densest regions within the molecular clouds. In run B5M0.5I0 the existence of dense cores ends after 
$t\approx 12\,$Myr, showing the complete lack of unstable cores after the compression by the bulk flows. Although a molecular cloud forms, it does not possess any region, which could possibly undergo gravitational contraction to form stars. 
However, the first evolutionary stages during the compression of the flows shows that the strong fields lead to higher masses of the dense gas due to 
the influence of the field. Inclining one WNM stream now shows striking difference. At first, the build up of dense cores starts out at later times as in the case of run B3M0.5I60. But the diffusive nature of this formation 
mechanism leads to the build--up of denser regions up to the end of the simulation, although there are stages where no dense cores exist. The evolutionary track of the dense gas is mainly influenced by the cloud 
motion and the magnetic forces. The whole dense material is thermally and magnetically highly stable, with the latter being the dominant aspect. Thus, although far more diffusive, the magnetic field is still able to 
suppress the formation of unstable cores and the subsequent star formation.\\
The stability of the dense regions is also indicated by the mass--to--flux ratio (see fig. \ref{fig15}). Only if the magnetic field is sufficiently weak, gravitational energy dominates over magnetic energy and the inner 
regions of the molecular clouds are rendered magnetically supercritical. In case of a 4$\mu$G field a great spread of mass--to--flux ratios is observed with all being subcritical at the stages shown. During the further 
evolution of the cloud, magnetically supercritical cores form. Additionally, by inclining one stream, low--mass cores are generated, which have approximately the same mass--to--flux 
ratio as in the case of head--on colliding streams, indicating that the magnetic field diffuses out of the regions.

\subsubsection{Analysis of the densest cores}
\begin{figure*}
 \includegraphics[width=0.3\textwidth,angle=-90]{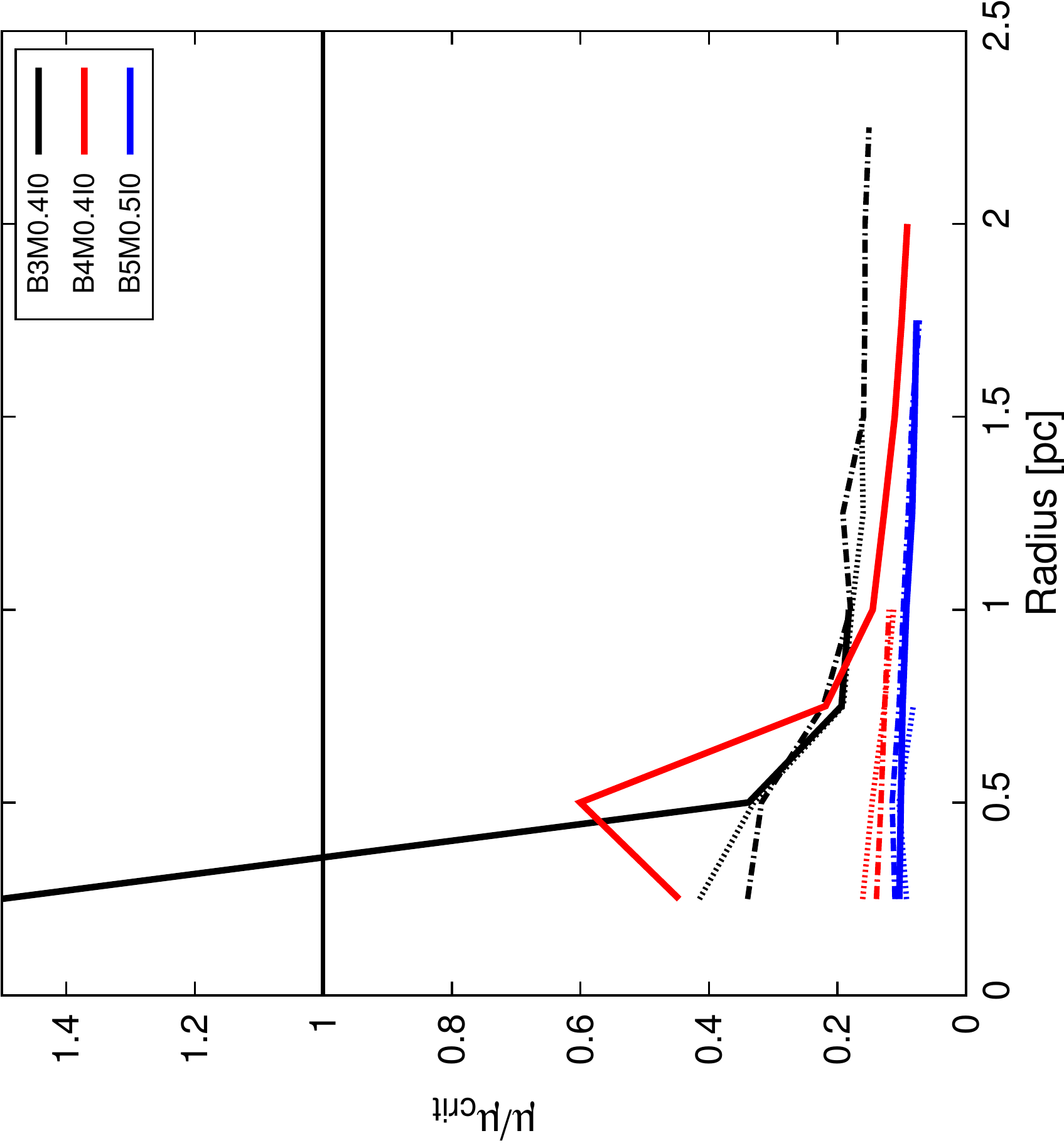}\includegraphics[width=0.3\textwidth,angle=-90]{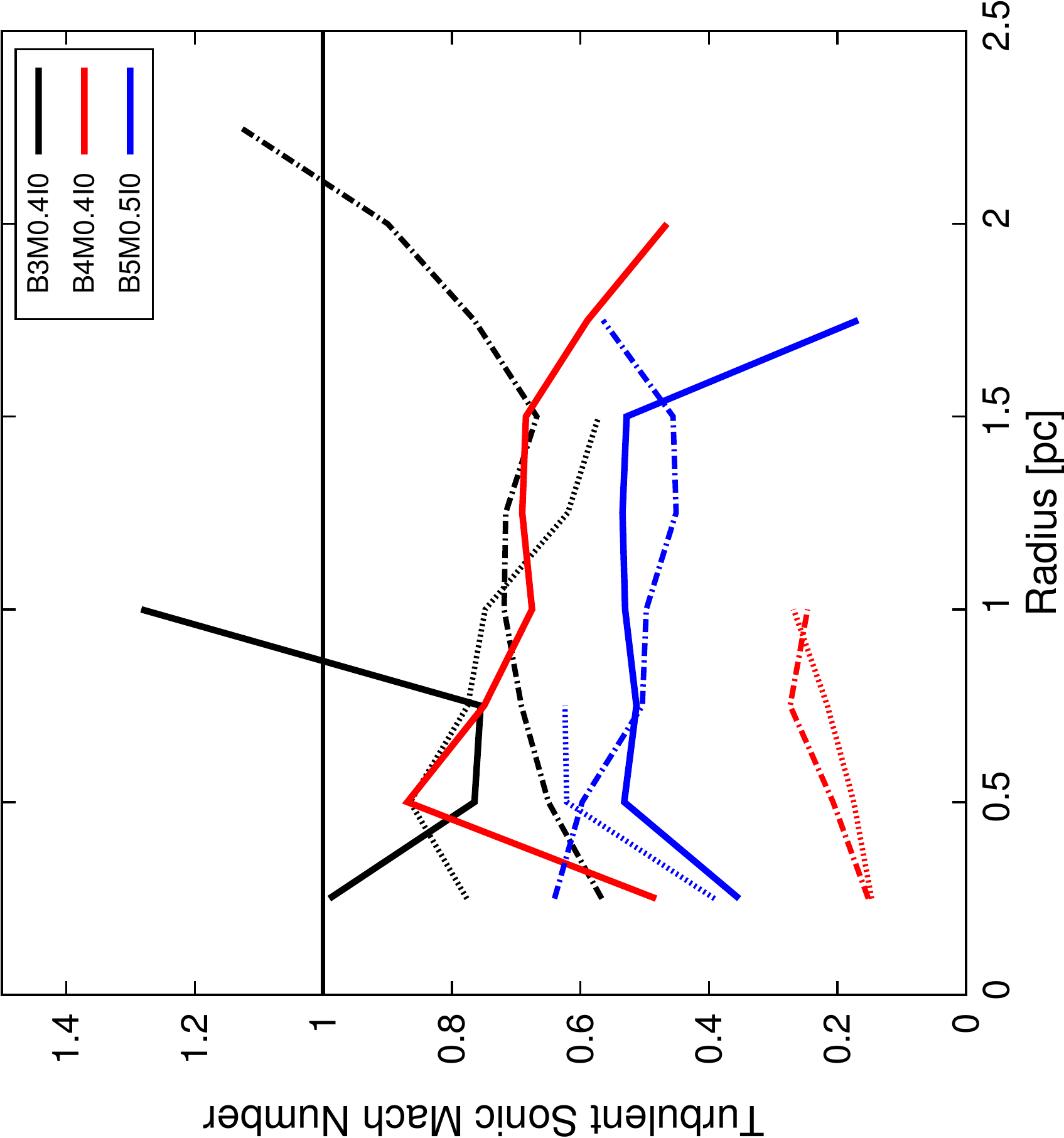}\includegraphics[width=0.3\textwidth,angle=-90]{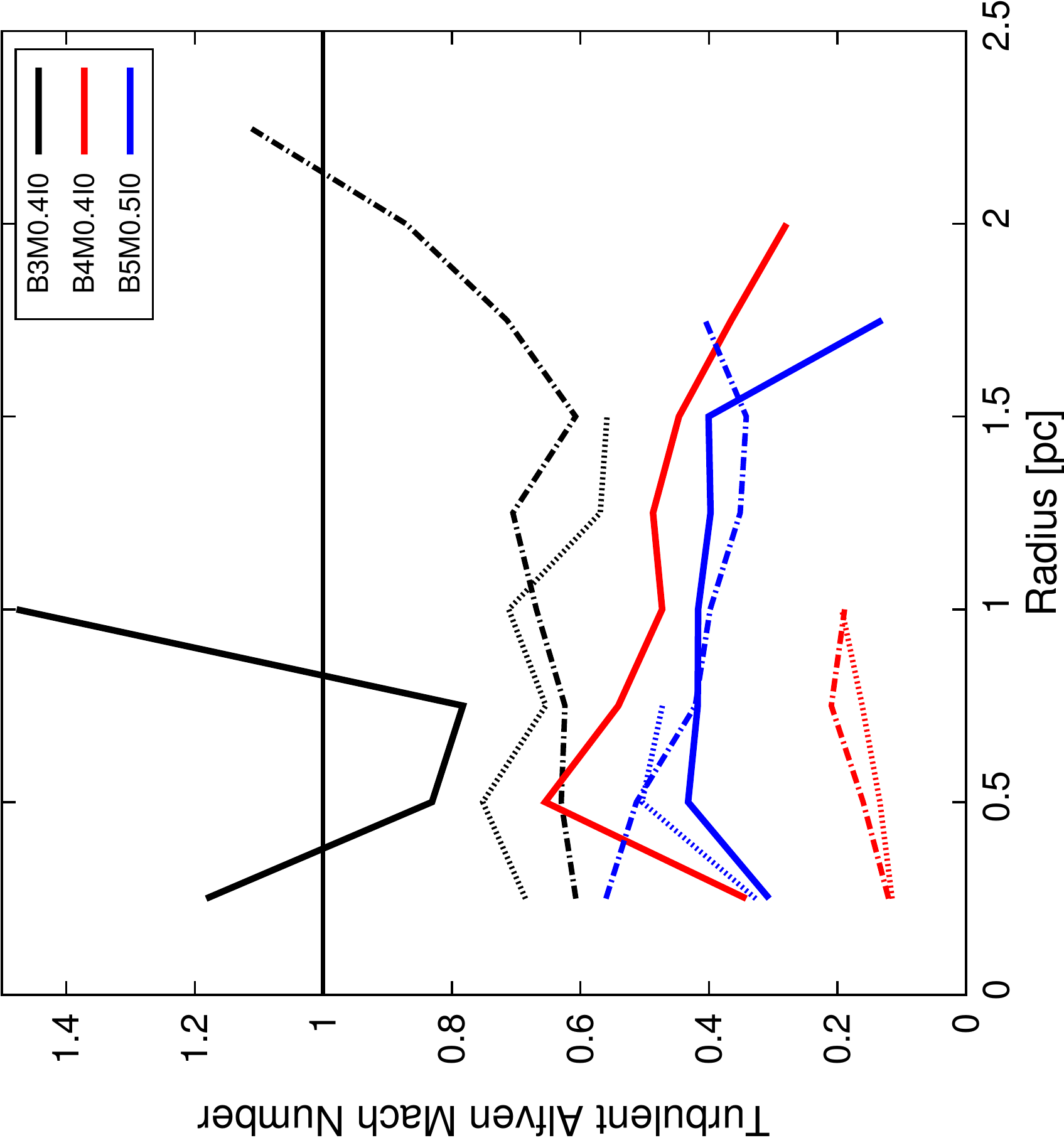}
 \caption{Radially averaged profiles of the three (indicated by different line styles) densest cores within the molecular clouds for runs B$\dots$M$\dots$I0. \ita{Left:} Mass--to--flux ratio. \ita{Middle:} Turbulent \ita{isothermal} Mach number. \ita{Right:} 
 Turbulent Alfv\'{e}n Mach number. The radial distance is evaluated with respect to the centre of mass. For runs B3M0.4I0 and B4M0.4I0, respectively, the data is shown shortly before the first star appears. The values for 
 the turbulent Mach numbers mimic the initial conditions, i.e. that the turbulence is still subsonic and subalfv\'{e}nic. Note, the masses of these cores range from $\approx$ 4\,M$_\odot$ (third massive in run B5M0.5I0) to 
 $\approx$ 250\,M$_\odot$ (most massive in run B3M0.4I0).}
 \label{fig16}
\end{figure*}

Now we analyse the three densest cores in more detail. For simplicity, 
we have assumed that the cores are spherical entities. Figure \ref{fig16} shows the normalised mass--to--flux ratio, the turbulent sonic and the turbulent Alfv\'{e}n Mach number as 
function of radial distance from the centre of mass.\\
For run B5M0.5I0 the mass--to--flux ratio is subcritical and constant throughout the whole core. This has two implications: 1) Magnetic support is sufficient to keep the core stable and 2) there is no 
evidence for accretion of matter along the field lines (which would increase the ratio locally). Besides being higher, the mass--to--flux ratio for run B3M0.4I0 shows some variation as function of radial distance for all 
three cores. The centre of the densest core (solid line) is seen to make a transition to a supercritical state surrounded by a subcritical halo. Although the other two cores are subcritical as a whole, they show the 
same signature. The cores in run B4M0.4I0 show a state between these of runs B3M0.4I0 and B5M0.5I0. As expected, 
the mass--to--flux ratio of the densest core decreases with increasing radius. The two other cores show 
only a roughly constant ratio. The mass--to--flux ratio of the first core is still subcritical, but the transition to a supercritical state is achieved at slightly later times. Note that the location of the maximum 
mass--to--flux ratio in this core does not coincide with the centre of mass.\\
The dynamics of the cores can be analysed by looking at the turbulent Mach numbers (see middle and right panel). The cores are \ita{subsonic} and \ita{subalfv\'{e}nic}, hence showing that 1) no strong compressions 
within the dense material occur and 2) the magnetic field prevents the gas from accumulating into denser unstable fragments. This is true for all clouds and can be interpreted as an imprint of the initial conditions. Only 
for the densest core in run B3M0.4I0, the outskirts are seen to be slighty supersonic and superalfv\'{e}nic, indicating turbulent accretion onto the core. The difference between the two Mach numbers is less for the 
weakest field, since it is not able to fully prevent collapse. If the field strength is higher, magnetic tension will accelerate the gas, while relaxing the field lines. This is why the sonic Mach number is slightly higher, 
but nevertheless the motions only reach subsonic or at most transsonic states. At these densities ($n\approx 10^3\,\mathrm{cm}^{-3}$), the flow seems to be mediated by the magnetic field lines, which in every case tends to suppress the build up of turbulent 
vortices.

\section{Summary \& Discussion}
\label{sec5}
In this study we have presented the results of MHD simulations of colliding flows with varying initial conditions. The strength of the turbulent velocity fluctuations, of the background magnetic field as well as the alignment 
of one of the WNM streams with the magnetic field was changed. We have
shown that dense clouds can form \ita{independent} of the initial conditions, but that their final 
mass and dynamics are mainly controlled by these (see
tab. \ref{tab:finstate}).
Increasing initial turbulence lead to lower cloud masses due to less
coherent gas streams. Oblique flows still lead to clouds with masses
comparable to what has been observed recently and stronger magnetic fields will generally lead 
to more massive molecular clouds. The first point seems at first a contradiction to \citet[][]{Inoue09} who stated that 
for larger inclined flows no dense, molecular clouds can form. However, here the cloud accretes mass and becomes 
molecular with time. As can be seen from fig. \ref{fig7}, the onset of the formation of dense gas is delayed with 
increasing inclination. This is indeed consistent with \citet[][]{Inoue09}, because the first few Myr are characterised by H\,I 
gas with densities below the threshold density of $n=100\,\mathrm{cm}^{-3}$. \\
Molecular clouds are able to condense out of the WNM, independent of
the magnetic field strength. However, only in the cases of fairly weak
initial magnetic fields, the formation of stars could be
initiated. 
Starting with subcritical HI flows, the magnetic  flux loss is in no cases sufficient to allow the build--up of supercritical cloud cores.
The tendency of the magnetic field to realign itself with the initial 
direction is a crucial factor for the overall evolution. In order to circumvent this problem, non--ideal MHD was resembled by 
means of tilted collisions. Increasing inclination leads to increased diffusivity of the magnetic field. The variation of the inclination as well as the flow dynamics showed no 
tendency for faster accumulation of gas or faster transition to thermally dominated regions, since the flow dynamics is rather controlled by the appearing shear flows than magnetic diffusion. \\ 
\begin{table*}
 \caption{Typical cloud parameters for gas with $n\geq100\,\mathrm{cm}^{-3}$ at the end of each simulation with equal turbulent and flow Mach number.}
 \begin{tabular}{p{0.5cm} p{0.5cm} p{1.cm} p{1.7cm} p{0.8cm} p{1.7cm} p{0.8cm} p{1.6cm} p{2.5cm} p{1.3cm}}
  \hline
  \hline
  $\phi$ 	&$\left|\vek{\mathrm{B}}\right|$ & Time &Cloud Mass	&SF? 	&Stellar Mass	&SFE	&E$_\mathrm{mag}$	&Velocity Dispersion 	&N$_\mathrm{jeans,mag}$\\
  
  $\left(^{\circ}\right)$ &$\left(\mu \mathrm{G}\right)$ &$\left(\mathrm{Myr}\right)$&$\left(10^{3}\mathrm{M}_\odot\right)$ &$\left(\mathrm{yes/no}\right)$ &$\left(\mathrm{M}_\odot\right)$ &$\left(\%\right)$ &$\left(10^{46}\mathrm{erg}\right)$ 
  &$\left(\mathrm{km/s}\right)$ &\\
  \hline
   0		&3 &21.79&40.81&yes&1346 &3.2&23.59&0.99&23.99\\
  30		&3 &18.29&21.84&yes&202  &0.9&12.28&1.29&12.51\\
  50		&3 &18.94&14.5 &yes&14.20&0.1&6.39 &0.73&9.86\\
  50$^{a}$	&3 &21.04&17.9 &yes&43.83&0.2&7.09 &0.84&14.28\\
  60		&3 &21.49&15.2 &yes&95.68&0.6&7.02 &1.45&10.68\\
  \hline
   0		&4 &39.52&18.33&yes&59.32&0.3&13.63&0.53&5.70\\
  30		&4 &27.13&23.93&no&------&------&21.53&1.21&6.30\\
  60		&4 &27.28&16.15&no&------&------&12.29&1.74&6.29\\
  \hline
   0		&5 &24.64&43.29&no&------&------&52.71&0.46&6.10\\	
  30		&5 &22.94&20.39&no&------&------&22.94&1.02&3.07\\
  40		&5 &37.93&11.56&no&------&------&12.08&1.07&2.16\\
  50		&5 &37.03&17.57&no&------&------&19.63&1.42&3.39\\
  60		&5 &31.93&7.31 &no&------&------&6.48 &1.79&1.55\\
  \hline
  \hline
 \end{tabular}\\
\label{tab:finstate}
\small{$a$: Different initial random seed for the turbulence.}
\end{table*}
We therefore stress the role of magnetic fields in the context of molecular cloud and star formation. We point 
out the complete lack of supercritical regions for realistic initial field strengths. As was shown in fig. \ref{fig15}, the 
normalised mass--to--flux ratio ranges from 0.3--1.7 ($B=3\mu\mathrm{G}$) and 0.3--0.7 ($B=4\mu\mathrm{G}$), 
respectively. At least the former case for weak fields compares well with the results of \citet[][$\mu/\mu_c\sim0.5-7.5$]{Chen14} as 
well as with observations \citep[$\mu/\mu_c\approx2$,][]{Troland08}. From the observational side, HI clouds may be supercritical as a whole, but their observed, dense subregions be subcritical.\\
The question remains, how clouds achieve the transition from sub-- to supercritical.\\

\section*{Acknowledgements}
BK and RB thank Enrique V\'{a}zquez--Semadeni for useful discussions and the referee for invaluable comments, 
which helped to improve the quality of this paper. BK acknowledges hospitality at Centro de Radioastronom\'{i}a y Astrof\'{i}sica, Universidad Nacional Aut\'{o}noma M\'{e}xico, 
during the initial stages of this study. 
The simulations were run on HLRN--III under project grant hhp00022. RB
acknowledges funding by the DFG via the Emmy-Noether grant BA
3706/1-1, the ISM-SPP 1573 grants BA 3706/3-1
and BA 3706/3-2, as well as for the grant BA 3706/4-1.
The software used in this work was in part developed by the DOE--supported ASC/Alliance Center for Astrophysical 
Thermonuclear Flashes at the University of Chicago.

\bibliographystyle{aa}
\bibliography{astro}

\label{lastpage}

\begin{appendix}
\section{Estimate of the effective magnetic diffusion}
\label{appa}
\begin{figure}
 \includegraphics[height=8cm,angle=-90]{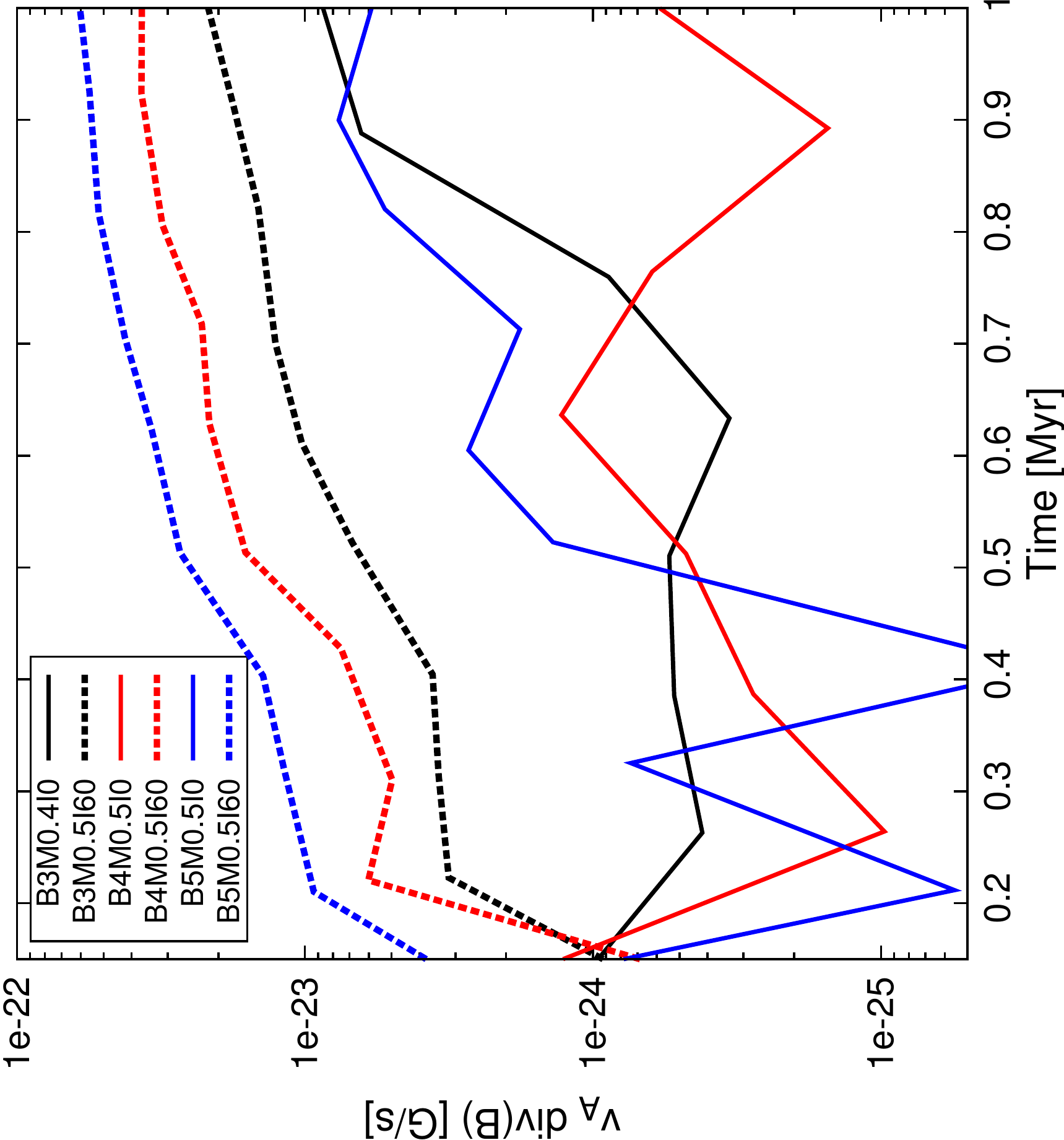}
 \caption{Shown is the deviation from the induction equation due to diffusion in the ideal MHD limit for inclinations of 0$^{\circ}$ (solid) and 60$^{\circ}$ (dashed). Different colours indicate different initial magnetic 
 field strengths.}
\label{figA1}
\end{figure}

We here give a simple estimate for the dependence of the ambipolar diffusion coefficient on the magnetic field 
strength and the inclination of the flow. Specific numerical values are not of special interest here. \\
The AD diffusion parameter is $\eta_\mathrm{AD}\propto B^2$.
The respective numerical diffusivity is given by the product of the grid size and the Alfv\'{e}n speed. The ratio 
$\eta_\mathrm{AD}/\eta_\mathrm{num}$ estimates the influence of magnetic to numerical diffusion. In order to receive the diffusion for 
tilted flows, we make use of the fact that inclined flows will generate tilted field lines. This tilting can be interpreted as impact of magnetic diffusion. The perturbed field $\delta B$ (which is here important for the 
diffusion coefficient) and the initial background field 
are then related by $\delta B=B_0sin\left(\varphi\right)$, where the tilting angle of the perturbed field, $\chi$, and the inclination of the flows, $\varphi$, are related by $\chi=90^{\circ}-\varphi$. The resultant ambipolar diffusion 
coefficient is then modified to $\eta_\mathrm{AD}\propto B_0^{2}\sin^{2}\left(\varphi\right)$, which gives zero diffusion for aligned flows and maximum diffusion for the perpendicular case (see fig. \ref{figA1}).

\end{appendix}

\end{document}